\documentclass{article}

\usepackage{amsmath,amssymb,amsthm}
\usepackage{dsfont}
\usepackage{graphicx}
\usepackage{geometry}
\usepackage[british]{babel}
\usepackage[utf8]{inputenc}
\usepackage{xcolor}
\usepackage{cite}
\usepackage{url}
\usepackage{diagbox}
\usepackage{caption}
\usepackage{subcaption}
\usepackage{rotating}

\allowdisplaybreaks

\def\vs{\lambda_S}
\def\vi{\lambda_I}
\def\vr{\lambda_R}
\def\ve{\lambda_E}
\def\va{\lambda_A}

\def\RR{\mathbb{R}}
\newcommand{\f}{\mathbf{f}}
\newcommand{\g}{\mathbf{g}}

\newcommand{\x}{\mathbf{x}}
\newcommand{\Q}{\mathbf{Q}}
\newcommand{\F}{\mathbf{F}}
\renewcommand{\S}{\mathbf{S}}

\newcommand{\SO}{S_u}
\newcommand{\IO}{I_u}
\newcommand{\EO}{E_u}
\newcommand{\AO}{A_u}
\newcommand{\RO}{R_u}
\newcommand{\DD}{D^u}

\newcommand{\U}{{\mathbf{U}}}

\renewcommand{\u}{{\bf{u}}}
\renewcommand{\v}{{\bf{v}}}

\newcommand{\E}{{\mathbf{E}}}
\newcommand{\D}{{\mathbf{D}}}
\newcommand{\J}{{\mathbf{J}}}

\begin{document}
\title{Spatial spread of COVID-19 outbreak in Italy using multiscale kinetic transport equations with uncertainty}

\author{Giulia Bertaglia$^*$, Walter Boscheri$^*$, Giacomo Dimarco$^*$, Lorenzo Pareschi\footnote{$^*$Department of Mathematics and Computer Science, University of Ferrara, Via Machiavelli 30 and Center for Modeling, Computing and Statistic CMCS, University of Ferrara, Via Muratori 9, 44121 - Ferrara, Italy}}

\maketitle
\begin{abstract}{In this paper we introduce a space-dependent multiscale model to describe the spatial spread of an infectious disease under uncertain data with particular interest in simulating the onset of the COVID-19 epidemic in Italy. While virus transmission is ruled by a SEIAR type compartmental model, within our approach the population is given by a sum of commuters moving on a extra-urban scale and non commuters interacting only on the smaller urban scale. A transport dynamic of the commuter population at large spatial scales, based on kinetic equations, is coupled with a diffusion model for non commuters at the urban scale. Thanks to a suitable scaling limit, the kinetic transport model used to describe the dynamics of commuters, within a given urban area coincides with the diffusion equations that characterize the movement of non-commuting individuals. Because of the high uncertainty in the data reported in the early phase of the epidemic, the presence of random inputs in both the initial data and the epidemic parameters is included in the model.
A robust numerical method is designed to deal with the presence of multiple scales and the uncertainty quantification process.
In our simulations, we considered a realistic geographical domain, describing the Lombardy region, in which the size of the cities, the number of infected individuals, the average number of daily commuters moving from one city to another, and the epidemic aspects are taken into account through a calibration of the model parameters based on the actual available data. 
The results show that the model is able to describe correctly the main features of the spatial expansion of the first wave of COVID-19 in northern Italy.
	}
\end{abstract}

\smallskip


\tableofcontents

\section{Introduction}
The advent of the COVID-19 pandemic has caused a strong commitment of many researchers acting in different fields with the scope of trying to understand and give explanations to the global crisis we are experiencing. From the mathematical modeling point of view, several progresses have been done concerning the development of epidemic models capable of taking into account the different facets of this terrible disease. In particular, many recent researches have been addressed to the search of control strategies \cite{APZ,APZ2,Tang,Gatto} to limit the spread and consequently hospitalizations and deaths, possibly reducing, at the same time, as much as possible the negative impact on the economy of the restrictive measures \cite{Ash, DPTZ}.

Most of the recently proposed model represent improvements at various levels of the seminal works on compartmental epidemiological modeling proposed originally by Kermack and McKendrick \cite{HWH00}. These approaches \cite{BGRCV,Buonomo, CV, Franco2020,Giordano,APZ,Tang,Loli} are typically focused on the epidemiological aspects of the virus spread under the hypothesis of global mixing of the population, hence, without taking into account the role of the spatial component in the evolution of an outbreak. 

Despite in many situations the above description is sufficient to delineate the global trend of an epidemic, there are cases in which the spatial homogeneity assumption does not hold true. In these situations, one seeks for local interventions in order to reduce the spread without eventually affecting regions where the incidence of the infection does not require special care, as for instance in the recent case of the COVID-19 in Italy \cite{legal}. Consequently, from the modelling point of view, the inclusion of the spatial dependence represents a key challenge \cite{REIM}.

Indeed, with the increasing amount of information on population mobility and the computational resources available today, the design and simulation of epidemic models based on partial differential equations (PDEs) that include the details of spatial dynamics can be considered a realistic goal. We recall that most of the existing epidemiological models taking into account spatial heterogeneities are based on reaction-diffusion equations \cite{ABLN, Cap, FMW, Liu, MWW,KGV, Sun, Veneziani2021, Veneziani2020, Wang2020}. Alternative modelling approaches are represented by the interaction of different homogeneous populations \cite{Gatto} or agent-based dynamics \cite{FWF}. 

Most of these models do not allow for a clear distinction about the possible spatial behaviors of the population inside a given compartment. Consequently, although capable of originating realistic spatial patterns in situations where individuals move indiscriminately through the domain, such approaches are likely to be less effective in the case where one is interested in studying the spread of a virus in the human population. Under these circumstances, it is more realistic to consider only commuting individuals moving in major and preferred directions, not considering overall mass migration between distinct urban areas. Indeed, not all individuals move indiscriminately in the region of interest, as most of the population only interacts at the urban scale. Furthermore, in contrast to the use of diffusion models, the propagation rate of the infection along the spatial domain is obviously finite. Recently, trying to overcome some of the above criticisms, such as the paradox of the infinite propagation speed, alternative models based on hyperbolic PDEs have been proposed \cite{BCV13, Bert,Bert2,Colombo,BDP}. 

In this work, following the approach introduced in \cite{BDP}, we consider a realistic compartmental structure for the description of the epidemic dynamic of commuters and non commuters individuals in presence of uncertain data. The model consists of a system of kinetic transport equations describing a large-scale (extra-urban) commuting population by a continuous density \cite{cs, CMPS, HS, Per} in a two-dimensional environment. This density can be interpreted as the probability for an individual to be at a given location and move in a given direction at a given instant of time, in analogy to particle flows in rarefied gases \cite{PT13,bellomo2020multiscale, Cer, Deli, PuSa, RS}. The above system is coupled to a second system consisting of a set of diffusion equations that characterize the movement of the non-commuting population on a small (urban) scale. The epidemic spread is ruled by a Susceptible, Exposed, Infected, Asymptomatic and Removed (SEIAR) compartmental structure, in which both commuters and non commuters can interact. Using a suitable scaling process \cite{LK, BDP}, the model allows us to highlight, firstly, the relationship with well-known existing approaches based on reaction-diffusion equations and, secondly, to pass naturally from a hyperbolic description in peripheral areas to a diffusive regime when reaching an urban area, with regard to the commuting population.

An important aspect concerns overcoming the limitation caused by standard deterministic models that rely on the assumption that the initial conditions, boundary conditions, and all involved epidemic and mobility parameters are known. However, as observed in the case of the COVID-19 epidemic, this assumption, especially in the early stages of the epidemic, is not reliable. For example, the initial conditions in terms of the number of infected and asymptomatic persons are certainly affected by uncertainty because data are limited and population screening cannot be error-free. Epidemic parameters, although normally estimated or calibrated, are also often candidates for being random variables in a realistic approach. Therefore, in order to take into account these limitations underlying deterministic models, in this paper we resort to a stochastic approach based on the introduction of random terms into the initial modeling \cite{APZ,APZ2, Bert2,Peirlinck}.
 
Once the model is set up, its numerical solution on a computational domain describing a realistic geographic scenario poses several difficulties.  In fact, the model consists of two coupled systems of PDEs, each characterized by five unknown functions living in a multidimensional space characterized by space, velocity, and stochastic variables. Additionally, we have to deal with the irregular shape of the spatial region and the multiscale nature of the dynamics (indeed, as previously stated, hyperbolic and parabolic behaviors coexist). Therefore, a particular care is devoted to the development of efficient and accurate numerical schemes. More precisely, a discretization of the system based on Gaussian quadrature points in velocity space \cite{GJL, JPT} and a finite volume approach on unstructured grids \cite{Dumbser2007693, ArepoTN} is considered. The adoption of asymptotic preserving time discretization techniques permits to avoid time step limitations introduced by the parabolic scaling without degradation of accuracy \cite{Bos1, Bos2, DP}. Finally, a non-intrusive stochastic collocation method, which guarantees spectral accuracy in the space of the uncertain parameters, is considered to deal with the uncertainty quantification process \cite{X, Bert2}.

The rest of the paper is organized as follows. In Section \ref{sect:mathmodel} the mathematical model is introduced. We first introduce the kinetic transport formulation for the epidemic compartments of commuters together with the corresponding diffusive dynamic of the non commuters in a deterministic setting. Subsequently, we link the two hyperbolic/parabolic dynamics through a formal passage to the limit for the system of commuters. A definition of the basic reproduction number of the epidemic for the resulting model is also reported. Next, we illustrate how to generalize the model in presence of uncertainty. The details of the numerical scheme used to approximate the resulting stochastic system are summarized in Appendix \ref{appendix:NM}. Section \ref{sect:numres} is devoted to present an application of the current modelling to the first outbreak of COVID-19 in Italy and its spread in the Lombardy Region. The capability of the model to accurately represent the first wave of the COVID-19 epidemic in Italy is discussed in detail through comparisons with recorded data reported by official sources \cite{prot_civile}. Conclusions and future perspectives are finally given in Section \ref{Conc}.

\section{A compartmental kinetic transport model}\label{sect:mathmodel}
Let $ \Omega\subset \RR^2$ characterize a two-dimensional geographical area of interest and assume that individuals have been separated into two different groups: a commuting population, typically moving over long distances (extra-urban), and a non-commuting population, moving only in small-scale urban areas. In the first part of this Section, for  ease of presentation, the  multiscale kinetic transport model is introduced in a deterministic setting. The relation between the current hyperbolic transport model and classical diffusion models is discussed in the second part. In the third part, details regarding the basic reproduction number associated with the model are given.  Finally, in the last part, we discuss the generalization of the deterministic model to the case where uncertainty is taken into account.
\subsection{Characterizing commuter and non commuter dynamics}
We consider a population of commuters at position $x\in\Omega$ moving with velocity directions $v \in \mathbb{S}^1$ and denote the respective kinetic densities of susceptible (individuals who may be infected by the disease) by $f_S=f_S(x,v,t)$, exposed (individuals in the latent period, which are not yet infectious) by $f_E=f_E(x,v,t)$, severe symptomatic infected by $f_I=f_I(x,v,t)$, mildly symptomatic or asymptomatic infected by $f_A=f_A(x,v,t)$ and removed (individuals healed or died due to the disease) by $f_R=f_R(x,v,t)$. 
The kinetic distribution of commuters is then given by
\[
f(x,v,t)=f_S(x,v,t)+f_E(x,v,t)+f_I(x,v,t)+f_A(x,v,t)+f_R(x,v,t),
\]
and their total density is obtained by integration over the velocity space 
\[
\rho(x,t)=\frac1{2\pi}\int_{\mathbb{S}^1} f(x,v_*,t)\,dv_*.
\]
As a consequence, one can recover the number of susceptible, exposed and recovered irrespective of their direction of displacement by integration over the velocity space. This gives
\[
S(x,t)=\frac1{2\pi}\int_{\mathbb{S}^1}  f_S(x,v,t)\,dv,\,\, E(x,t)=\frac1{2\pi}\int_{\mathbb{S}^1}  f_E(x,v,t)\,dv,\,\, R(x,t)=\frac1{2\pi}\int_{\mathbb{S}^1}  f_R(x,v,t)\,dv,
\]
which we refer to as the density fractions of non-infectious individuals, whereas
\[
I(x,t)=\frac1{2\pi}\int_{\mathbb{S}^1}  f_I(x,v,t)\,dv,\,\, A(x,t)=\frac1{2\pi}\int_{\mathbb{S}^1}  f_A(x,v,t)\,dv,
\]
are the density fractions of infectious individuals.

In this setting, the kinetic densities of the commuters satisfy the transport equations
\begin{eqnarray}
\nonumber
\frac{\partial f_S}{\partial t} + v_S \cdot \nabla_x f_S &=& -F_I(f_S, I_T)-F_A(f_S, A_T) +\frac1{\tau_S}\left(S-f_S\right)\\
\nonumber
\frac{\partial f_E}{\partial t} + v_E \cdot\nabla_x f_E &=&  F_I(f_S, I_T)+F_A(f_S, A_T)-a f_E+\frac1{\tau_E}\left(E-f_E\right)\\
\label{eq:kineticc}
\frac{\partial f_I}{\partial t} + v_I \cdot\nabla_x f_I &=& a\sigma f_E -\gamma_I f_I+\frac1{\tau_I}\left(I-f_I\right)\\
\nonumber
\frac{\partial f_A}{\partial t} + v_A \cdot\nabla_x f_A &=& a(1-\sigma) f_E -\gamma_A f_A+\frac1{\tau_A}\left(A-f_A\right)\\
\nonumber
\frac{\partial f_R}{\partial t} + v_R \cdot\nabla_x f_R &=& \gamma_I f_I+\gamma_A f_A+\frac1{\tau_R}\left(R-f_R\right)
\end{eqnarray}
where the total densities of non infected individuals are defined by
\[
S_T(x,t)=S(x,t)+\SO(x,t),\,\,\, E_T(x,t)=E(x,t)+\EO(x,t),\,\,\, R_T(x,t)=R(x,t)+\RO(x,t),
\]
and similarly the total densities of infected by
\[
I_T(x,t)=I(x,t)+\IO(x,t),\,\,\, A_T(x,t)=A(x,t)+\AO(x,t).
\]
In the above equations, $\SO(x,t)$, $\EO(x,t)$, $\IO(x,t)$, $\AO(x,t)$, $\RO(x,t)$ are the density fractions of non-commuters who, by assumption, move only on an urban scale. These densities satisfy a diffusion dynamic acting only at the same local scale
\begin{eqnarray}
\nonumber
\frac{\partial \SO}{\partial t} &=& -F_I(\SO, I_T) -F_A(\SO, A_T) + \nabla_x\cdot ({\DD_S}\nabla_x \SO) \\
\nonumber
\frac{\partial \EO}{\partial t} &=& F_I(\SO, I_T) + F_A(\SO, A_T) -a\EO + \nabla_x \cdot({\DD_E}\nabla_x \EO) \\
\label{eq:diffuse}
\frac{\partial \IO}{\partial t}  &=&  a\sigma\EO-\gamma_I \IO+\nabla_x\cdot ({\DD_I}\nabla_x \IO)\\
\nonumber
\frac{\partial \AO}{\partial t}  &=&  a(1-\sigma)\EO-\gamma_A \AO+\nabla_x\cdot ({\DD_A}\nabla_x \AO)\\
\nonumber
\frac{\partial \RO}{\partial t}  &=& \gamma_I \IO+\gamma_A\AO+\nabla_x\cdot ({\DD_R}\nabla_x \RO).
\end{eqnarray}
In the resulting model \eqref{eq:kineticc}-\eqref{eq:diffuse} that couples the commuting and non-commuting dynamics, the velocities $v_i=\lambda_i(x) v$  in \eqref{eq:kineticc}, as well as the diffusion coefficients $\DD_i=\DD_i(x)$ in \eqref{eq:diffuse}, with $i\in\{S,E,I,A,R\}$, are designed to take into account the heterogeneity of geographical areas, and are thus chosen dependent on the spatial location. Similarly, also the relaxation times $\tau_i=\tau_i(x)$, $i\in\{S,E,I,A,R\}$ are space dependent. 
The quantities $\gamma_I=\gamma_I(x)$ and $\gamma_A=\gamma_A(x)$ are the recovery rates of symptomatic and asymptomatic infected (inverse of the infectious periods), respectively, while $a(x)$ represents the inverse of the latency period and $\sigma(x)$ is the probability rate of developing severe symptoms \cite{Tang,Gatto,Buonomo}.

The transmission of the infection is governed by the incidence functions $F_I(\cdot,I_T)$ and $F_A(\cdot,A_T)$. We assume local interactions to characterize the nonlinear incidence functions \cite{KM05, CS78}
\begin{equation}
F_I(g,I_T)=\beta_I \frac{g I_T^p}{1+\kappa_I I_T},\qquad F_A(g,A_T)=\beta_A \frac{g A_T^p}{1+\kappa_A A_T},
\label{eq:incf}
\end{equation}
where the classic bi-linear case corresponds to $p = 1$, $k_I=k_A=0$. 
Parameters $\beta_I=\beta_I(x,t)$ and $\beta_A=\beta_A(x,t)$ characterize the contact rates of highly symptomatic and mildly symptomatic/asymptomatic infectious individuals, accounting for both the number of contacts and the probability of transmission. Hence, they may vary based on the effects of government control actions, such as wearing of masks, shutdown of specific activities or lockdowns \cite{HWH00,Giordano,APZ}. On the other hand, parameters $\kappa_I=\kappa_I(x,t)$ and $\kappa_A=\kappa_A(x,t)$ are the incidence damping coefficients based on the self-protective behavior assumed by the individuals due to the awareness of the epidemic risk \cite{Bert2,Franco2020,Wang2020}. 

Alternative incidence functions are given by
\begin{equation}
	F_I(g,I_T)=\beta_I \frac{g I_T^p}{1+\kappa_I \int_{\bar\Omega}I_T \, dx},\qquad F_A(g,A_T)=\beta_A \frac{g A_T^p}{1+\kappa_A \int_{\bar\Omega} A_T\, dx},
	\label{eq:incf1}
\end{equation}
where $\bar\Omega$ is a chosen portion of the domain which permits to take into account the fact that social distancing may depend on the average level of infection of the region rather than only on the local situation. 
The resulting model \eqref{eq:kineticc}-\eqref{eq:diffuse} will be referred to as multiscale kinetic SEIAR (MK-SEIAR) model. Note that, because of the presence of two populations acting at different scales, the model allows a more realistic description of the typical commuting dynamic involving only a fraction of the population and distinguishes it from the epidemic process affecting the entire population.

\subsection{Commuters behavior in urban areas}
The hyperbolic transport model for the commuters deserves some remarks. In fact, while it is clear that a hyperbolic description permits to describe correctly the daily extra-urban commuting part, the same individuals when moving inside the urban area are better described by a traditional diffusion model. 
A remarkable feature of the transport model \eqref{eq:kineticc}, is that it permits to recover a classical diffusion behavior under the hypothesis that the relaxation times $\tau_{S,I,R}$ tend to zero while keeping finite the diffusion coefficients
\begin{equation}
	D_S=\frac12\lambda_S^2\tau_S,\quad D_E=\frac12\lambda_E^2\tau_E,\quad D_I=\frac12\lambda_I^2\tau_I,\quad D_A=\frac12\lambda_A^2\tau_A,\quad D_R=\frac12\lambda_R^2\tau_R.
	\label{eq:diffcf}
\end{equation}
More precisely, 
let us introduce the flux functions
\[
\begin{split}
J_S=\frac{\lambda_S}{2\pi} \int_{\mathbb{S}^1} & v f_S(x,v,t)\,dv,\quad J_E=\frac{\lambda_E}{2\pi}\int_{\mathbb{S}^1}  v f_E(x,v,t)\,dv,\quad J_I=\frac{\lambda_I}{2\pi}\int_{\mathbb{S}^1}  v f_I(x,v,t)\,dv\\
&J_A=\frac{\lambda_A}{2\pi} \int_{\mathbb{S}^1}  v f_A(x,v,t)\,dv,\quad J_R=\frac{\lambda_R}{2\pi}\int_{\mathbb{S}^1}  v f_R(x,v,t)\,dv.
\end{split}
\]
Then, integrating system \eqref{eq:kineticc} in $v$, we get the following set of equations for the macroscopic densities of commuters
\begin{eqnarray}
\nonumber
\frac{\partial S}{\partial t} + \nabla_x\cdot J_S &=& -F_I(S, I_T)-F_A(S, A_T)\\
\nonumber
\frac{\partial E}{\partial t} + \nabla_x\cdot J_E &=& F_I(S, I_T)+F_A(S, A_T)-aE\\
\label{eq:density}
\frac{\partial I}{\partial t} + \nabla_x\cdot J_I &=& a\sigma E -\gamma_I I\\
\nonumber
\frac{\partial A}{\partial t} + \nabla_x\cdot J_A &=& a(1-\sigma) E -\gamma_A A\\
\nonumber
\frac{\partial R}{\partial t} + \nabla_x\cdot J_R &=& \gamma_I I+\gamma_A A
\end{eqnarray}
whereas the flux functions satisfy
\begin{eqnarray}
\nonumber
\frac{\partial J_S}{\partial t} +  \frac{\vs^2}{2\pi} \int_{\mathbb{S}^1}  (v\cdot \nabla_x f_S)v\,dv &=& -F_I(J_S, I_T)-F_A(J_S,A_T)-\frac1{\tau_S} J_S\\
\nonumber
\frac{\partial J_E}{\partial t} +  \frac{\ve^2}{2\pi} \int_{\mathbb{S}^1}  (v\cdot \nabla_x f_E)v\,dv &=& \frac{\lambda_E}{\lambda_S}\left(F_I(J_S, I_T)+F_A(J_S,A_T)\right)-a J_E -\frac1{\tau_E} J_E\\
\label{eq:flux}
\frac{\partial J_I}{\partial t} +  \frac{\vi^2}{2\pi} \int_{\mathbb{S}^1}  (v\cdot \nabla_x f_I)v\,dv &=& \frac{\lambda_I}{\lambda_E}a\sigma J_E - \gamma_I J_I-\frac1{\tau_I} J_I\\
\nonumber
\frac{\partial J_A}{\partial t} +  \frac{\va^2}{2\pi} \int_{\mathbb{S}^1}  (v\cdot \nabla_x f_A)v\,dv &=& \frac{\lambda_A}{\lambda_E}a(1-\sigma) J_E - \gamma_A J_A-\frac1{\tau_A} J_A\\
\nonumber
\frac{\partial J_R}{\partial t} +  \frac{\vr^2}{2\pi} \int_{\mathbb{S}^1}  (v\cdot \nabla_x f_R)v\,dv &=& \frac{\lambda_R}{\lambda_I} \gamma_I J_I+\frac{\lambda_R}{\lambda_A} \gamma_A J_A
-\frac1{\tau_R} J_R.
\end{eqnarray}
Clearly, the above system is not closed because the evolution of the fluxes in \eqref{eq:flux} involves higher order moments of the kinetic densities. The diffusion limit can be formally recovered, by introducing the space dependent diffusion coefficients \eqref{eq:diffcf}
and letting $\tau_{S,I,R} \to 0$. We get from the r.h.s. in \eqref{eq:kineticc}
\[
\begin{split}
&f_S=S,\quad f_E=E,\quad f_I=I ,\quad f_A=A,\quad f_R=R,
\end{split}
\]
and, consequently, from \eqref{eq:flux} we recover Fick's law
\begin{equation}
J_S = -{D_S} \nabla_x S,\quad  J_E = -{D_E}\nabla_x E,\quad J_I = -{D_I}\nabla_x I,\quad J_A = -{D_A}\nabla_x A,\quad J_R = -{D_R} \nabla_x R,
\label{eq:flick}
\end{equation}
since
\[
\int_{\mathbb{S}^1}  (v\cdot \nabla_x S)v\,dv =
\int_{\mathbb{S}^1}  (v\otimes v)\,dv \nabla_x S  = \pi \nabla_x S\,,
\]
and similarly for the other densities. Thus, substituting \eqref{eq:flick} into \eqref{eq:density} we get the diffusion system for the population of commuters \cite{MWW, Sun, Webb}
\begin{eqnarray}
\nonumber
\frac{\partial S}{\partial t} &=& -F_I(S, I_T) -F_A(S, A_T) + \nabla_x\cdot ({D_S}\nabla_x S) \\
\nonumber
\frac{\partial E}{\partial t} &=& F_I(S, I_T) + F_A(S, A_T) -a E + \nabla_x\cdot ({D_E}\nabla_x E) \\
\label{eq:diff}
\frac{\partial I}{\partial t}  &=&  a\sigma E-\gamma_I I+\nabla_x\cdot ({D_I}\nabla_x I)\\
\nonumber
\frac{\partial A}{\partial t}  &=&  a(1-\sigma) E-\gamma_A A+\nabla_x\cdot ({D_A}\nabla_x A)\\
\nonumber
\frac{\partial R}{\partial t}  &=& \gamma_I I+\gamma_A A+\nabla_x\cdot ({D_R}\nabla_x R)
\end{eqnarray}
which is coupled with system \eqref{eq:diffuse} for the non-commuting counterpart. 
The capability of the model to account for different regimes, hyperbolic or parabolic, accordingly to the space dependent relaxation times $\tau_i$, $i\in\{S,E,I,A,R\}$, makes it suitable for describing the dynamics of human beings. Indeed, it is clear that the daily routine is a complex mixing of individuals moving at the scale of a city and individuals moving among different urban centers. 
In this situation, due to the lack of microscopic information and the high complexity of the dynamics, it is reasonable to avoid describing the details of movements within an urban area and model this through a diffusion operator. On the other hand, commuters when moving from one city to another follow well-established connections for which a description via transport operators is more appropriate.  

\subsection{Basic reproduction number}
The standard threshold of epidemic models is the well-known reproduction number $R_0$, which defines the average number of secondary infections produced when one infected individual is introduced into a host population in which everyone is susceptible \cite{HWH00} during its entire period of infectiousness. This number determines when an infection can invade and persist in a new host population. For many deterministic epidemic models, an infection begins in a fully susceptible population if and only if $R_0 > 1$. 
 Its definition in the case of spatially dependent dynamics is not straightforward, particularly when considering its spatial dependence. In the following, assuming no inflow/outflow boundary conditions in $\Omega$, integrating over velocity and space, we derive the following definition for the average reproduction number value on the domain $\Omega$
\begin{equation}
\begin{aligned}
R_0(t) &= \frac{\int_\Omega F_I(S_T,I_T) \,dx}{\int_\Omega \gamma_I(x) I_T(x,t) \,dx} \cdot \frac{\int_\Omega a(x)\sigma (x) E_T(x,t) \,dx}{\int_\Omega a(x) E_T(x,t) \,dx} \\&+ \frac{\int_\Omega F_A(S_T,A_T) \,dx}{\int_\Omega \gamma_A(x) A_T(x,t)\, dx} \cdot \frac{\int_\Omega a(x)(1-\sigma(x)) E_T(x,t) \,dx}{\int_\Omega a(x) E_T(x,t)\, dx} \,.
\end{aligned}
\label{eq.R0_2}
\end{equation}
The derivation of the above expression for $R_0(t)$, computed following the \textit{next-generation matrix} approach \cite{Diekmann}, is presented in detail in \cite{Bert2} using a suitable linearization of the corresponding nonlinear process for the space averaged quantities.

It is worth to underline that from definition \eqref{eq.R0_2} it can be deduced that it is a combination of the growth of $E_T,I_T$ and $A_T$ that determines the persistence of the epidemic, not solely the growth of $E_T$ in time, neither the growth of the simple sum $E_T+I_T+A_T$.
If, additionally, compartments $I$ and $A$ are considered homogeneously mixed in a unique compartment, allowing $\beta_I = \beta_A = \beta$, $\kappa_A=\kappa_I=\kappa$ and $\gamma_I = \gamma_A = \gamma$, we recover a SEIR-type compartmental model and the reproduction number results as in \cite{BDP}:
\begin{equation}
R_0(t)=\frac{\int_{\Omega} F(S_T,I_T)\,dx}{\int_{\Omega} \gamma(x) I_T(x,t)\,dx}.
\label{eq:R0b}
\end{equation}

Let us finally observe that, under the same no inflow/outflow boundary conditions, integrating in $\Omega$ equations \eqref{eq:kineticc}  and \eqref{eq:diffuse}   yields respectively the conservation of the total populations of commuters and non-commuters
\[
\begin{split}
&\frac{\partial}{\partial t} \int_{\Omega} (S(x,t)+E(x,t)+I(x,t)+A(x,t)+R(x,t))\,dx =0,\\ 
&\frac{\partial}{\partial t} \int_{\Omega} (\SO(x,t)+\EO(x,t)+\IO(x,t)+\AO(x,t)+\RO(x,t))\,dx =0. 
\end{split}
\] 

\subsection{Including data uncertainty}
To extend the model \eqref{eq:kineticc}-\eqref{eq:diffuse} to the case in which uncertainties are taken into account, let us suppose that the population of commuters depend on an additional random vector  $\boldsymbol{z}  = (z_1,\ldots,z_{d})^T \in \Omega_{\boldsymbol{z}} \subseteq \mathbb{R}^{d}$, where $z_1,\ldots,z_{d}$ are independent random variables. This vector is used to characterize possible sources of uncertainty in the physical system due to lack of information on the actual number of infected or specific epidemic characteristics of the infectious disease. 

Thus, in the system we have the following high-dimensional unknowns
\[f_S=f_S(x,v,t,\boldsymbol{z}),\ f_E=f_E(x,v,t,\boldsymbol{z}),\ f_I=f_I(x,v,t,\boldsymbol{z}),\ f_A=f_A(x,v,t,\boldsymbol{z}), f_R=f_R(x,v,t,\boldsymbol{z}).\]
The same considerations apply to the non-commuter population, yielding
\[\SO=\SO(x,t,\boldsymbol{z}),\ \EO=\EO(x,t,\boldsymbol{z}),\ \IO=\IO(x,t,\boldsymbol{z}),\AO=\AO(x,t,\boldsymbol{z}),\RO=\RO(x,t,\boldsymbol{z}).\] 
Notice that besides the introduction of a new vector of variables, the structure of the model \eqref{eq:kineticc}-\eqref{eq:diffuse} does not change, i.e. there is no direct variation of the unknowns with respect to $\boldsymbol{z}$, which, instead, have to be intended as parameters into the equations. We will further assume that also some epidemic parameters are affected by uncertainty. Therefore, for instance, parameters acting inside the incidence function may have an additional dependence of the form
\[\beta_I=\beta_I(x,t,\boldsymbol{z}), \ \beta_A=\beta_A(x,t,\boldsymbol{z}).
\]
\[k_I=k_I(x,t,\boldsymbol{z}), \ k_A=k_A(x,t,\boldsymbol{z}),\]

In the next section, we discuss several numerical examples based on the model \eqref{eq:kineticc}-\eqref{eq:diffuse} with uncertainty. We will consider that the initial number of detected infected $I$, derived from the data at disposal \cite{istat,prot_civile}, represents only a lower bound while the true values are not known but affected by uncertainty. Consequently, also initial conditions of exposed $E$, asymptomatic $A$ and susceptible $S$ will contain a stochastic dependence. 




\section{Application to COVID-19 spread in Italy}\label{sect:numres}
To validate the proposed methodology in a realistic geographical and epidemiological scenario, a numerical test reproducing the epidemic outbreak of COVID-19 in the Lombardy Region of Italy, from February 27, 2020 to March 22, 2020, is designed, taking into account the uncertainty underlying initial conditions of infected individuals. 
We underline that the discretization of the multiscale system of PDEs \eqref{eq:kineticc}-\eqref{eq:diffuse} presented in this work is not trivial and requires the construction of a specific numerical method able to correctly describe the transition from a convective to a diffusive regime in realistic geometries. Additionally the method should be capable to deal efficiently with the uncertainty characterized by the stochastic nature of the model. 
Details on these numerical aspects are given in Appendix \ref{appendix:NM} together with the tables of the data used for population mobility in Appendix \ref{appendix:tables}.

\begin{figure}[t!]
\centering
\begin{subfigure}{0.49\textwidth}
\includegraphics[width=1\linewidth]{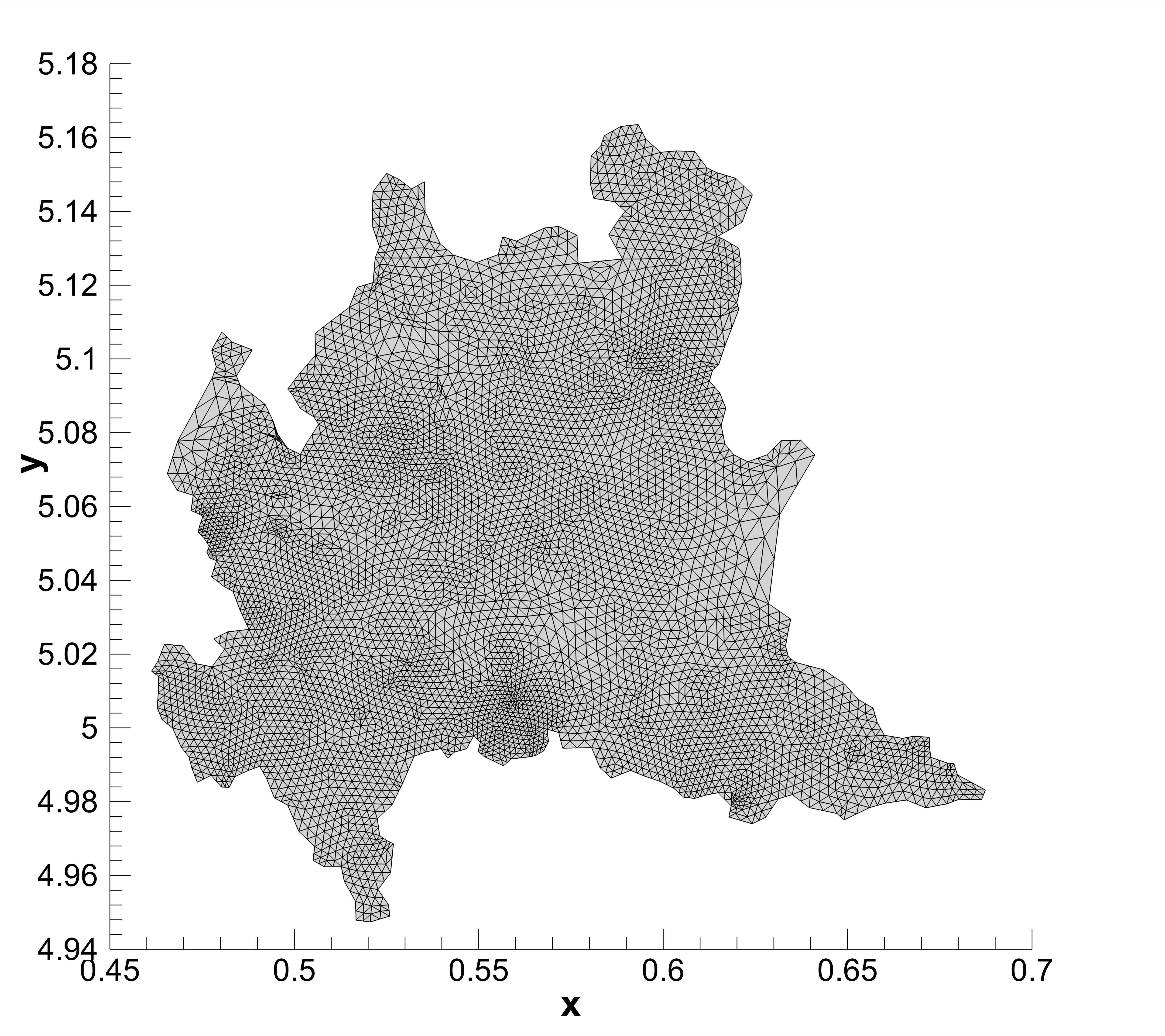}
\caption{mesh grid}
\label{mesh_grid}
\end{subfigure}
\begin{subfigure}{0.49\textwidth}
\includegraphics[width=1\linewidth]{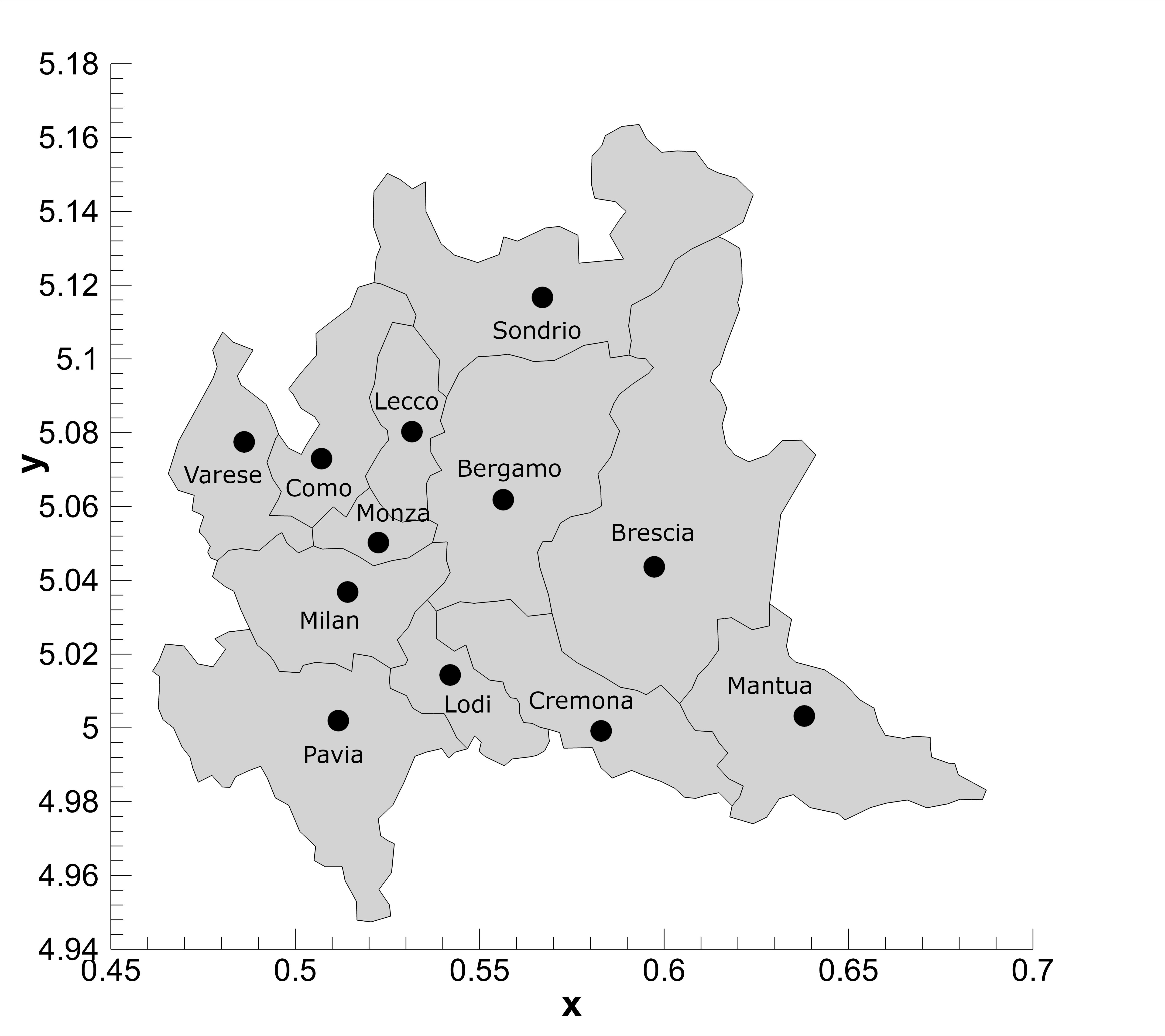}
\caption{provinces}
\label{provinces}
\end{subfigure}
\begin{subfigure}{0.49\textwidth}
\includegraphics[width=1\linewidth]{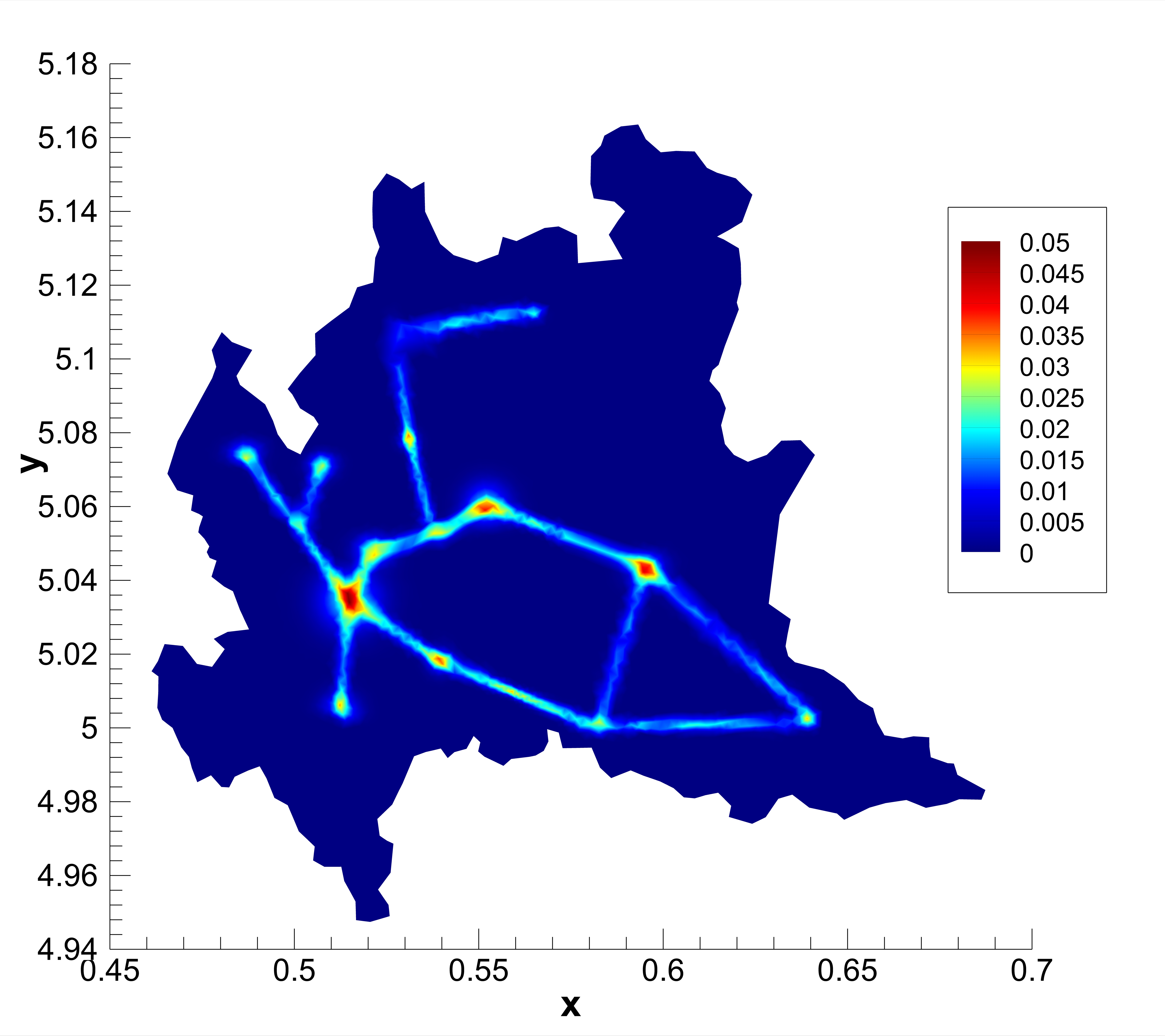}
\caption{$\lambda_i$, $i\in\{S,E,A,R\}$}
\label{lambda_distribution}
\end{subfigure}
\begin{subfigure}{0.49\textwidth}
\includegraphics[width=1\linewidth]{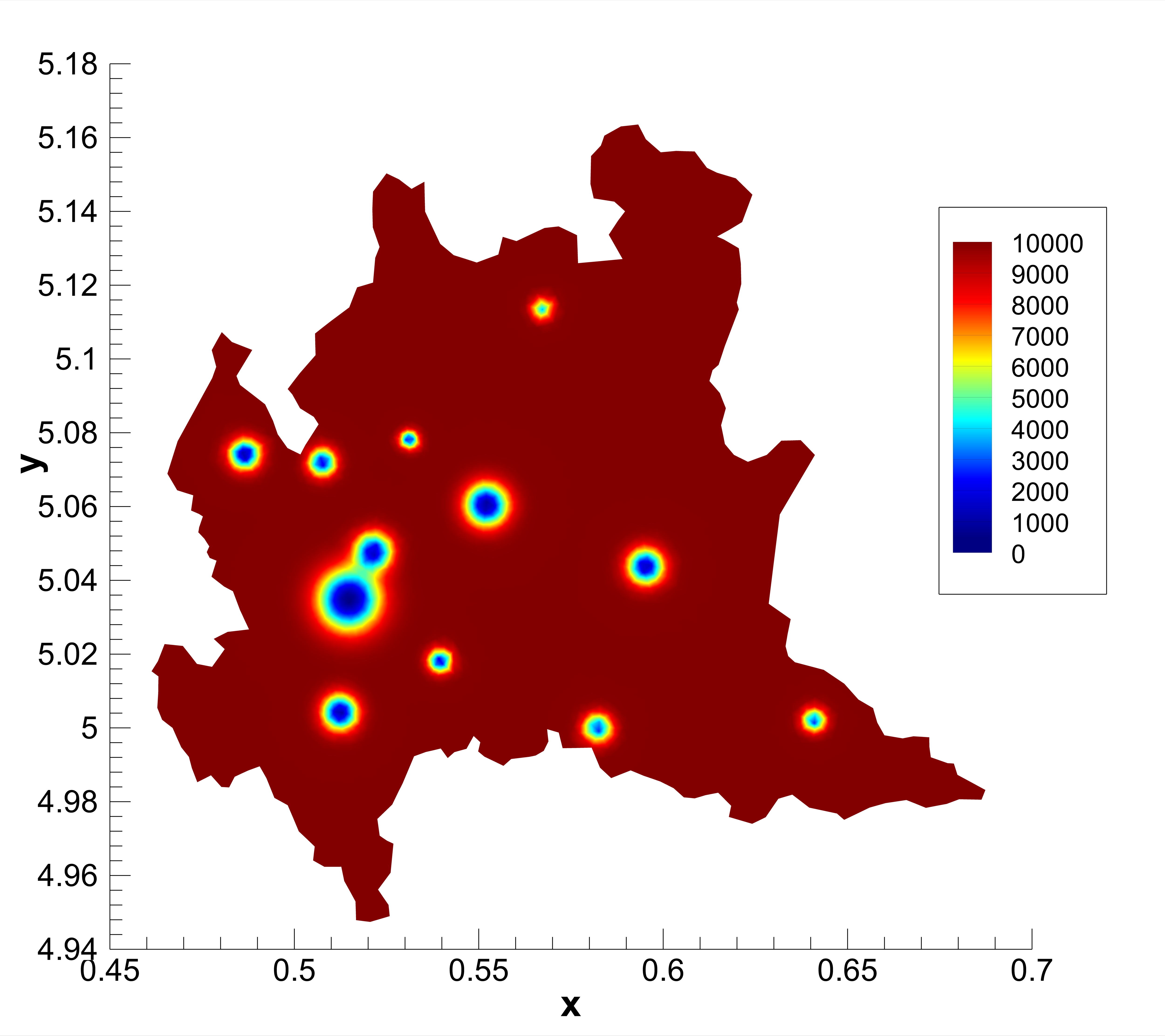}
\caption{$\tau_i$, $i\in\{S,E,I,A,R\}$}
\label{tau_distribution}
\end{subfigure}
\caption{Top: unstructured computational mesh used to discretize the Lombardy Region (a) and identification of the provinces (b). Bottom: initial condition imposed for characteristic speeds $\lambda$ (c) and relaxation times $\tau$ (d).}
\label{fig.computational_domain}
\end{figure}

\subsection{Computational setup}
The computational domain is defined in terms of the boundary that circumscribes the Lombardy Region, which is available in \cite{istat} as a list of georeferenced points in the ED50/UTM Zone 32N reference coordinate system. To avoid ill-conditioned reconstruction matrices and other related problems that arise when dealing with large numbers in finite arithmetic, all coordinates are re-scaled by a factor of $10^6$. The resulting computational grid is composed of $N_E = 10792$ triangular control volumes. The mesh grid is presented in Figure \ref{mesh_grid}. No-flux boundary conditions are imposed in the whole boundary of the domain, assuming that the population is not moving from/to the adjacent Regions. The domain is then subdivided in the $\mathcal{N}_c = 12$ provinces of Lombardy: Pavia (PV), Lodi (LO), Cremona (CR), Mantua (MN), Milan (MI), Bergamo (BG), Brescia (BS), Varese (VA), Monza-Brianza (MB), Como (CO), Lecco (LC) and Sondrio (SO). The identification of these cities is shown in Figure \ref{provinces}.

The units of measure chosen for this numerical test can be summarized as follows:
\begin{equation*}
\mathrm{1\, km = 10^{-3} \,L\,, \quad 1\, person \approx 10^{7} \,P\,, \quad 1\, day = 2\,T\,,}
\end{equation*}
with [L], [P] and [T] being the length, person and time units used in the simulation, respectively. Notice that the normalization of the population is made with respect to the total number of individuals of the Region (taken from \cite{ISTATdemo}), which is $M = 10.027.602$, to properly work in a context in which the total population is equal to the unit.

To avoid the mobility of the population in the entire territory and to simulate a more realistic geographical scenario in which individuals travel along the main traffic paths of the Region, different values of propagation speeds are assigned in the domain which reflect, as close as possible, the real characteristics of the territory. Along the main connections of the Region, a mean value of $\lambda = 0.04$ is prescribed for compartments $S,E,A$ and $R$, which ensures a maximum travel distance of 80 km within a day; while, for the same compartments, $\lambda = 0.02$ is ulterior fixed in the urbanized circles. A spatial width of $h=0.5$ km is assigned to the traveling paths. On the other hand, assuming that highly infectious subjects are mostly detected in the most optimistic scenario, being subsequently quarantined or hospitalized, the speed assigned to compartment $I$ is set null. However, the infected people, even if limited by quarantines and social distancing, can still contribute to the spread the disease via the diffusion process at the urban scale (mimicking for instance the still possible infections happening at the family level). A null value $\lambda = 0$ is set in the rest of the computational domain for all the epidemiological compartments. The resulting distribution of the characteristic speeds is visible from Figure \ref{lambda_distribution}.

The relaxation time is set $\tau_i = 10^4$ for $i\in\{S,E,I,A,R\}$, so that the model recovers a hyperbolic regime in the entire region, apart from the main cities, where a parabolic setting is prescribed to correctly capture the diffusive behavior of the disease spreading which typically occurs in highly urbanized zones. Hence, in the urban area of each city the following relaxation time $\tau_{c,i}$ is prescribed:
\begin{equation*}
\tau_{c,i} = \tau_i + (\tau_0 - \tau_i) \sum_{c=1}^{\mathcal{N}_c} e^{-\frac{(x-x_c)^2 +(y-y_c)^2}{2r_c^2}},
\end{equation*}
with a diffusive relaxation time chosen to be $\tau_0 =10^{-4}$. 
The resulting distribution of the relaxation times is presented in Figure \ref{tau_distribution}.

\begin{figure}[t!]
\centering
\begin{subfigure}{0.49\textwidth}
\includegraphics[width=1\linewidth]{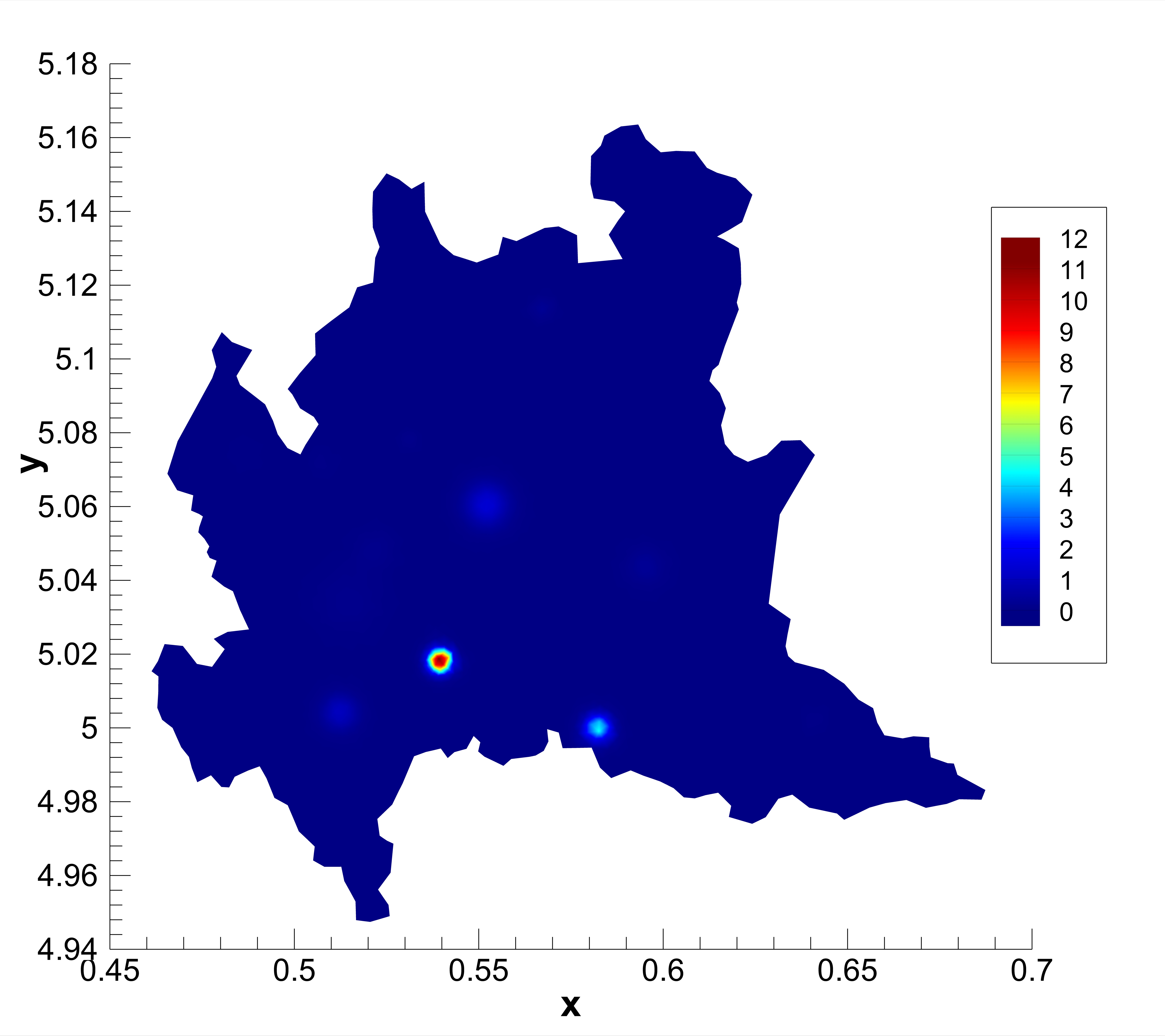}
\subcaption{$E_{T,0}(x,y)+I_{T,0}(x,y)+A_{T,0}(x,y)$}
\label{EIA}
\end{subfigure}
\begin{subfigure}{0.49\textwidth}
\includegraphics[width=1\linewidth]{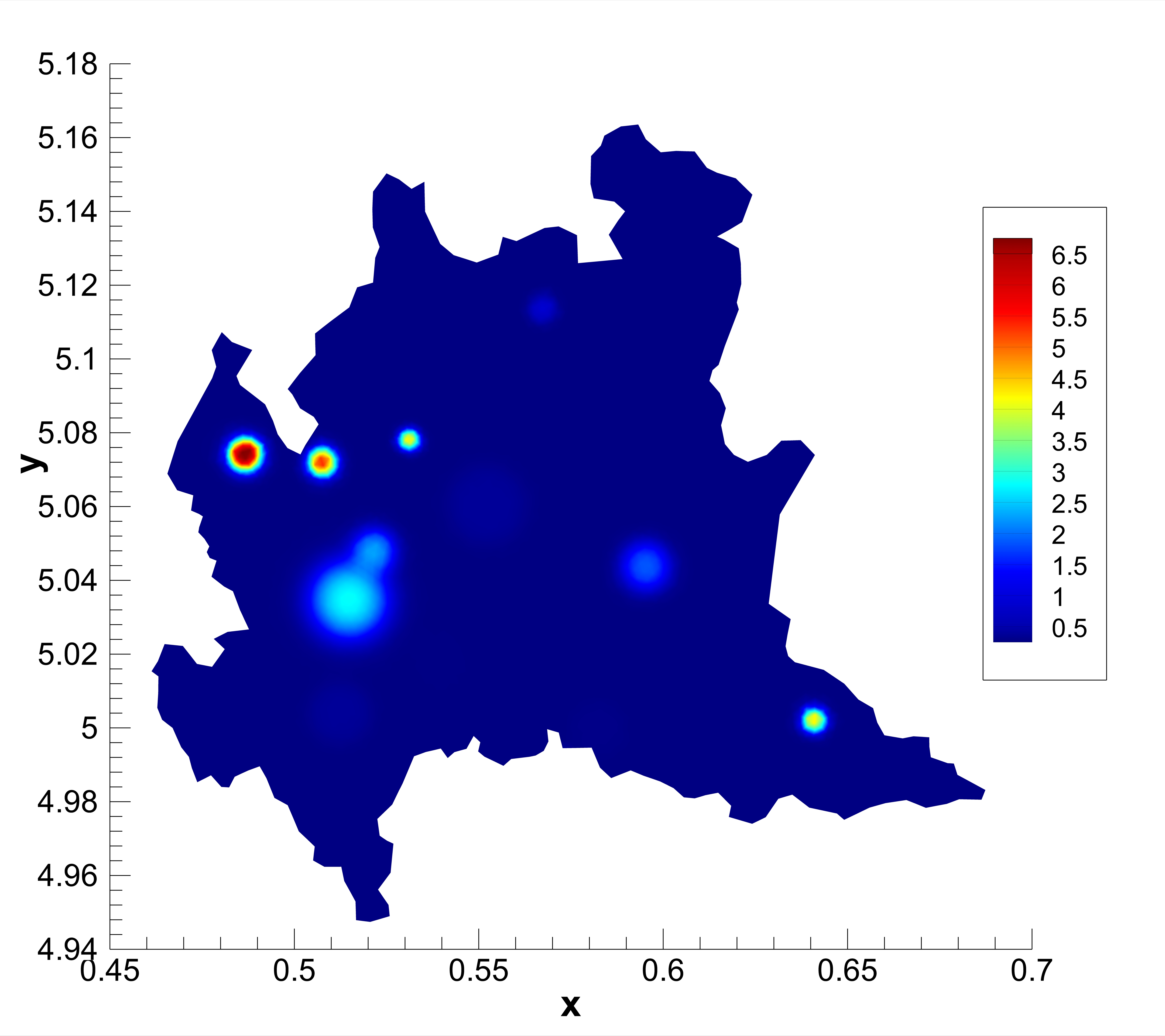}
\subcaption{$R_0(0)$}
\label{R0}
\end{subfigure}
\caption{Initial distribution (on February 27, 2020) of the infected population $E_{T,0}+I_{T,0}+A_{T,0}$ (a) and of the reproduction number $R_0(0)$ (b) in the Lombardy Region.}
\label{fig.IC}
\end{figure}

\subsection{Uncertain initial data and epidemic parameters}
Considering $p_c$ the number of citizens of a generic city (province) denoted with subscript $c$, the initial spatial distribution of the generic population $f(x,y)$ is assigned, for each province and each epidemiological compartment, as a multivariate Gaussian function with the variance being the radius of the urban area $r_c$:
\begin{equation*}
f(x,y) = \frac{1}{2\pi r_c} e^{-\frac{(x-x_c)^2 +(y-y_c)^2}{2r_c^2}} p_c \,,
\end{equation*}
with $(x_c, y_c)$ representing the coordinates of a generic city center. The initial population setting, for each province of the Lombardy Region, is taken from \cite{ISTATdemo} and reported in Table \ref{tab:IC_lombardia},  Appendix \ref{appendix:tables}. Note that, the radius $r_c$ associated to each city, defined in Table \ref{tab:IC_lombardia} (first column), permits to exactly re-obtain the population $p_c$ when integrating over the computational domain the initial spatially distributed population.

Since at the beginning of the pandemic, tracking of positive individuals in Italy was very scarce, in this numerical test we consider that the initial amount of infected people is the leading quantity affected by uncertainty. To this aim, we introduce a single source of uncertainty $z$ having uniform distribution, $z \sim \mathcal{U}(0, 1)$ so that the initial conditions for compartment $I$, in each control volume, are prescribed as
\begin{equation*}
I_T(0,z) = I_{T,0} (1+\mu z)\,,
\end{equation*}
with $I_{T,0}$ initial amount of highly infectious corresponding to the values reported by February 27, 2020 in the GitHub repository \cite{prot_civile} daily updated by the Civil Protection Department of Italy for each city, listed in the last column of Table \ref{tab:IC_lombardia}.
This choice effectively associates all infected individuals detected with the $I$ compartment, as a result of the screening policy adopted during February--March 2020 in Italy. In fact, tests to assess the presence of SARS-CoV-2 were performed almost exclusively on patients with consistent symptoms and fever at the beginning of this pandemic. Regarding the uncertainty of these data, we impose $\mu=1$, assuming that at least half of the actual highly symptomatic infected were detected at the beginning of the pandemic outbreak.
For all the cities with zero infected detected by February 27, 2020 (e.g., Mantua, Varese, Como and Lecco), we choose to fix $I_{T,0}=1$ in order to assign an effective uncertainty.

Based on the estimations reported in \cite{Bert2}, the expected initial amount of exposed $E_{T,0}$ and asymptomatic/mildly symptomatic individuals $A_{T,0}$ is imposed so that $E_{T,0} = 10\, I_{T,0}$ and $A_{T,0}= 9\, I_{T,0}$ in each location. Therefore, also initial conditions for compartments $E$, $A$ and $S$ become stochastic, depending on the initial amount of severe infectious at each location:
\[E_T(0,z) = 10\,I_T(0,z) \, , \qquad A_T(0,z) = 9\,I_T(0,z)\, , \]
\[ S_T(0,z) = N - E_T(0,z) - I_T(0,z) - A_T(0,z)\,.\]
Finally, removed individuals are initially set null everywhere in the network, $R_T(0,z)=0.0$.

To properly subdivide the population in commuters and non-commuters, regional mobility data are considered. In particular, the matrix of commuters presented in Table \ref{tab:matrix_lombardia}, Appendix \ref{appendix:tables}, reflects mobility data provided by the Lombardy Region for the regional fluxes of year 2020 \cite{opendataLombardia}, which is in agreement with the one derived from ISTAT data released in October, 2011, as also confirmed in \cite{Gatto}. Therefore, to each control volume, the total percentage of commuters referred to the province where it is located is assigned, and non-commuting individuals are computed as a result of conservation principles.
From Table \ref{tab:matrix_lombardia} it can be noticed that some connections are not taken into account simply because the amount of commuters along these routes is negligible if compared to the amount of individuals traveling in the other paths and with respect to the dimension of the populations. 

Concerning epidemiological parameters, accordingly with values reported in \cite{Gatto,Buonomo,Bert2}, we set $\gamma_I~=~1/14$, $\gamma_A = 2\gamma_I = 1/7$, $a = 1/3$, considering these parameters as clinical ones and therefore deterministic. We also assume the probability rate of developing severe symptoms $\sigma = 1/12.5$, as in \cite{Buonomo,Bert2}. 

From the first day simulated in this test, the population was aware of the risk associated with COVID-19 and recommendations such as washing hands often, not touching their faces, avoiding handshakes and keeping distances had already been disseminated by the government, hence, we initially fix coefficients $k_I=k_A=50$.

The initial value of the transmission rate of asymptomatic/mildly infectious people is calibrated as the result of a least square problem with respect to the observed cumulative number of infected $I_T(t)$, through a deterministic SEIAR ODE model set up for the whole Lombardy Region, which provide the estimate $\beta_A = 0.58\times 10^{-3}$. As previously mentioned, since we are assuming that highly infectious subjects are mostly detected in the most optimistic scenario, being subsequently quarantined or hospitalized, the transmission rate of $I$ is set $\beta_I = 0.03\,\beta_A$, as in \cite{Gatto,Buonomo,Bert2}. Finally, in the incidence function we fix $p=1$.

With this parametric setup, we obtain an initial expected value of the basic reproduction number for the Region, evaluated as from definition \eqref{eq.R0_2}, $R_0(0)~=~3.2$, which is in agreement with estimations reported in \cite{Gatto,Buonomo,vollmer2020,Bert2}. Nevertheless, with the proposed methodology it is possible to present the heterogeneity underlying the basic reproduction number at the local scale, as shown in Figure \ref{fig.IC} for $\mu=0$, together with the initial global amount of infected people $E_{T,0}(x,y)+I_{T,0}(x,y)+A_{T,0}(x,y)$ present in the domain.

To model the effects of the lockdown imposed by the government from March 9, 2020 in north of Italy \cite{legal}, from that day the transmission rate $\beta_A$ is reduced by 50\%, also increasing the coefficients $k_I = k_A = 80$ as a result of the population becoming increasingly aware of the epidemic risks. In addition, the percentage of commuting individuals is reduced by 60\% for each compartment according to mobility data tracked through GPS systems of mobile phones and made temporarily available by Google \cite{aktay2020,vollmer2020}.

\begin{figure}[p!]
\centering
\begin{subfigure}{0.32\textwidth}
\includegraphics[width=1\linewidth]{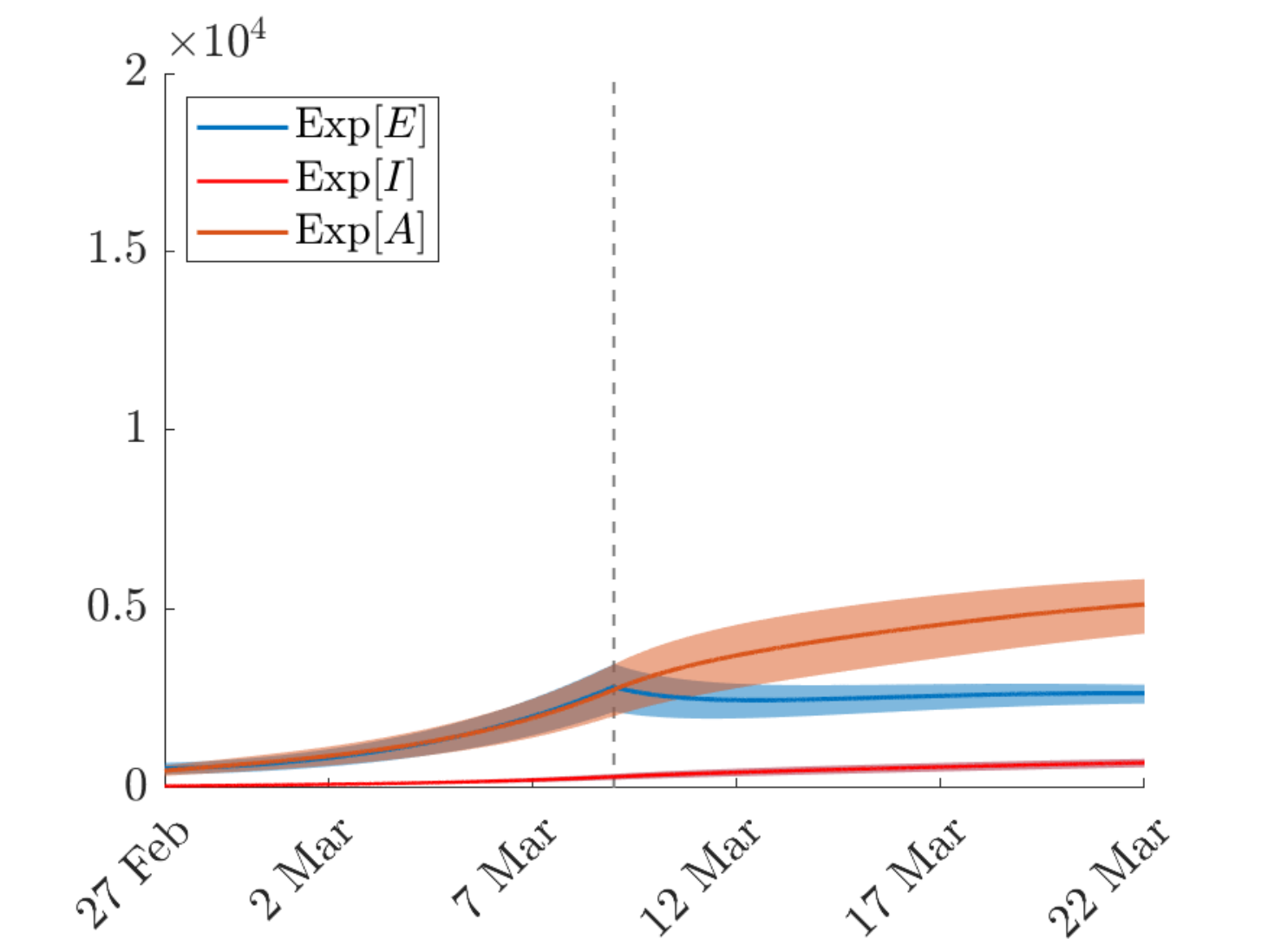}
\caption{Pavia}
\label{fig.PV}
\end{subfigure}
\begin{subfigure}{0.32\textwidth}
\includegraphics[width=1\linewidth]{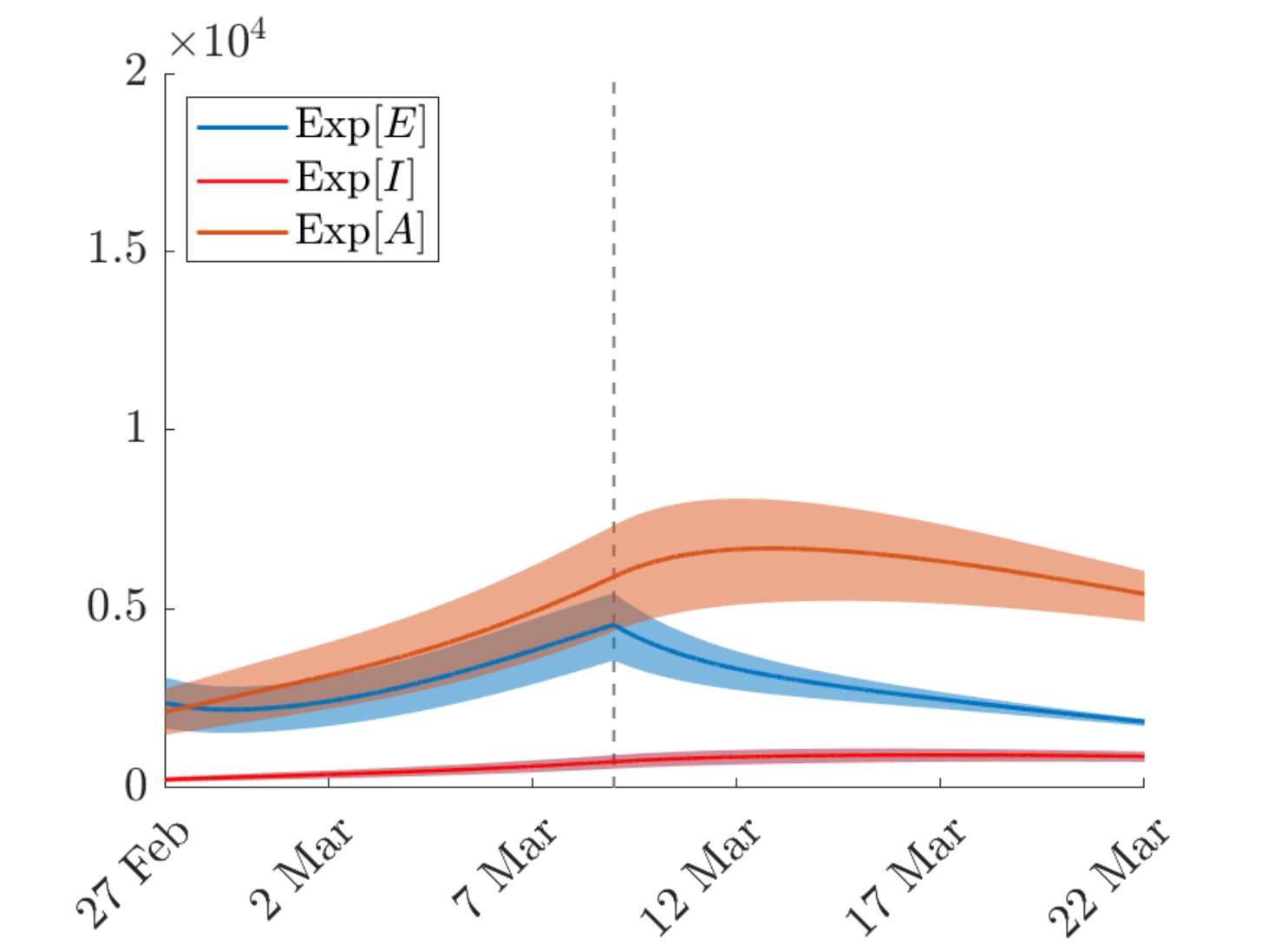}
\caption{Lodi}
\label{fig.LO}
\end{subfigure}
\begin{subfigure}{0.32\textwidth}
\includegraphics[width=1\linewidth]{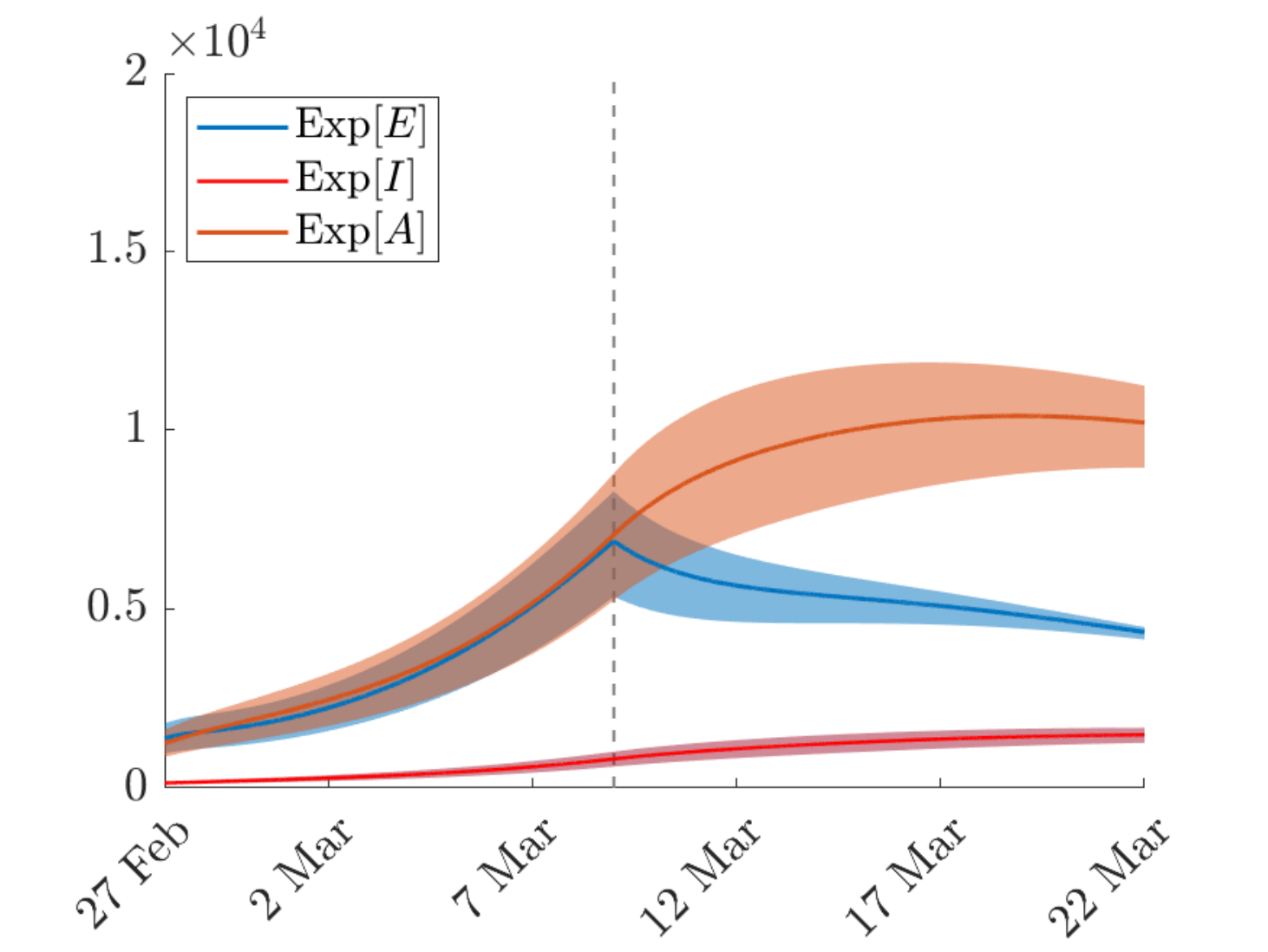}
\caption{Cremona}
\label{fig.CR}
\end{subfigure}
\begin{subfigure}{0.32\textwidth}
\includegraphics[width=1\linewidth]{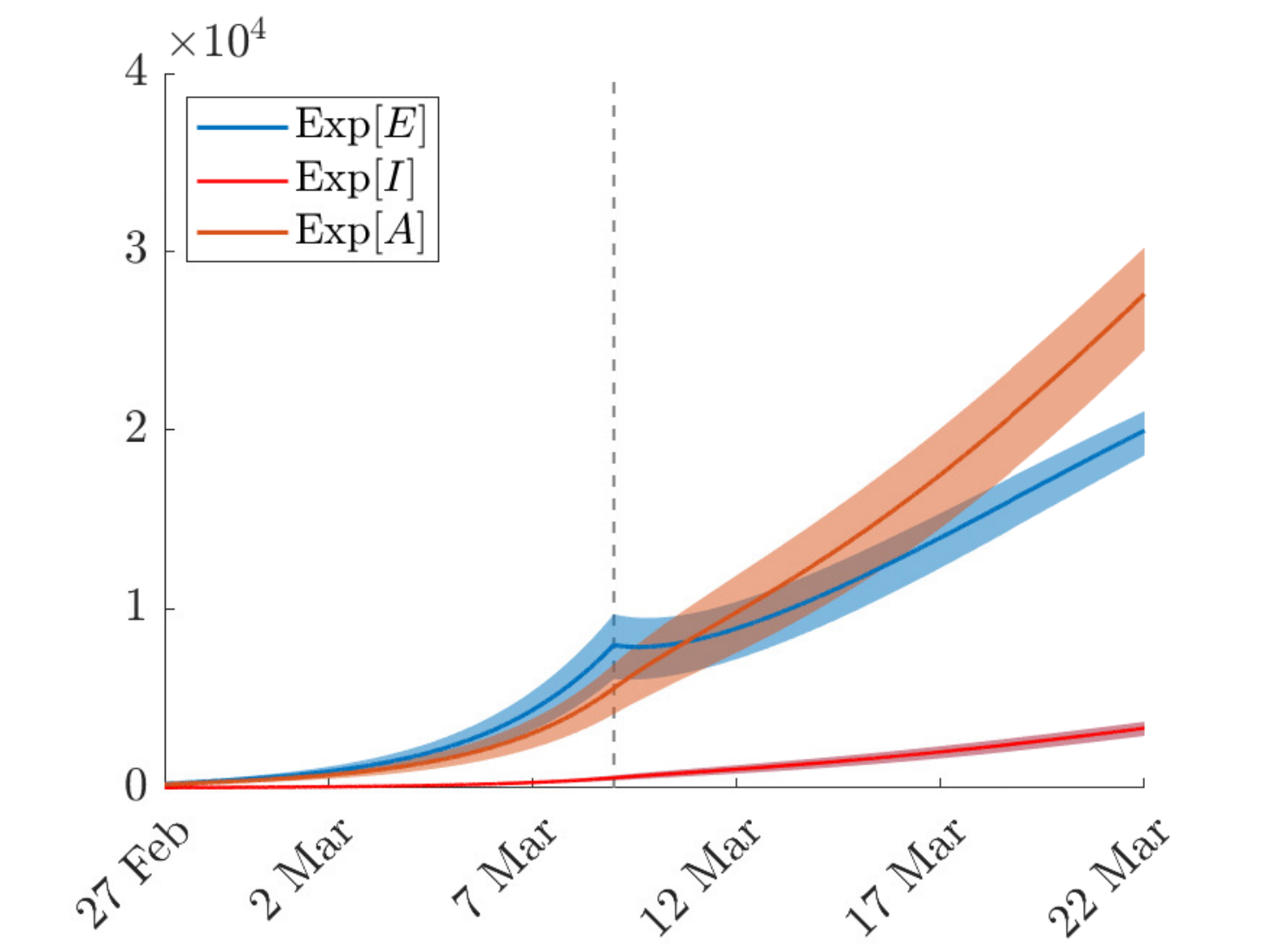}
\caption{Milan}
\label{fig.MI}
\end{subfigure}
\begin{subfigure}{0.32\textwidth}
\includegraphics[width=1\linewidth]{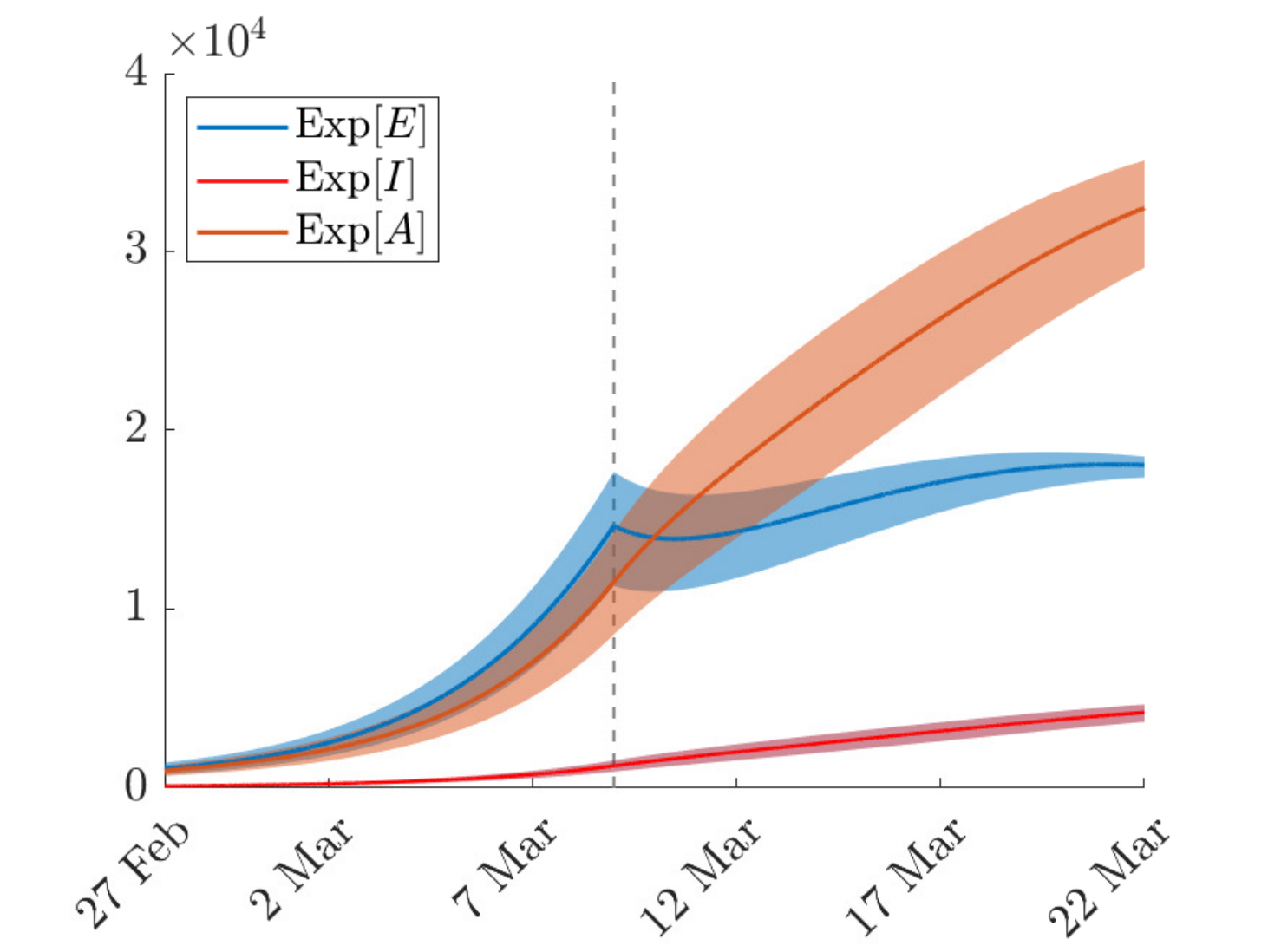}
\caption{Bergamo}
\label{fig.BG}
\end{subfigure}
\begin{subfigure}{0.32\textwidth}
\includegraphics[width=1\linewidth]{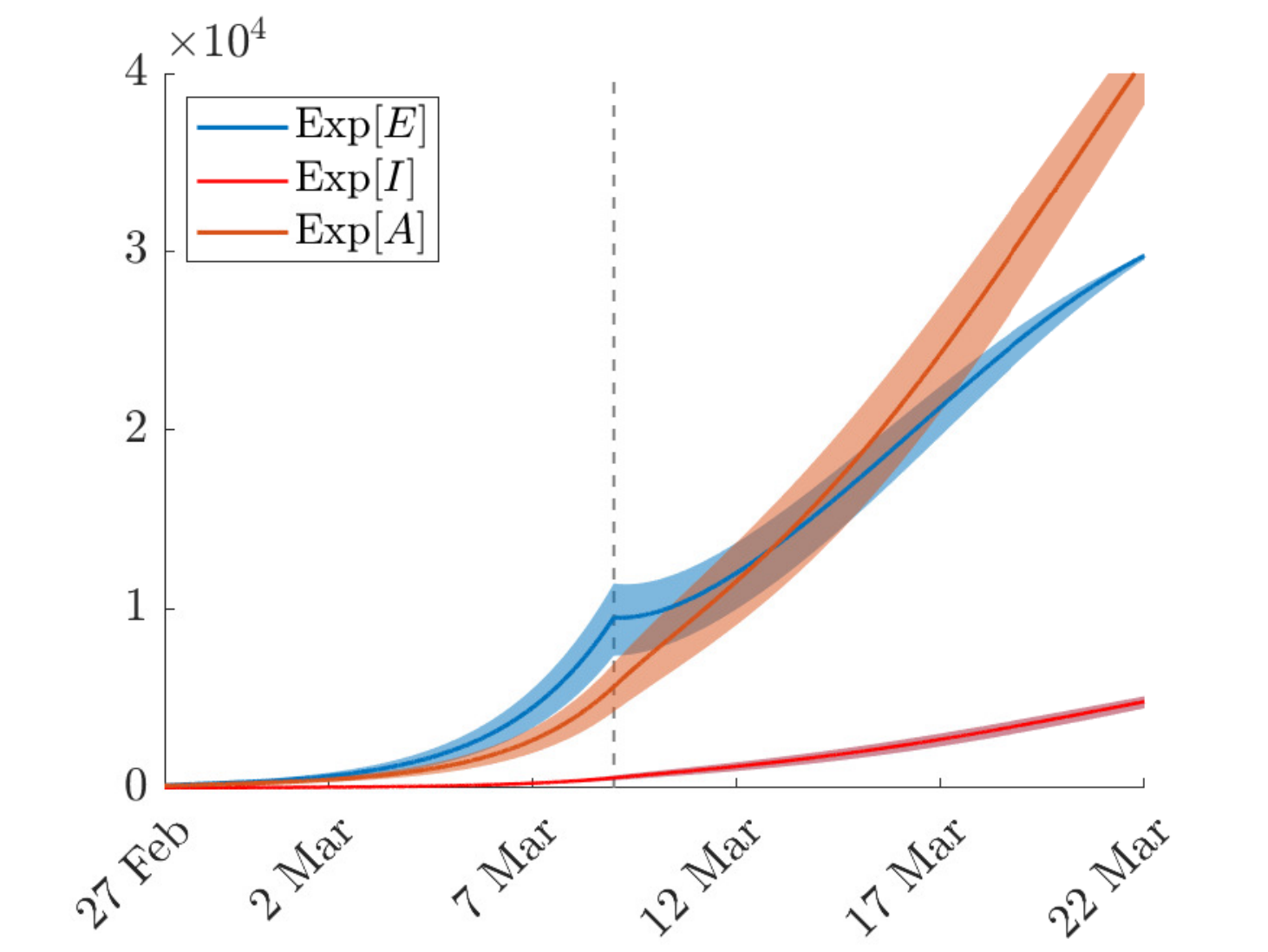}
\caption{Brescia}
\label{fig.BS}
\end{subfigure}
\begin{subfigure}{0.32\textwidth}
\includegraphics[width=1\linewidth]{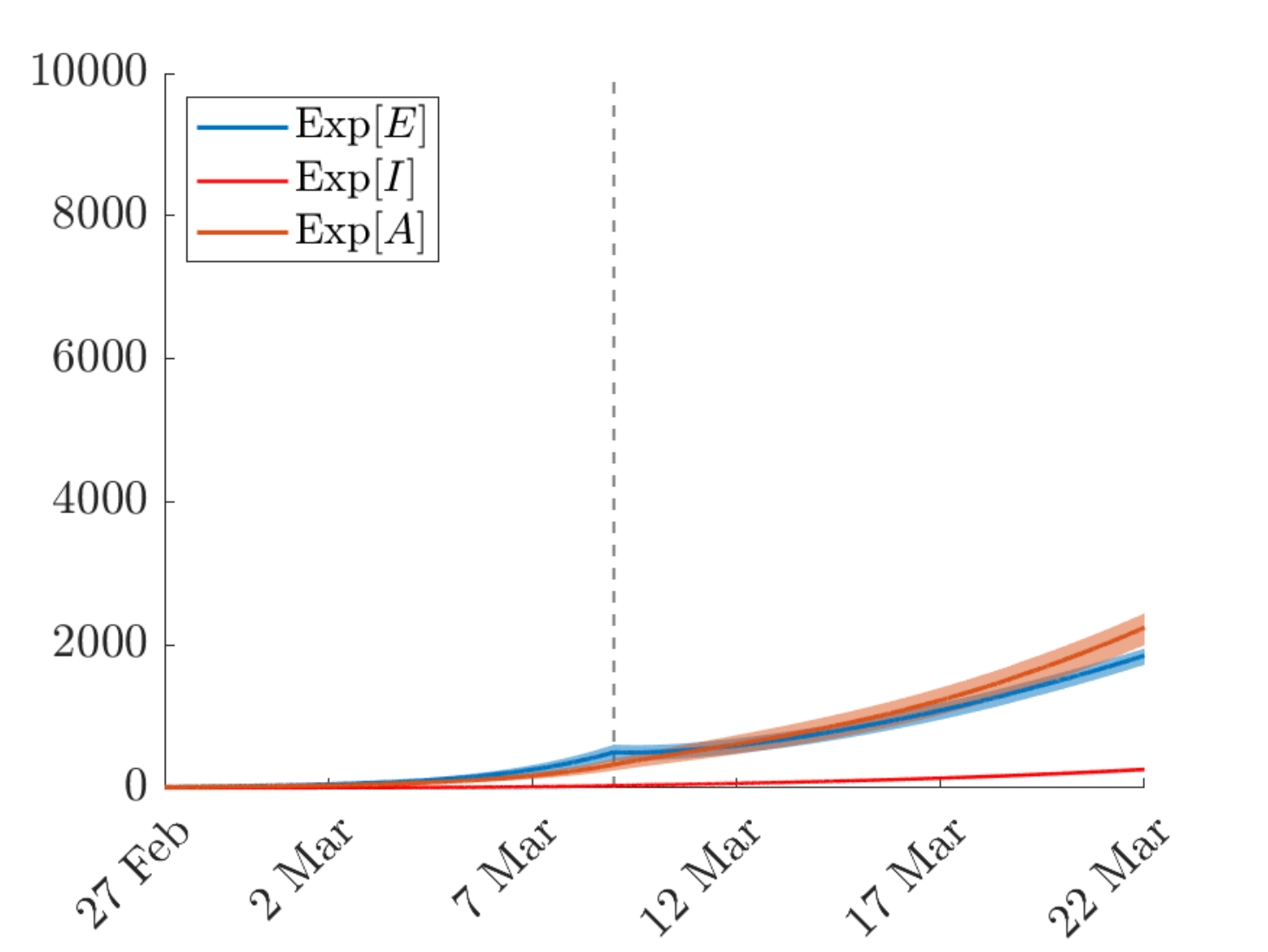}
\caption{Varese}
\label{fig.VA}
\end{subfigure}
\begin{subfigure}{0.32\textwidth}
\includegraphics[width=1\linewidth]{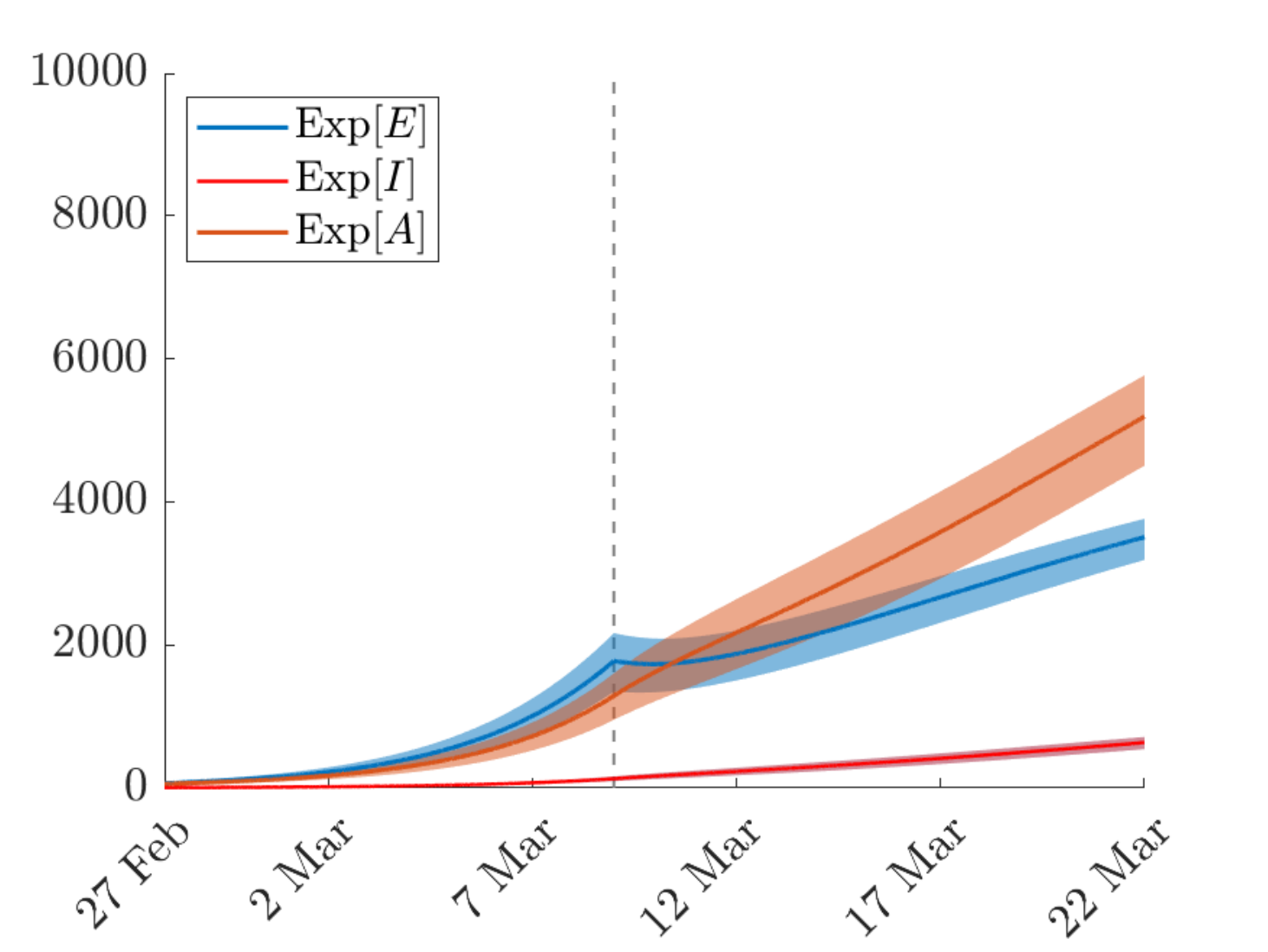}
\caption{Monza-Brianza}
\label{fig.MB}
\end{subfigure}
\begin{subfigure}{0.32\textwidth}
\includegraphics[width=1\linewidth]{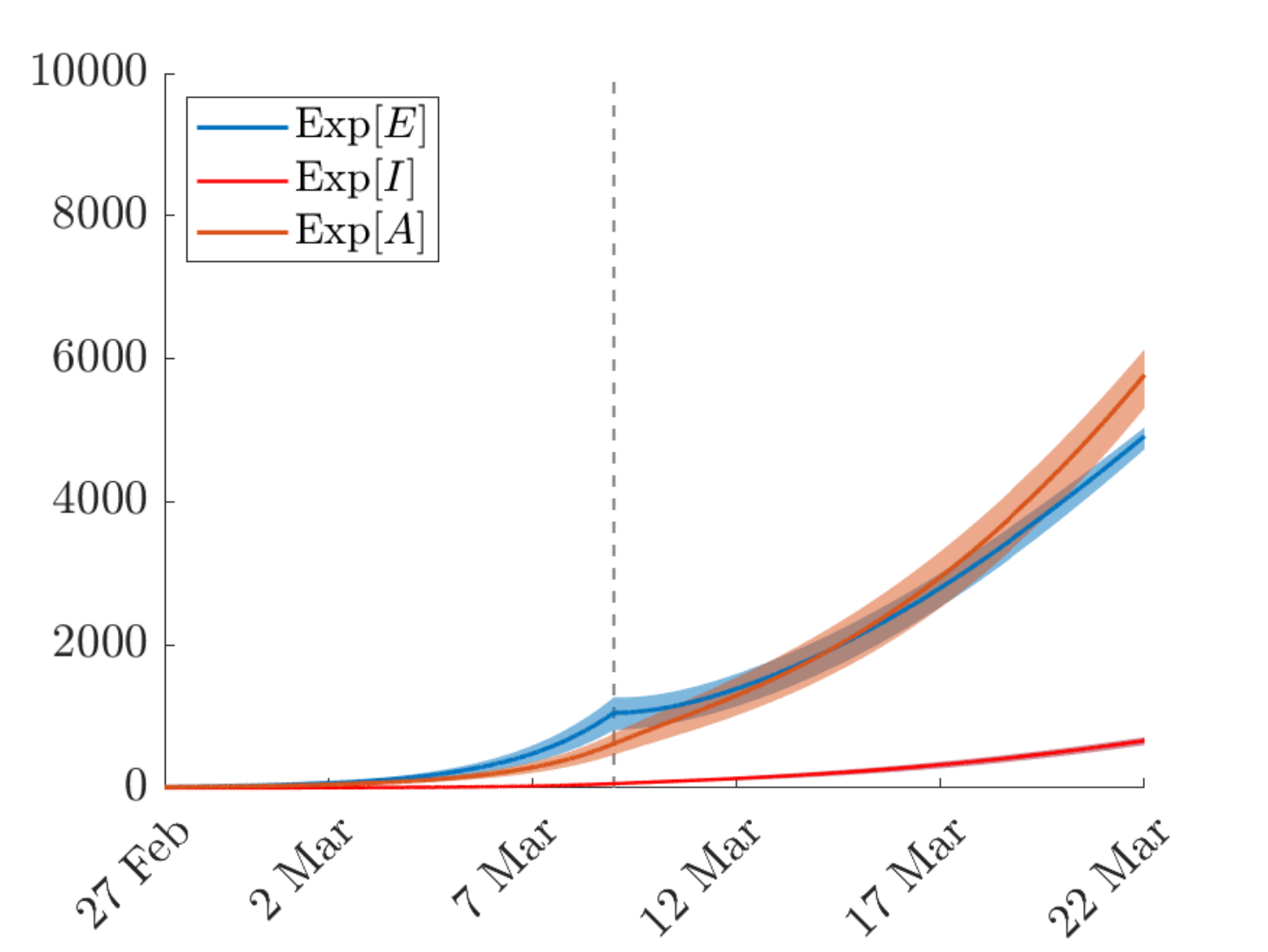}
\caption{Mantua}
\label{fig.MN}
\end{subfigure}
\begin{subfigure}{0.32\textwidth}
\includegraphics[width=1\linewidth]{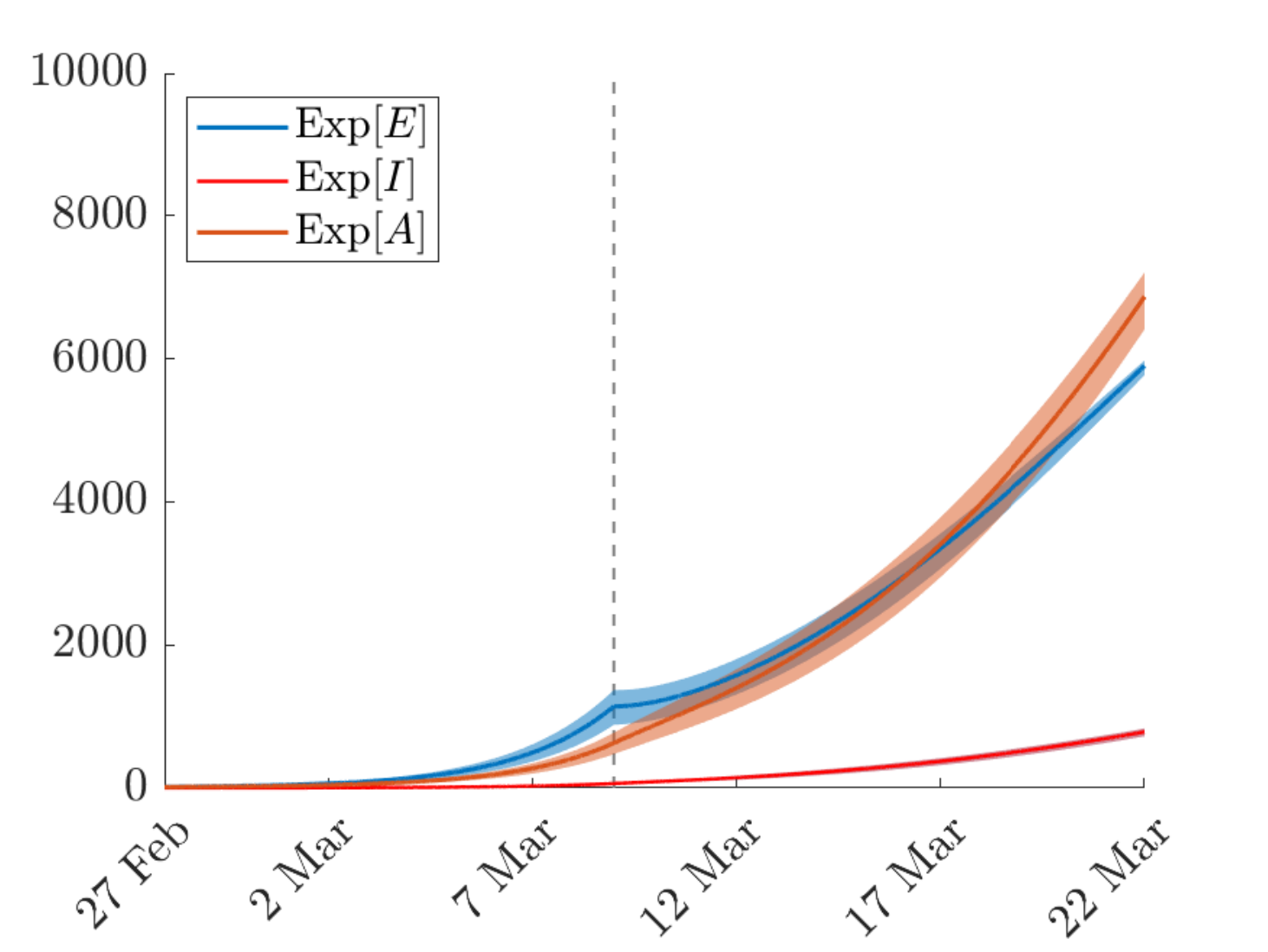}
\caption{Lecco}
\label{fig.LC}
\end{subfigure}
\begin{subfigure}{0.32\textwidth}
\includegraphics[width=1\linewidth]{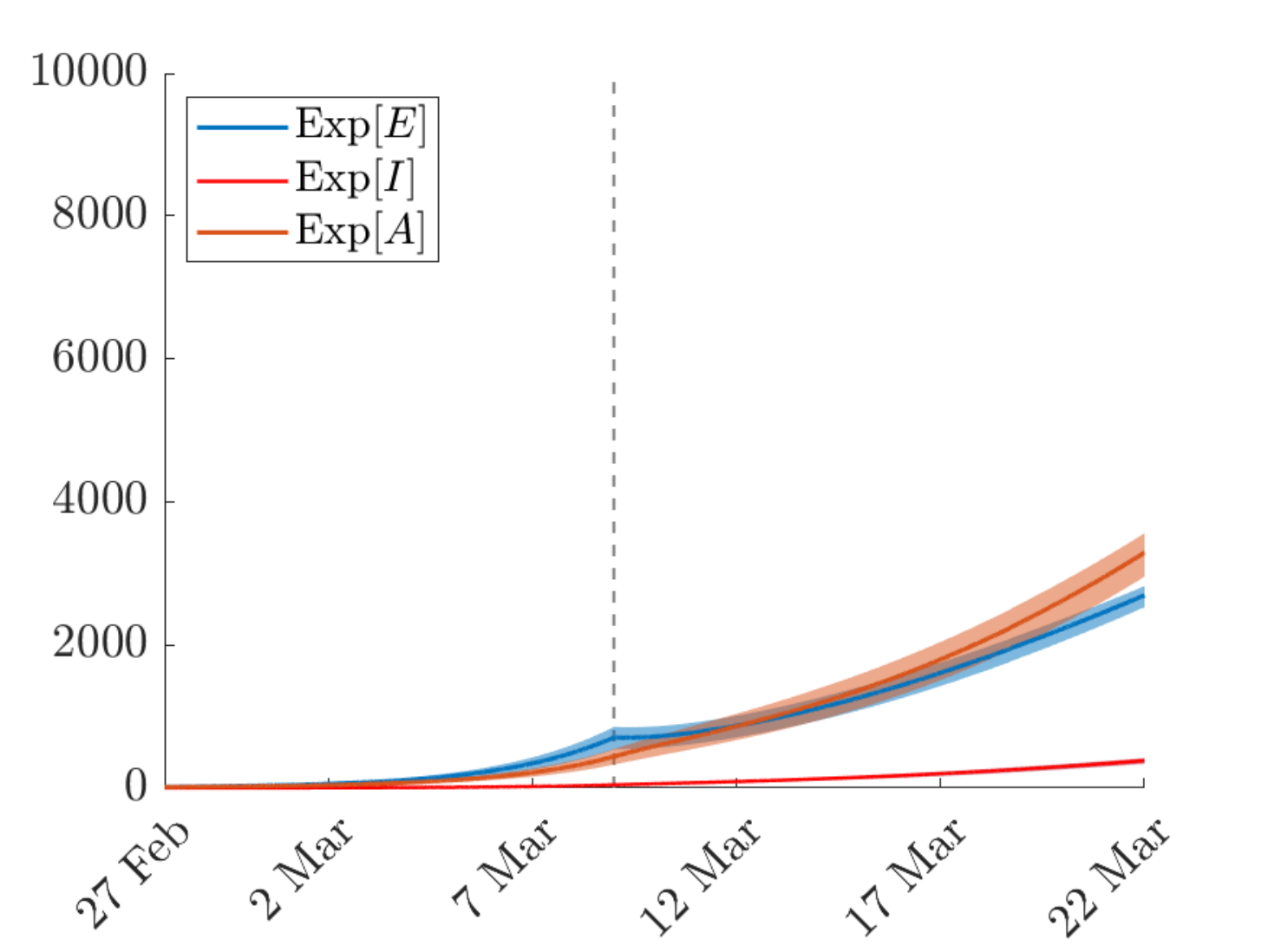}
\caption{Como}
\label{fig.CO}
\end{subfigure}
\begin{subfigure}{0.32\textwidth}
\includegraphics[width=1\linewidth]{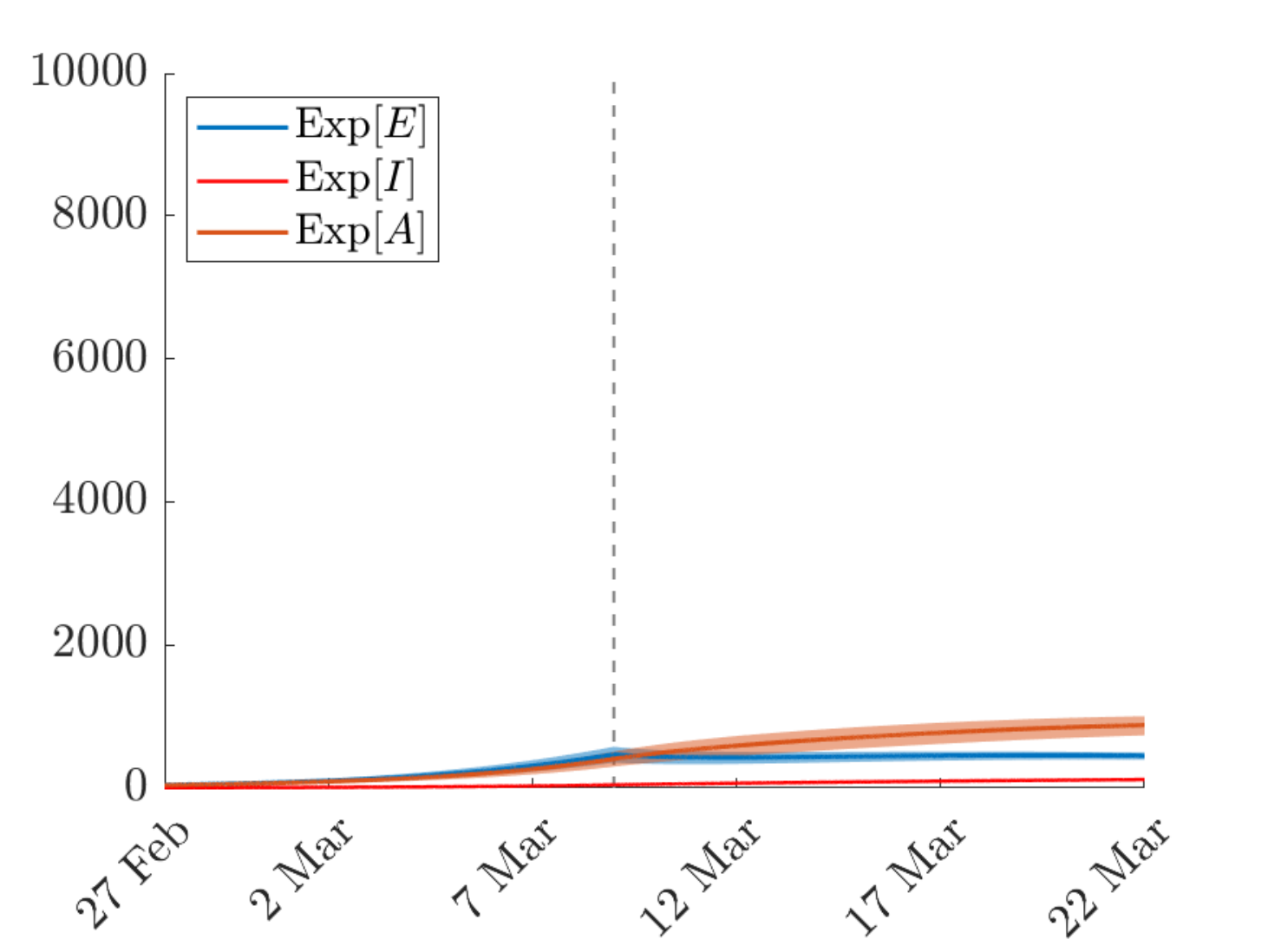}
\caption{Sondrio}
\label{fig.SO}
\end{subfigure}
\caption{Numerical results, with 95\% confidence intervals, of the simulation of the first outbreak of COVID-19 in Lombardy, Italy. Expected evolution in time of compartments $E$, $A$, $I$. Vertical dashed lines identify the onset of governmental lockdown restrictions.}
\label{fig.results_Lombardy}
\end{figure}
\begin{figure}[p!]
\centering
\begin{subfigure}{0.32\textwidth}
\includegraphics[width=1\linewidth]{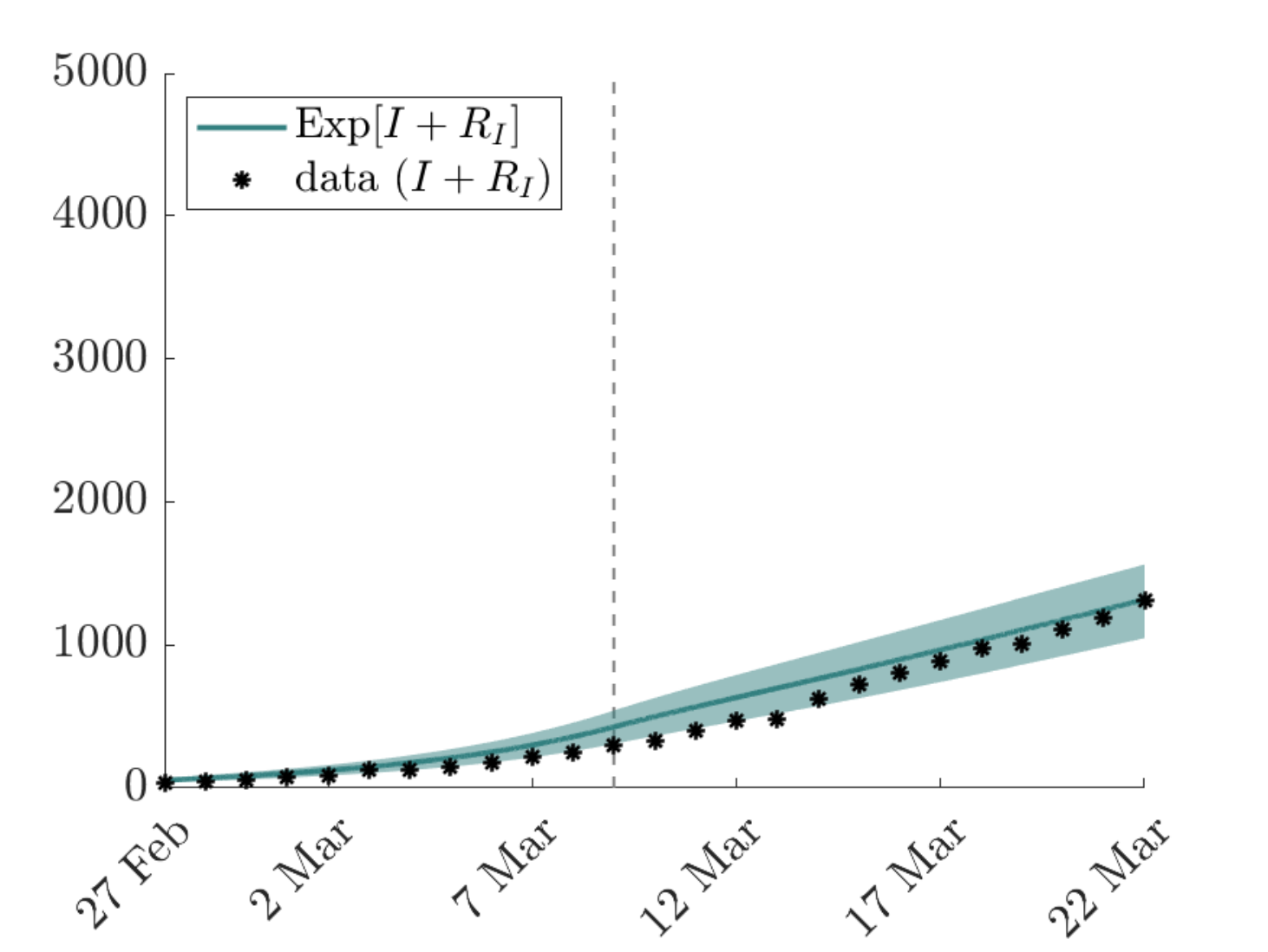}
\caption{Pavia}
\label{fig.PV_RI}
\end{subfigure}
\begin{subfigure}{0.32\textwidth}
\includegraphics[width=1\linewidth]{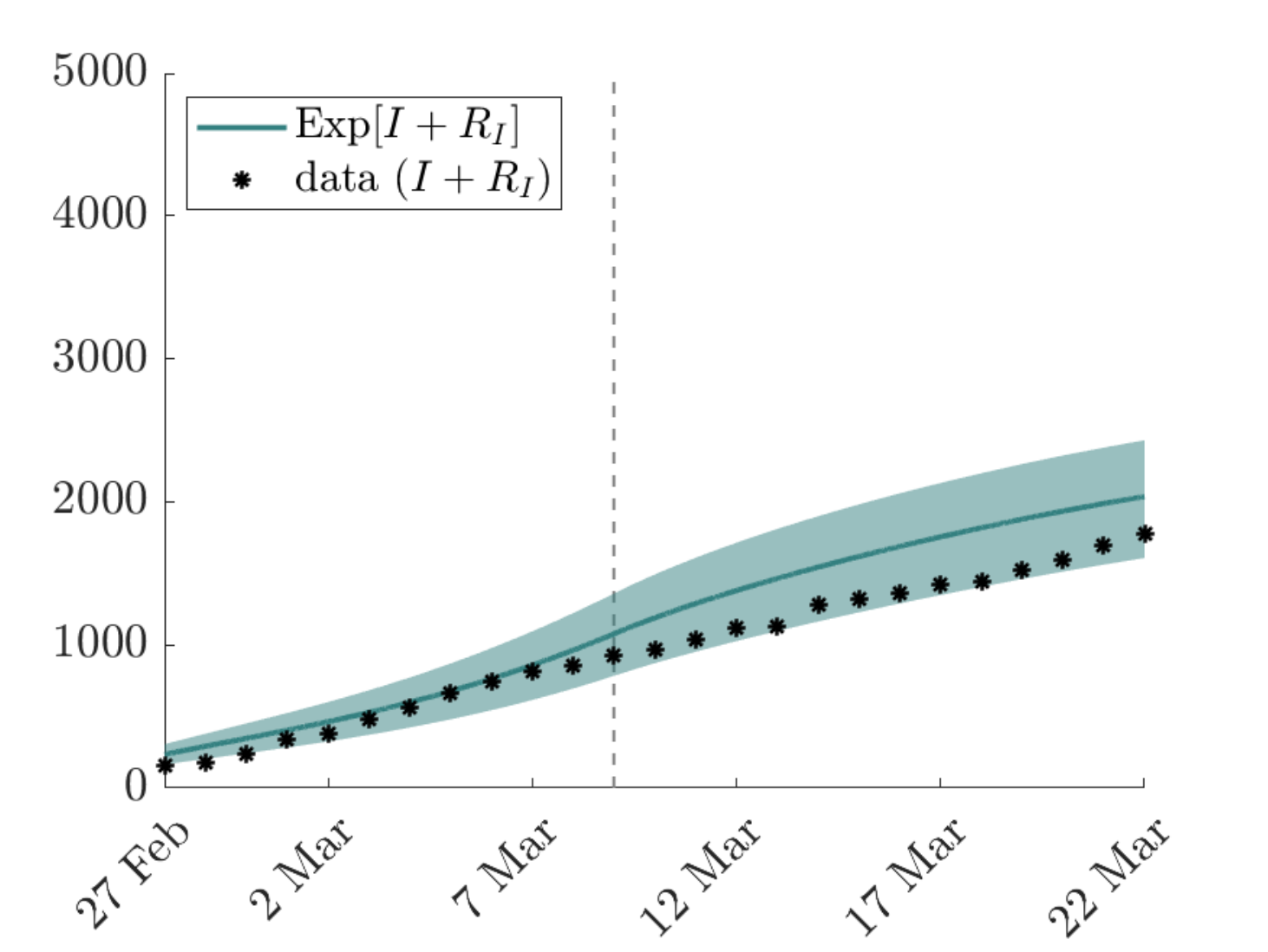}
\caption{Lodi}
\label{fig.LO_RI}
\end{subfigure}
\begin{subfigure}{0.32\textwidth}
\includegraphics[width=1\linewidth]{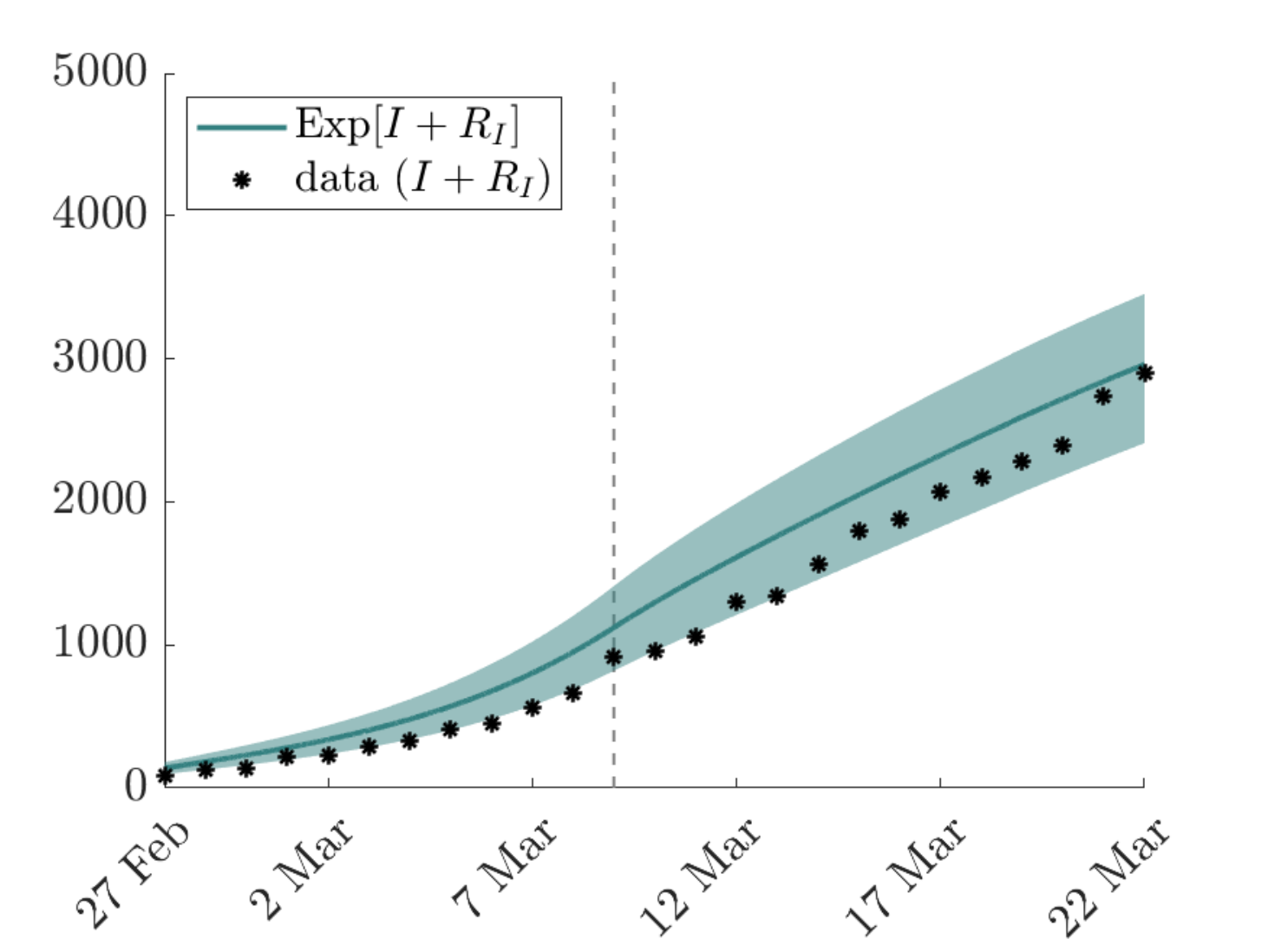}
\caption{Cremona}
\label{fig.CR_RI}
\end{subfigure}
\begin{subfigure}{0.32\textwidth}
\includegraphics[width=1\linewidth]{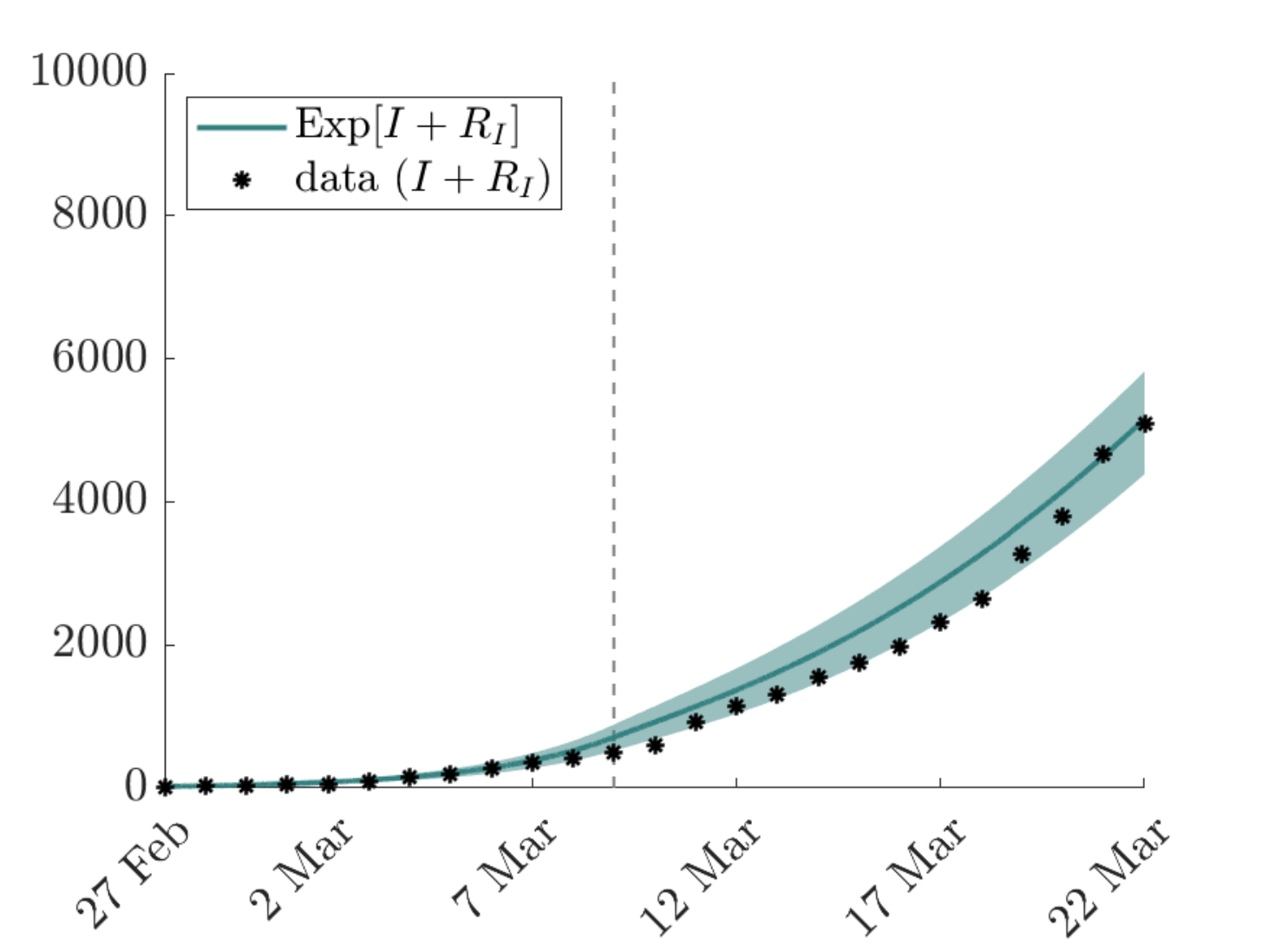}
\caption{Milan}
\label{fig.MI_RI}
\end{subfigure}
\begin{subfigure}{0.32\textwidth}
\includegraphics[width=1\linewidth]{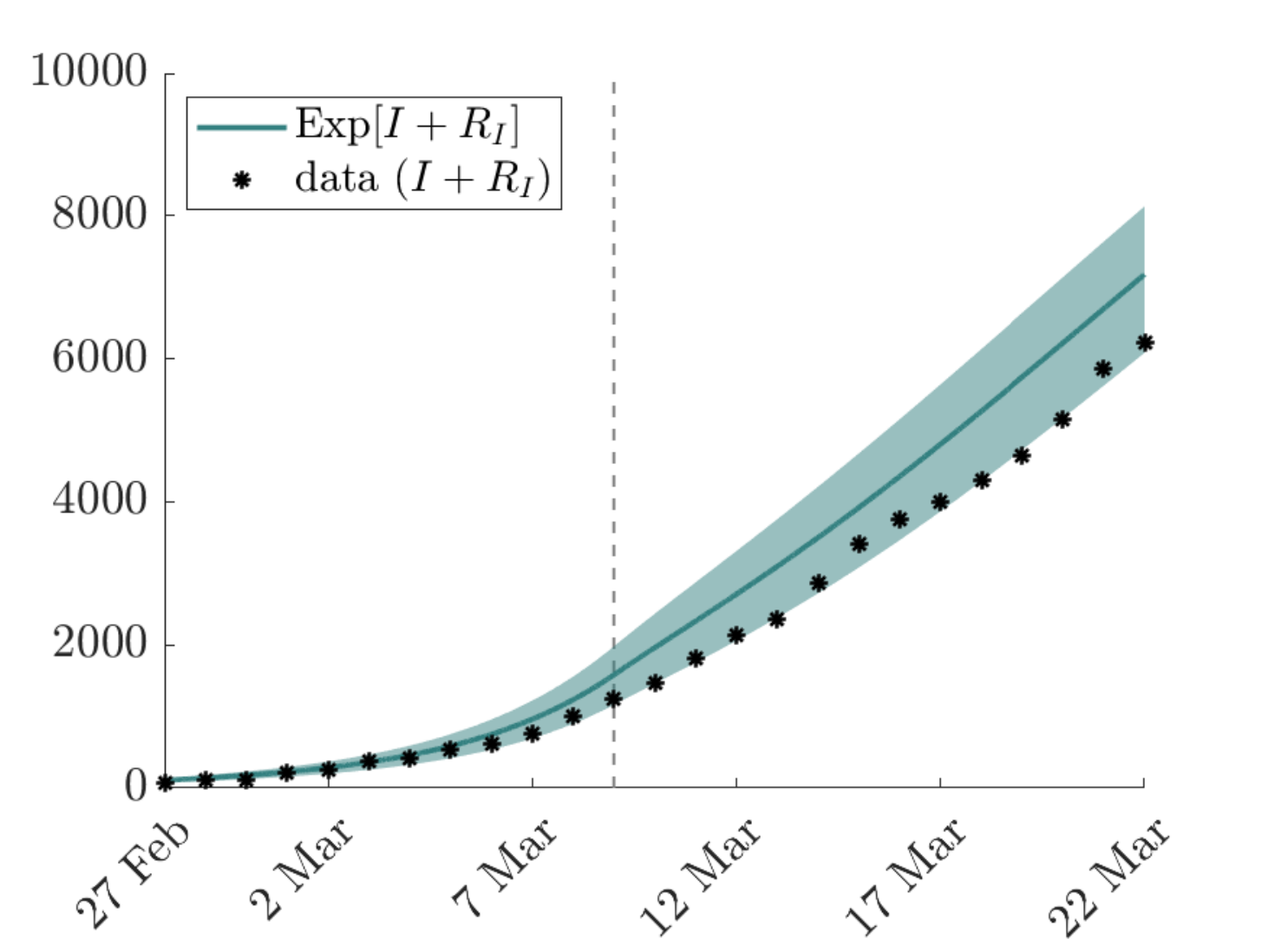}
\caption{Bergamo}
\label{fig.BG_RI}
\end{subfigure}
\begin{subfigure}{0.32\textwidth}
\includegraphics[width=1\linewidth]{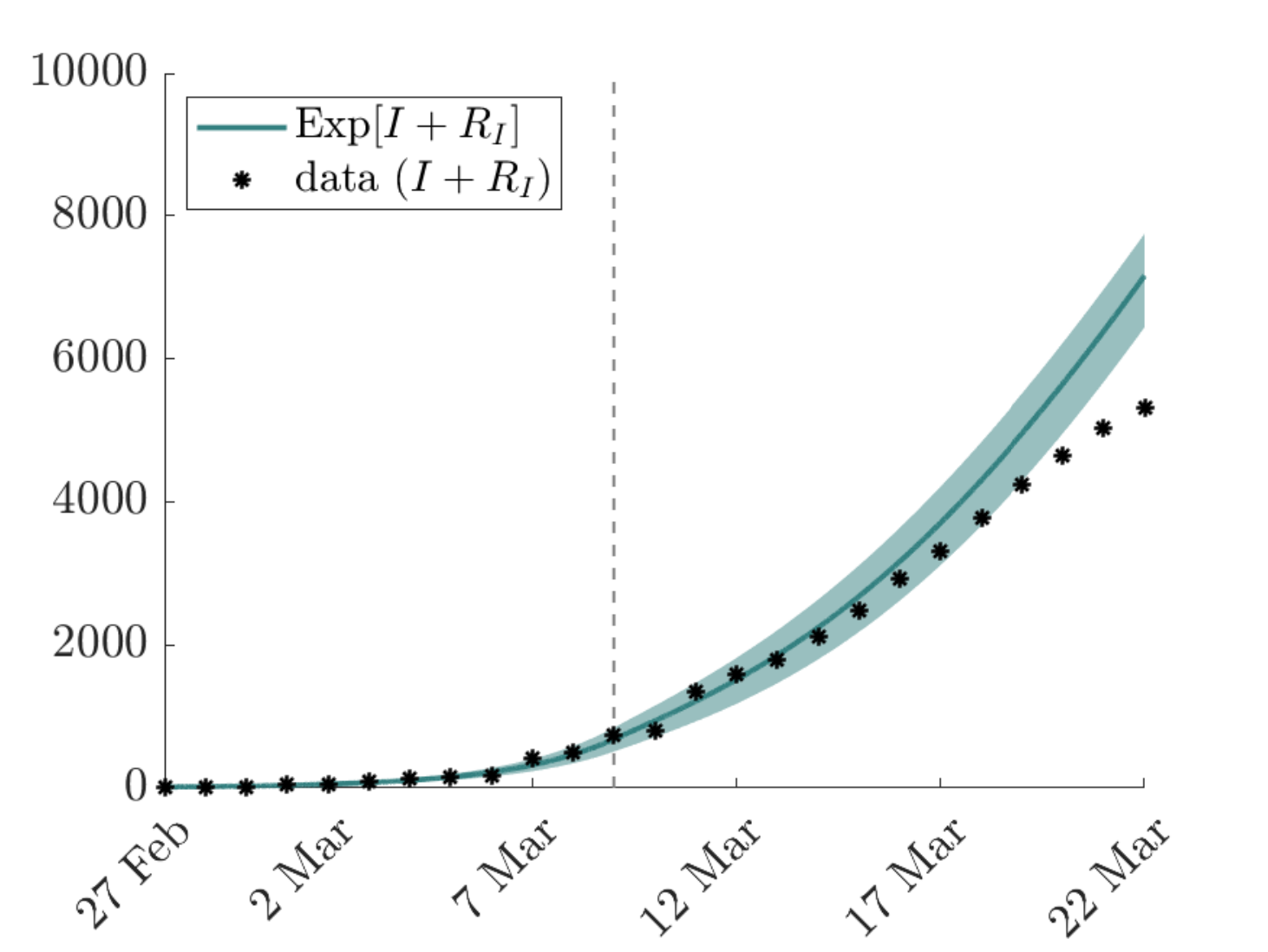}
\caption{Brescia}
\label{fig.BS_RI}
\end{subfigure}
\begin{subfigure}{0.32\textwidth}
\includegraphics[width=1\linewidth]{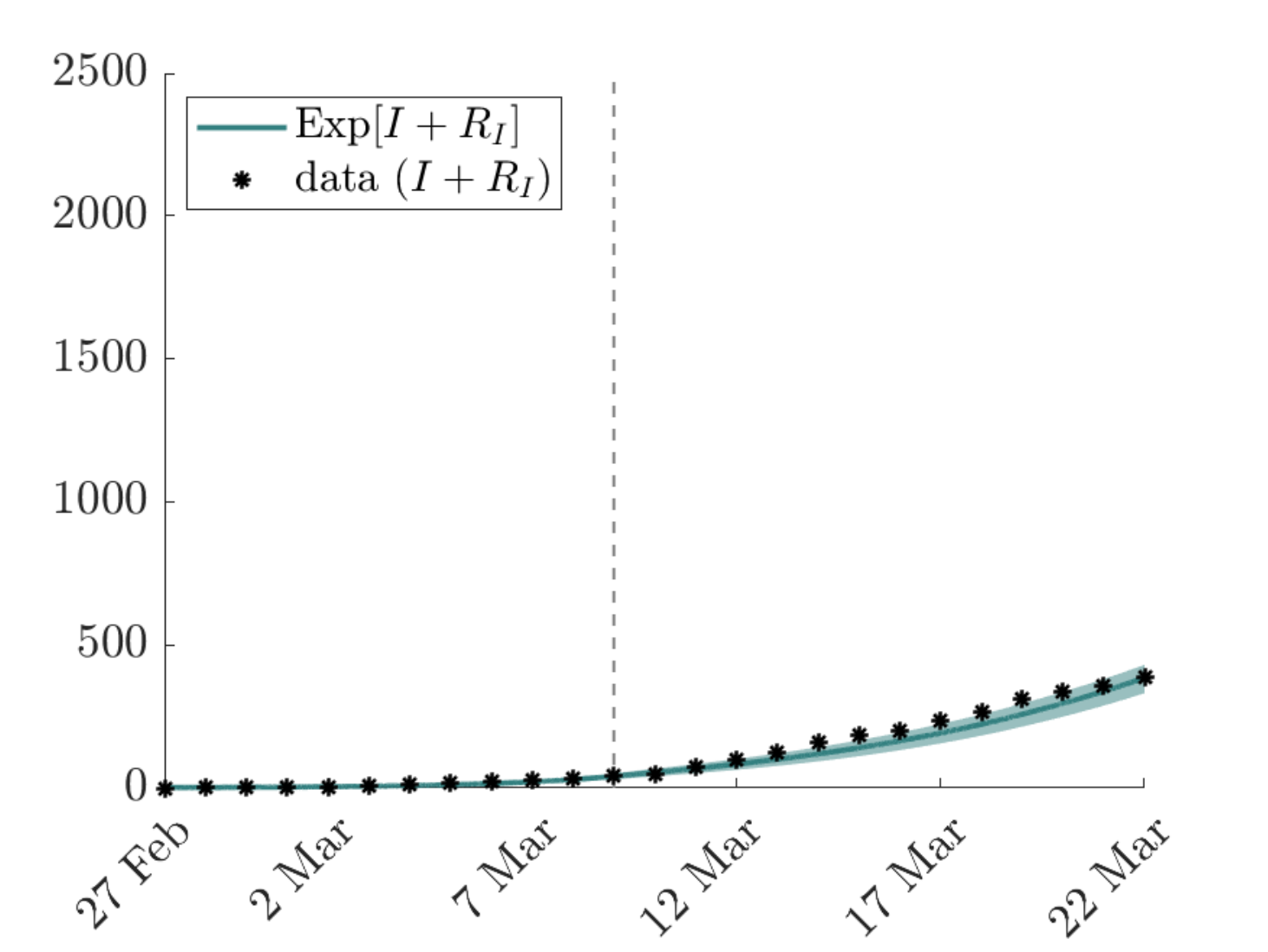}
\caption{Varese}
\label{fig.VA_RI}
\end{subfigure}
\begin{subfigure}{0.32\textwidth}
\includegraphics[width=1\linewidth]{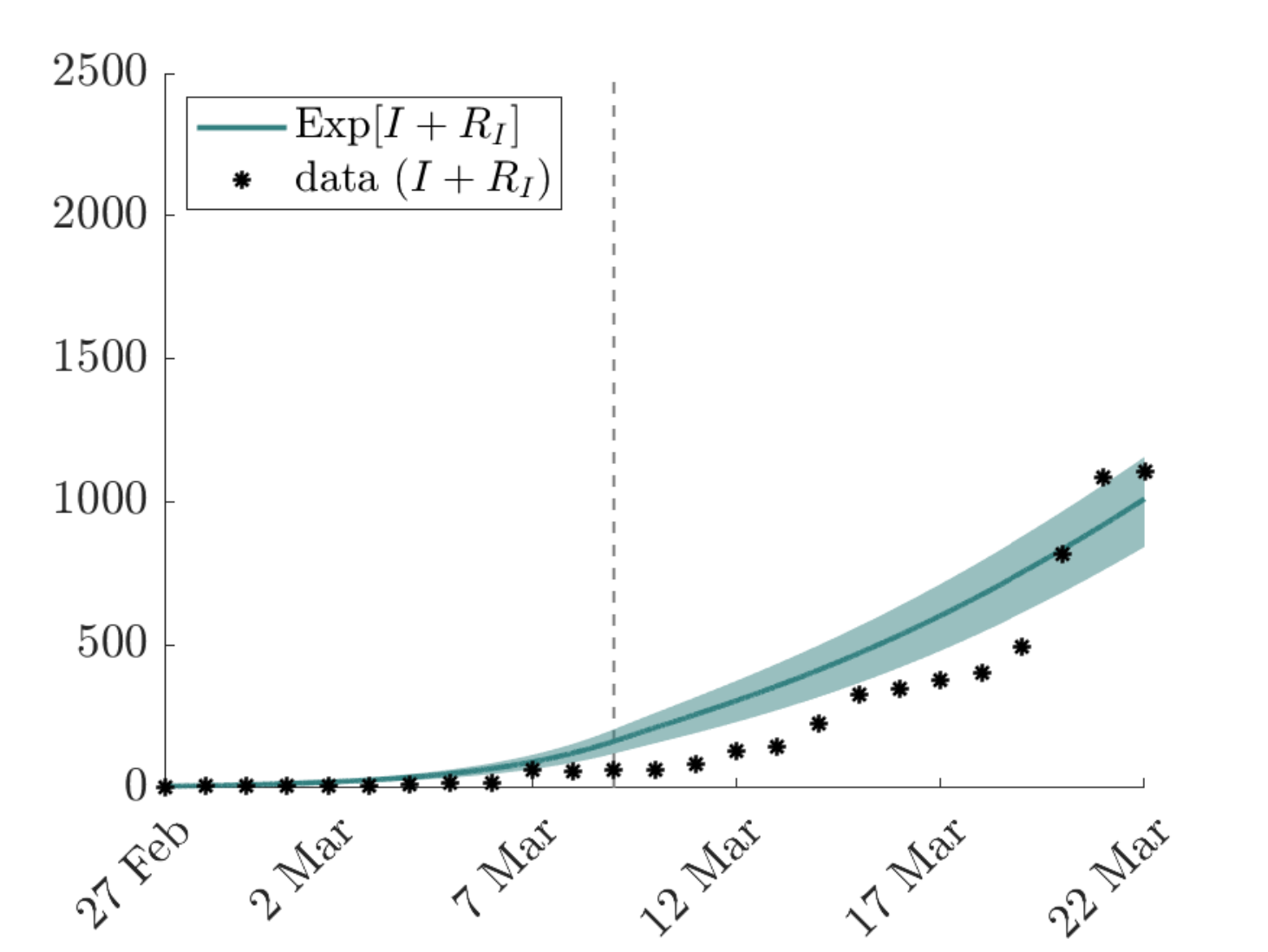}
\caption{Monza-Brianza}
\label{fig.MB_RI}
\end{subfigure}
\begin{subfigure}{0.32\textwidth}
\includegraphics[width=1\linewidth]{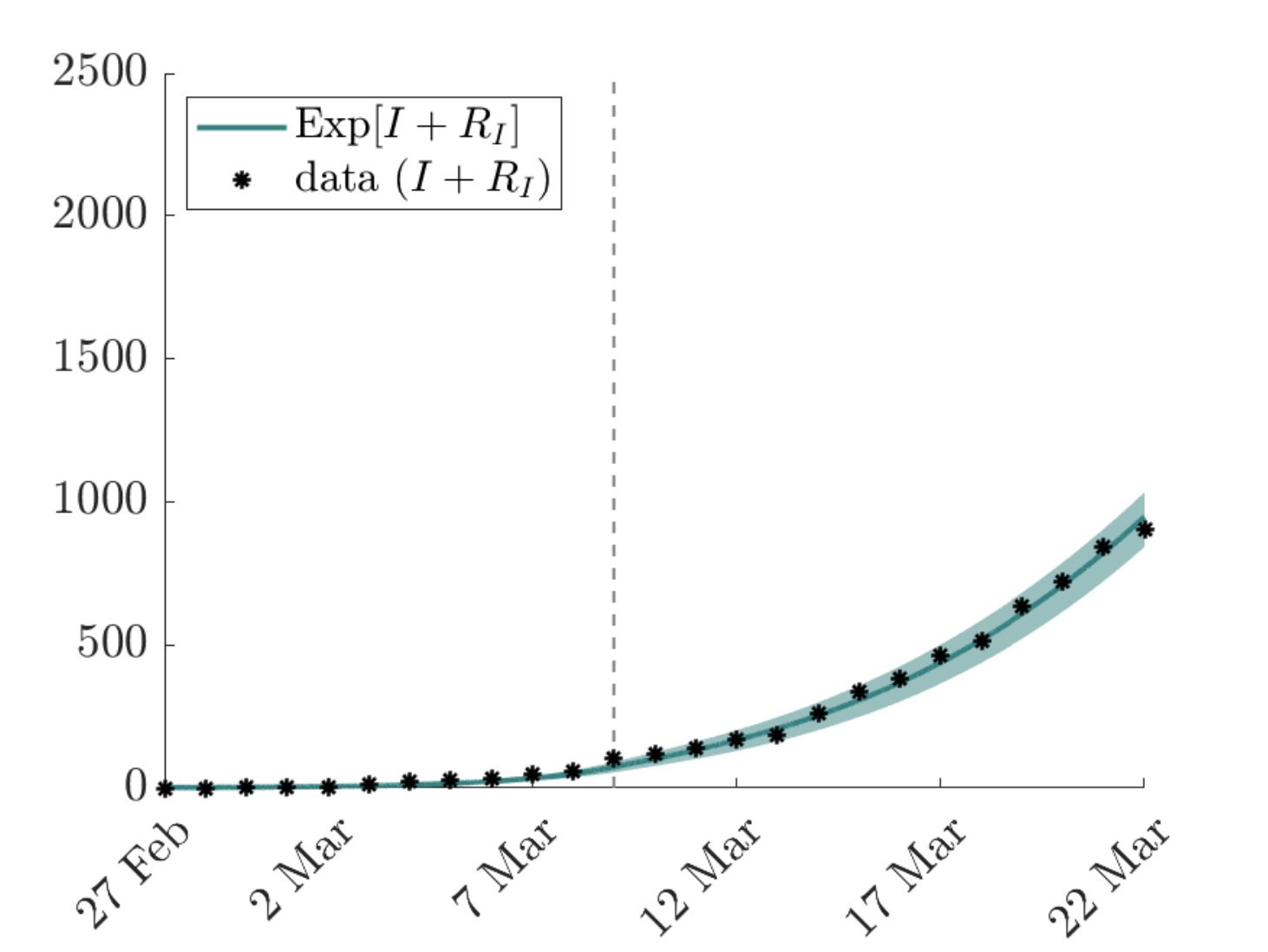}
\caption{Mantua}
\label{fig.MN_RI}
\end{subfigure}
\begin{subfigure}{0.32\textwidth}
\includegraphics[width=1\linewidth]{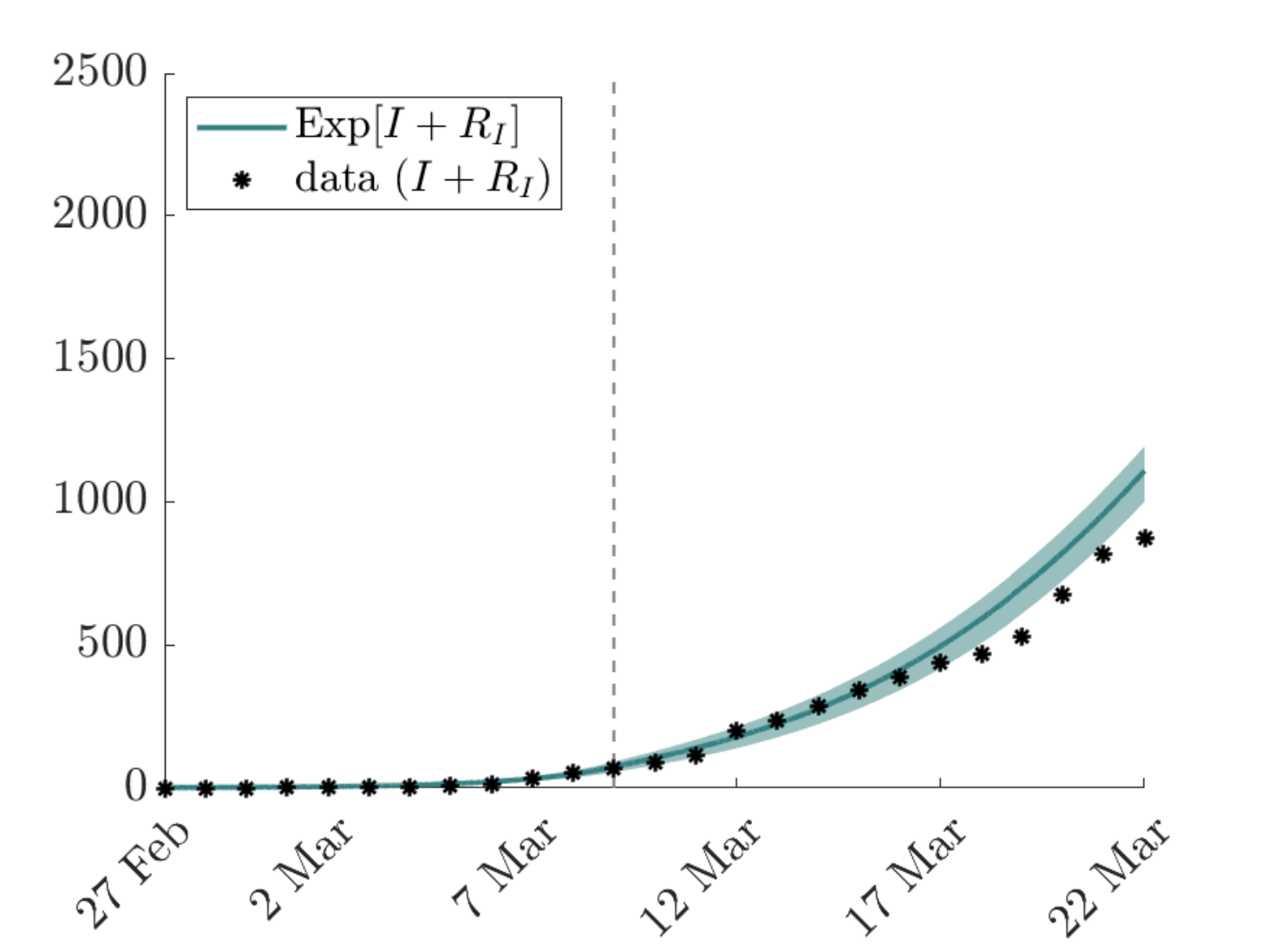}
\caption{Lecco}
\label{fig.LC_RI}
\end{subfigure}
\begin{subfigure}{0.32\textwidth}
\includegraphics[width=1\linewidth]{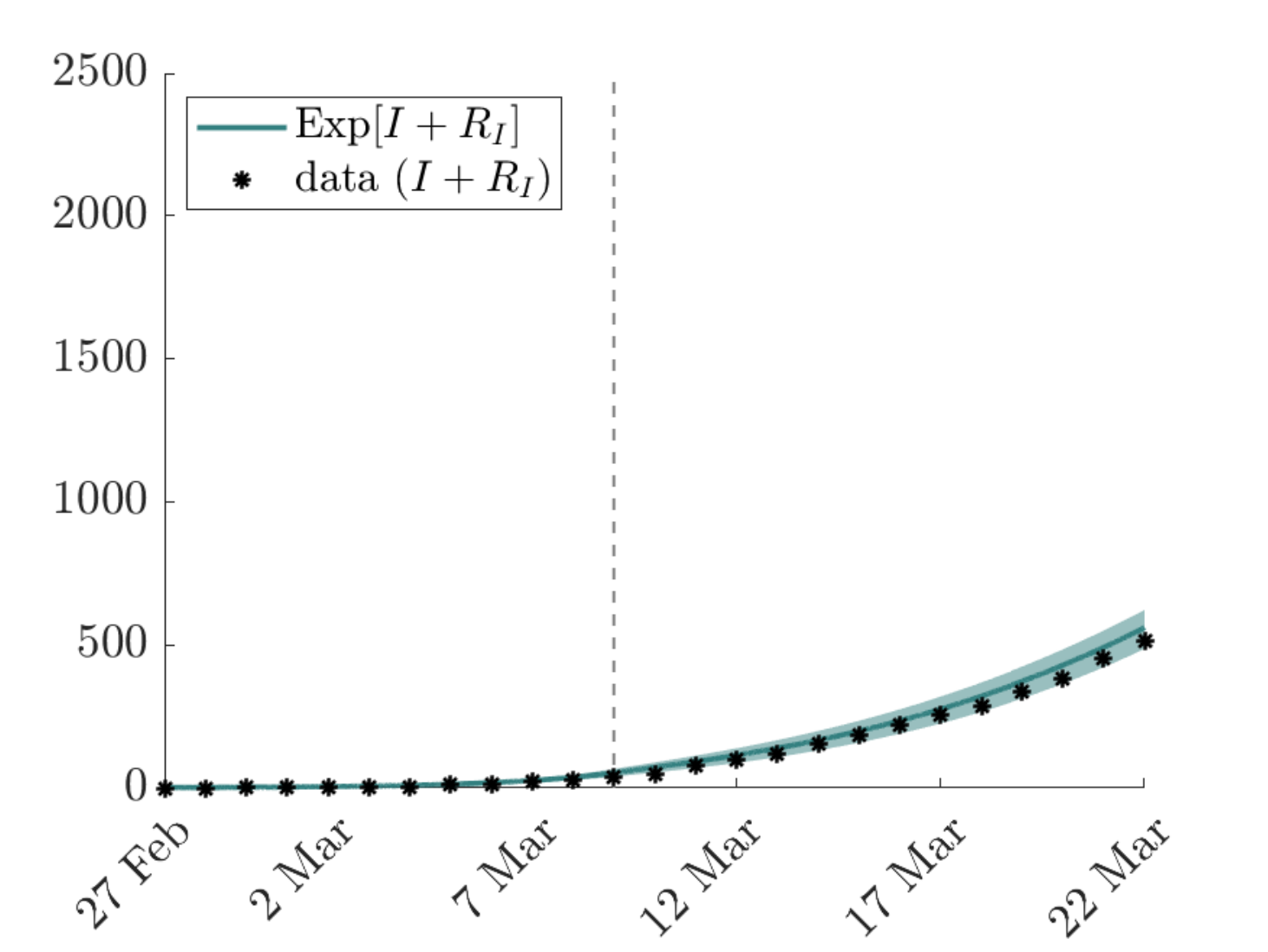}
\caption{Como}
\label{fig.CO_RI}
\end{subfigure}
\begin{subfigure}{0.32\textwidth}
\includegraphics[width=1\linewidth]{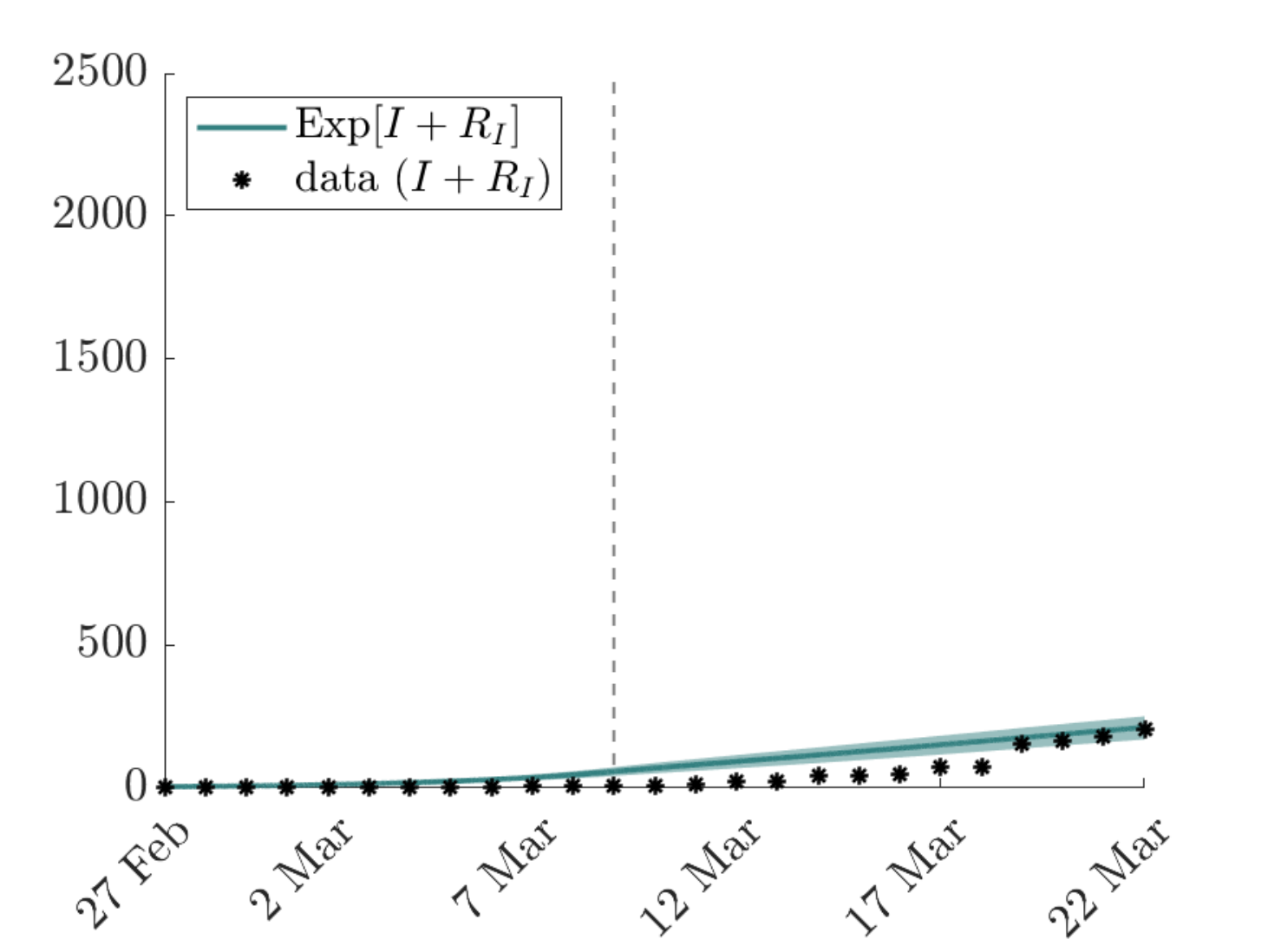}
\caption{Sondrio}
\label{fig.SO_RI}
\end{subfigure}
\caption{Numerical results, with 95\% confidence intervals, of the simulation of the first outbreak of COVID-19 in Lombardy, Italy. Expected evolution in time of the cumulative amount of severe infectious ($I+R_I$) compared with data of cumulative infectious taken from the COVID-19 repository of the Civil Protection Department of Italy \cite{prot_civile}. Vertical dashed lines identify the onset of governmental lockdown restrictions.}
\label{fig.results_Lombardy_RI}
\end{figure}
\begin{figure}[p!]
\centering
\begin{subfigure}{0.32\textwidth}
\includegraphics[width=1\linewidth]{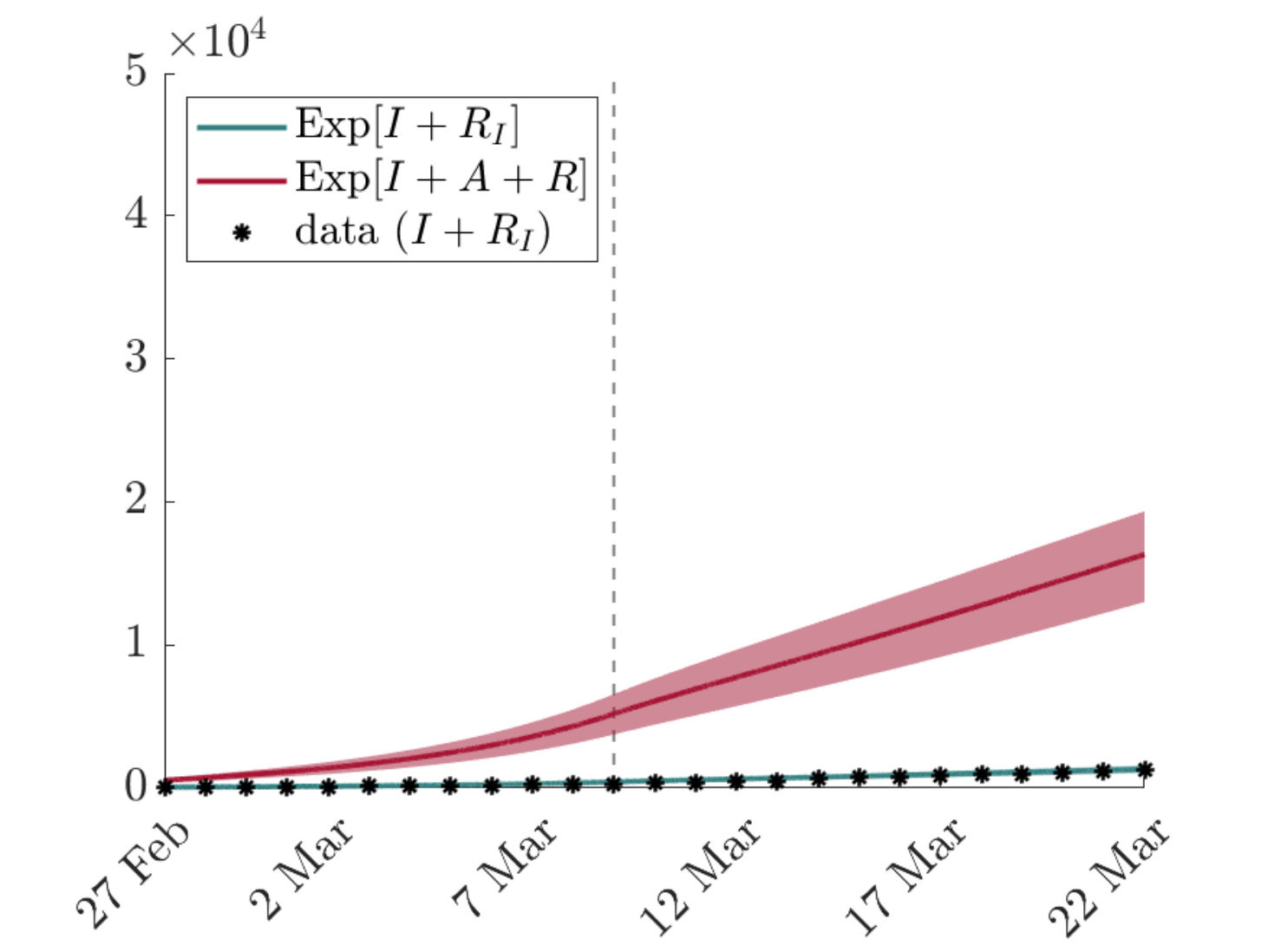}
\caption{Pavia}
\label{fig.PV_RIA}
\end{subfigure}
\begin{subfigure}{0.32\textwidth}
\includegraphics[width=1\linewidth]{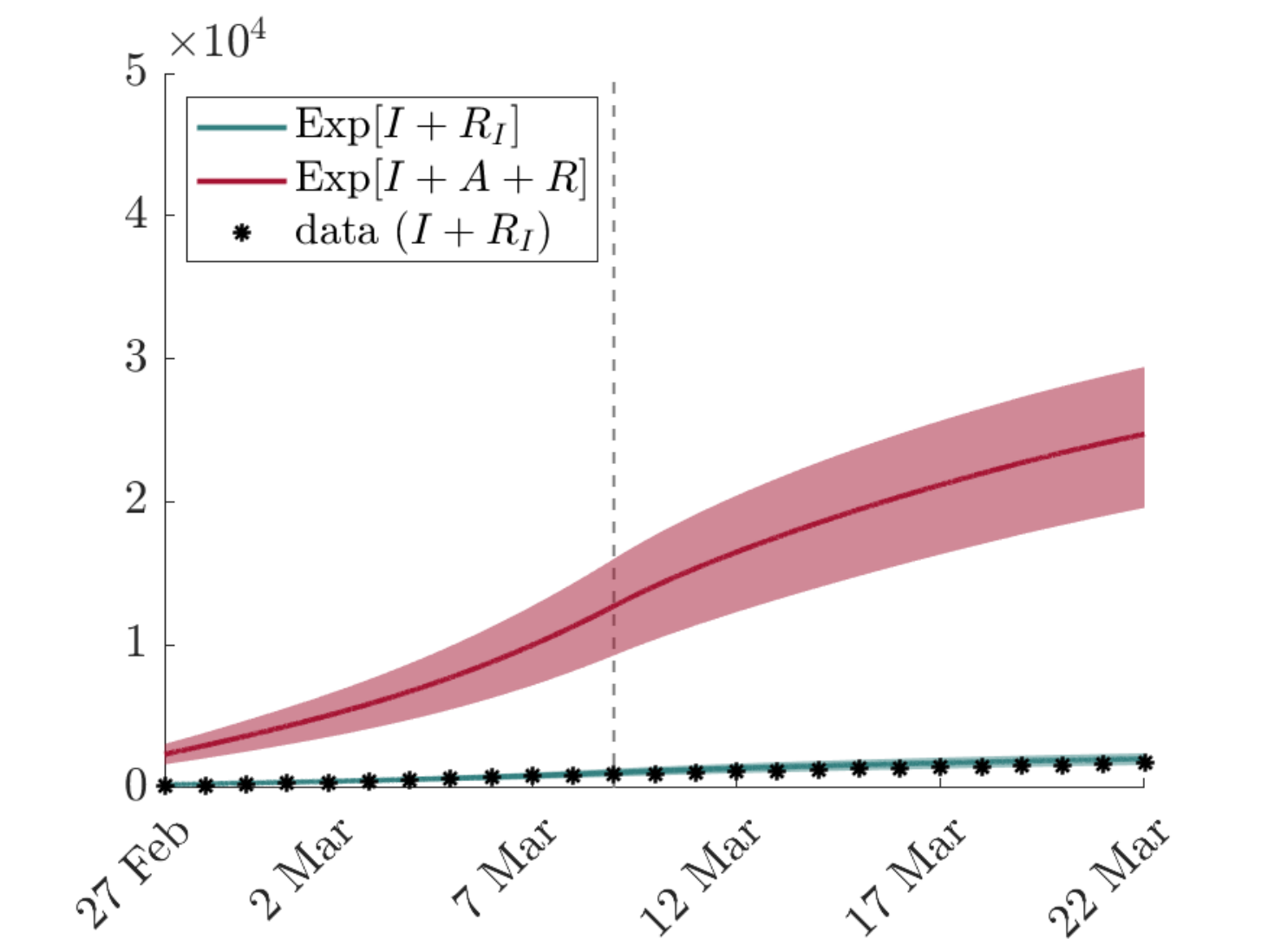}
\caption{Lodi}
\label{fig.LO_RIA}
\end{subfigure}
\begin{subfigure}{0.32\textwidth}
\includegraphics[width=1\linewidth]{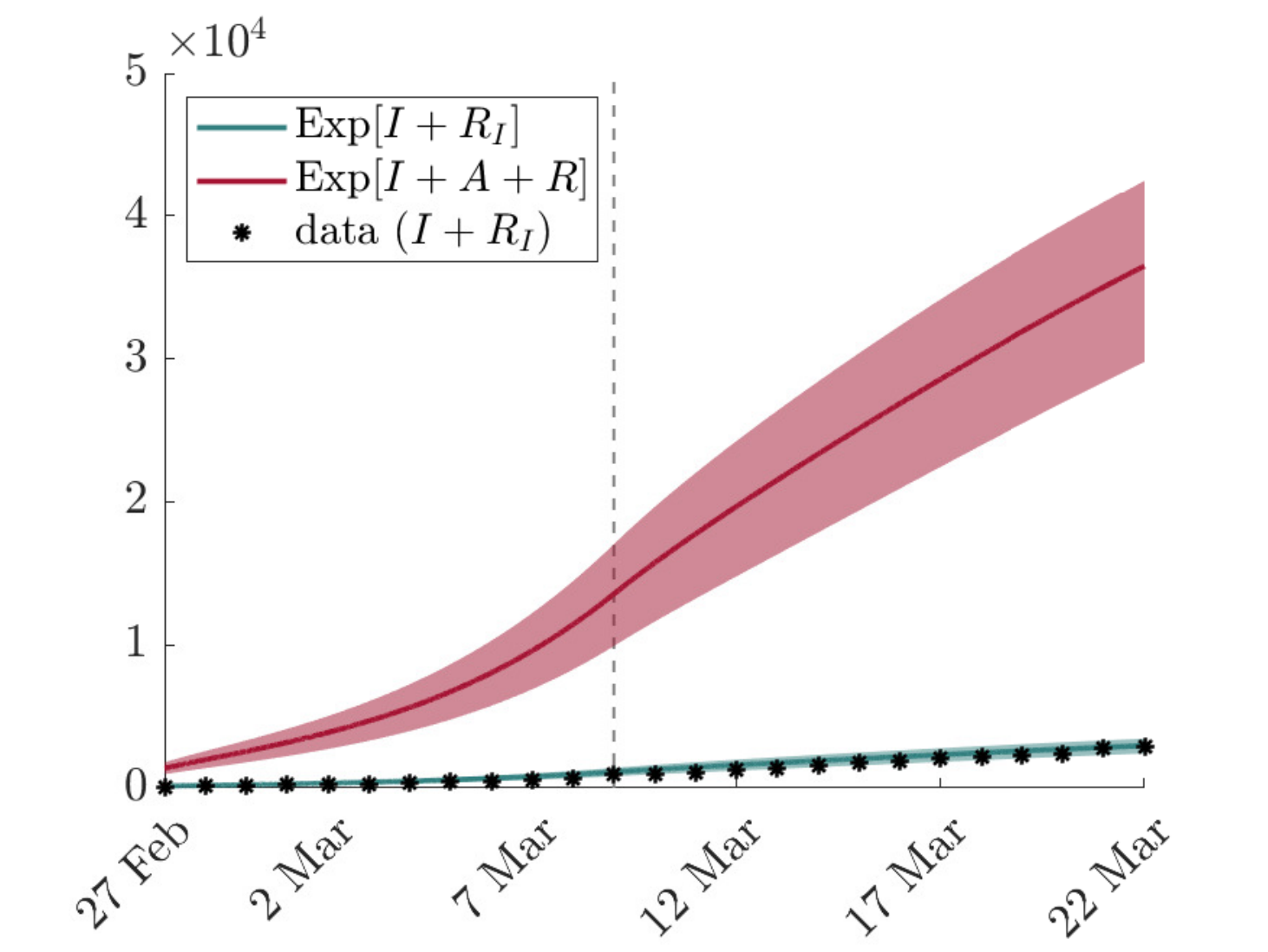}
\caption{Cremona}
\label{fig.CR_RIA}
\end{subfigure}
\begin{subfigure}{0.32\textwidth}
\includegraphics[width=1\linewidth]{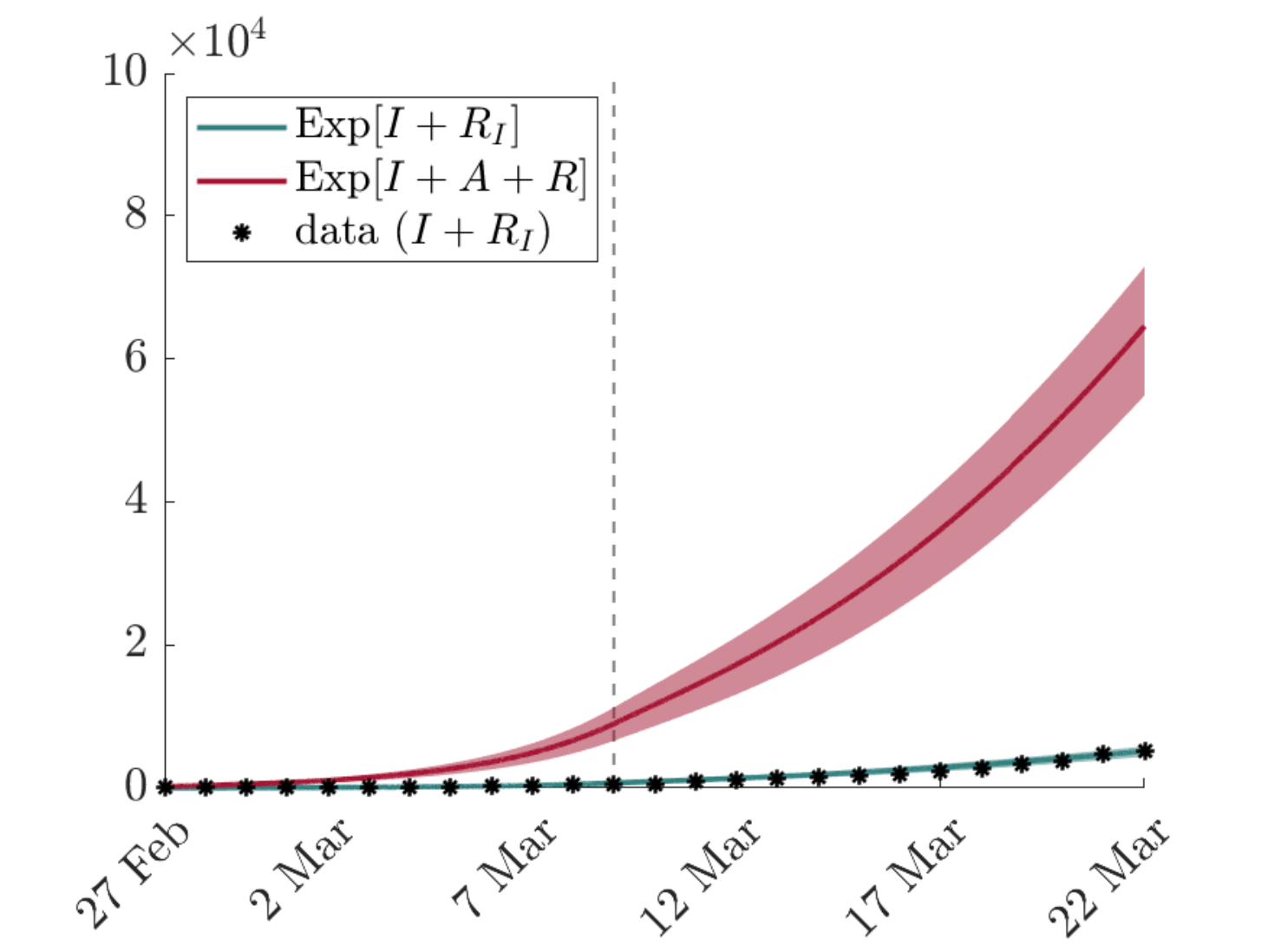}
\caption{Milan}
\label{fig.MI_RIA}
\end{subfigure}
\begin{subfigure}{0.32\textwidth}
\includegraphics[width=1\linewidth]{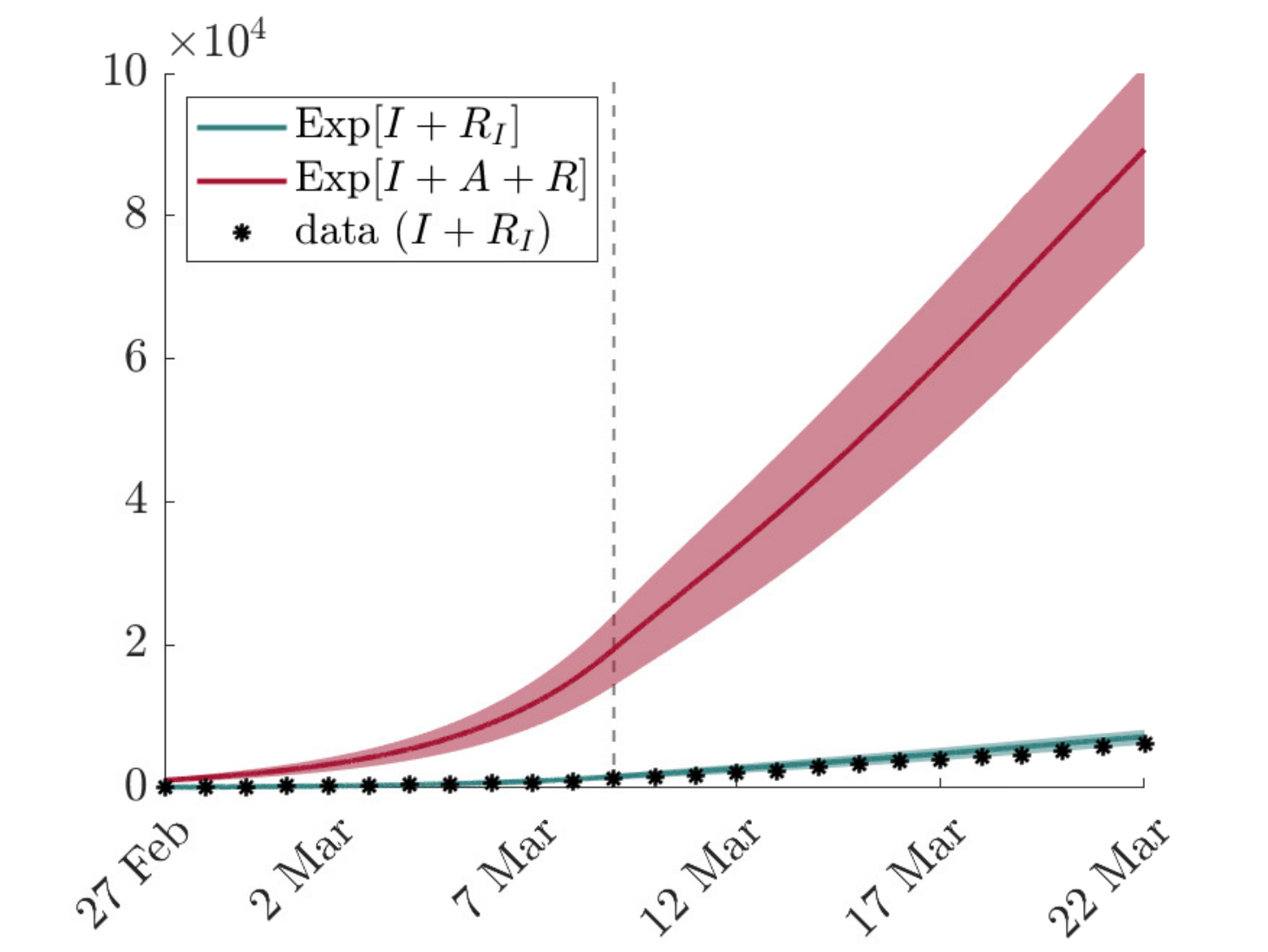}
\caption{Bergamo}
\label{fig.BG_RIA}
\end{subfigure}
\begin{subfigure}{0.32\textwidth}
\includegraphics[width=1\linewidth]{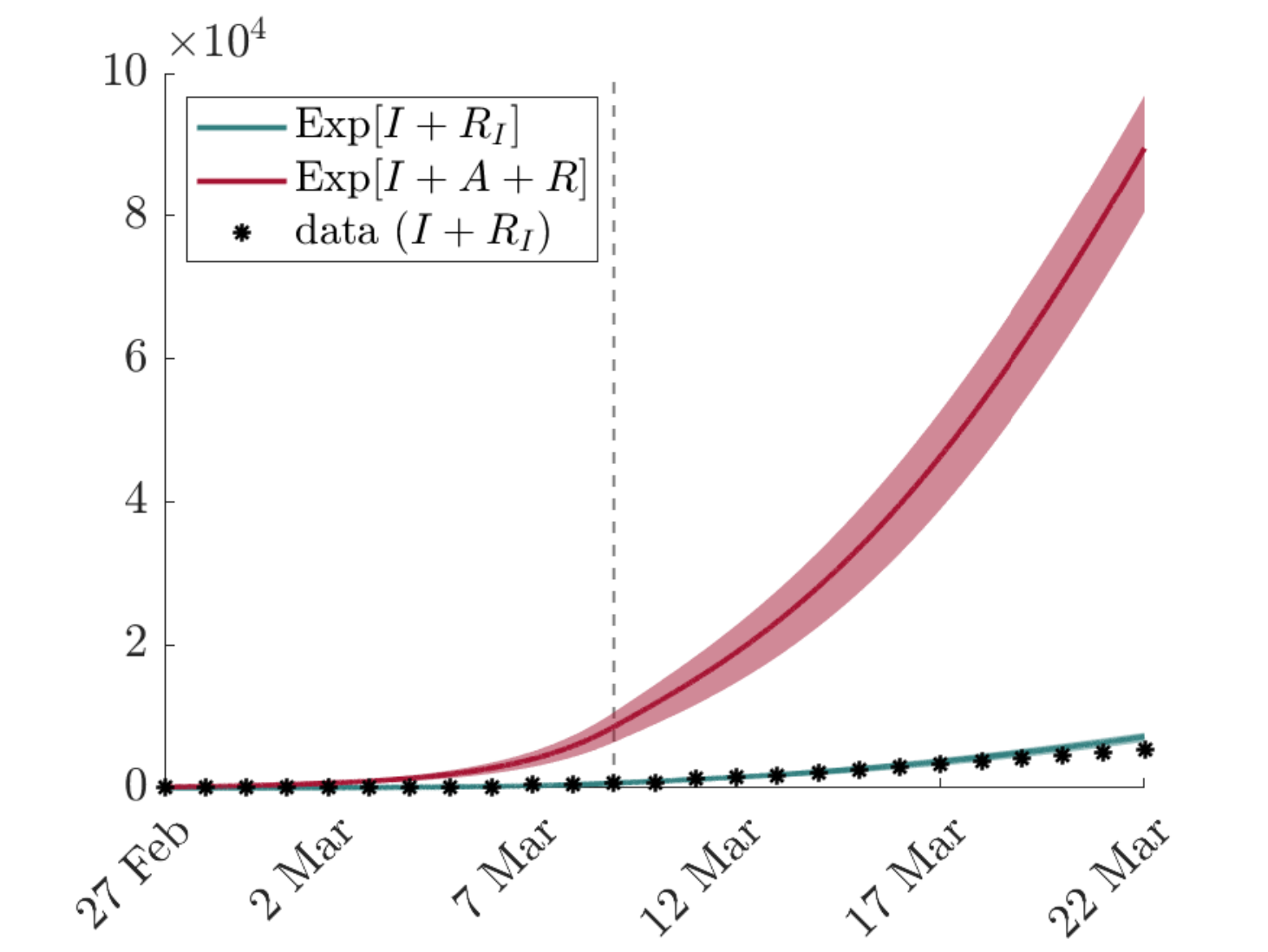}
\caption{Brescia}
\label{fig.BS_RIA}
\end{subfigure}
\begin{subfigure}{0.32\textwidth}
\includegraphics[width=1\linewidth]{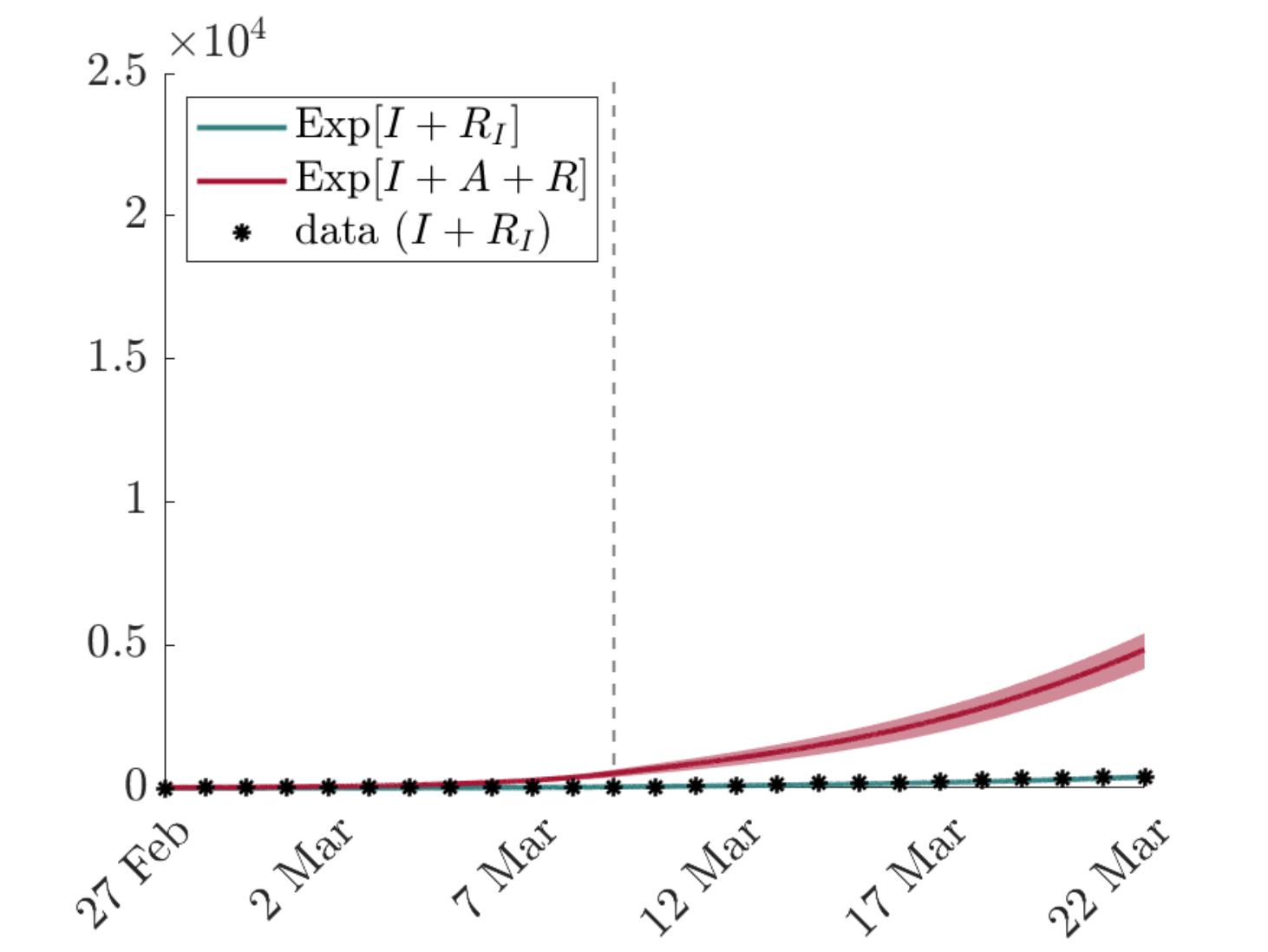}
\caption{Varese}
\label{fig.VA_RIA}
\end{subfigure}
\begin{subfigure}{0.32\textwidth}
\includegraphics[width=1\linewidth]{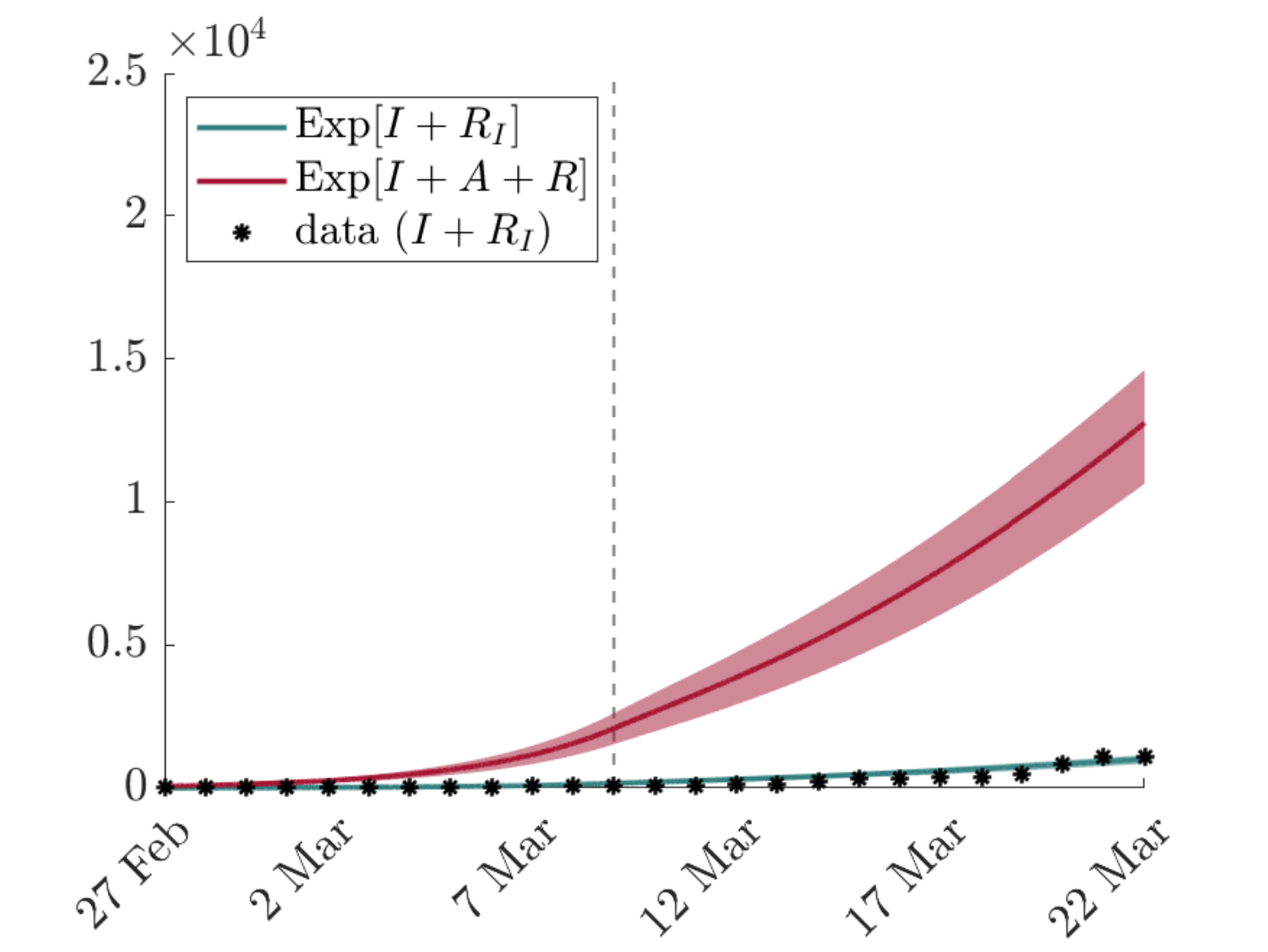}
\caption{Monza-Brianza}
\label{fig.MB_RIA}
\end{subfigure}
\begin{subfigure}{0.32\textwidth}
\includegraphics[width=1\linewidth]{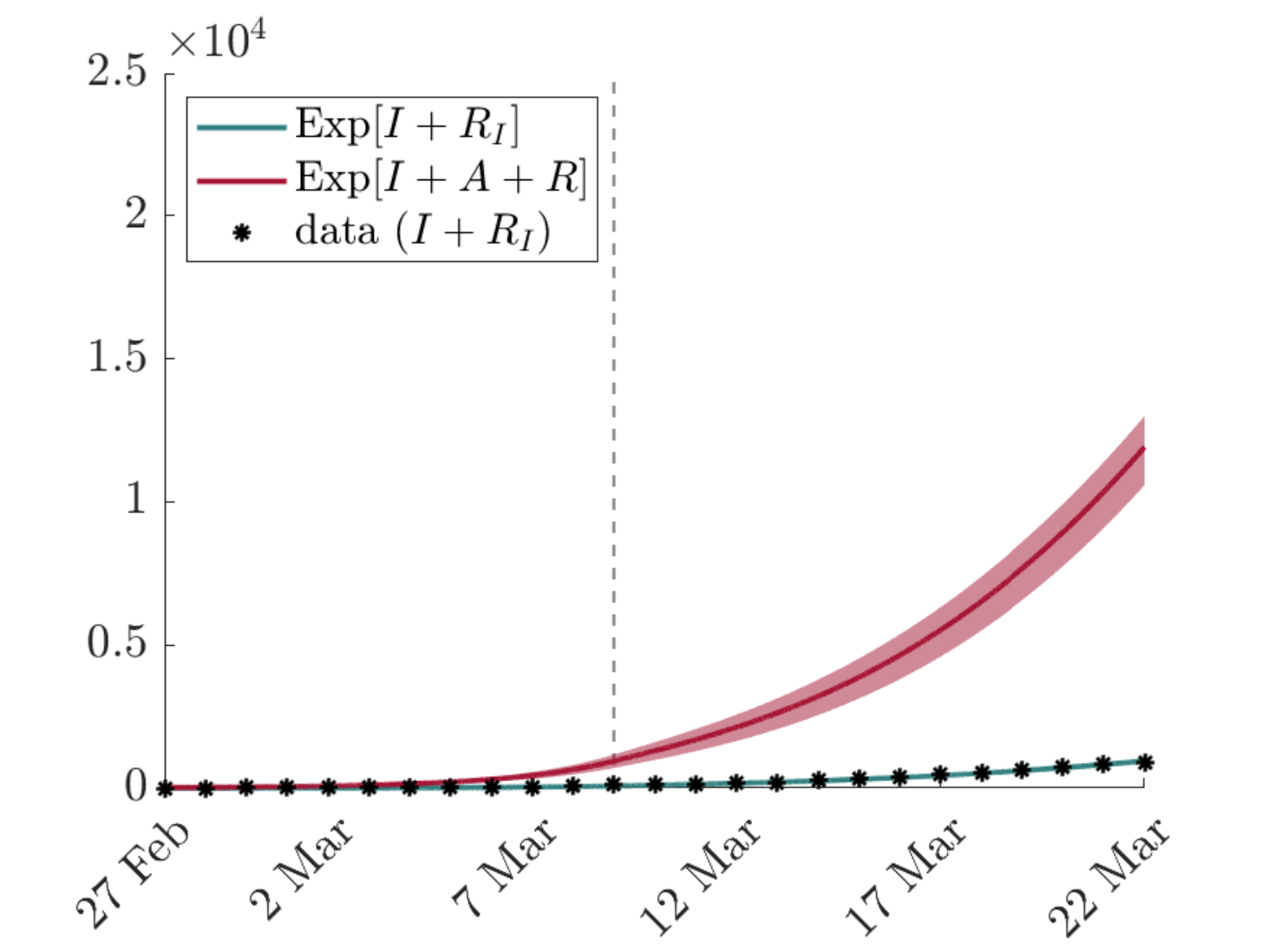}
\caption{Mantua}
\label{fig.MN_RIA}
\end{subfigure}
\begin{subfigure}{0.32\textwidth}
\includegraphics[width=1\linewidth]{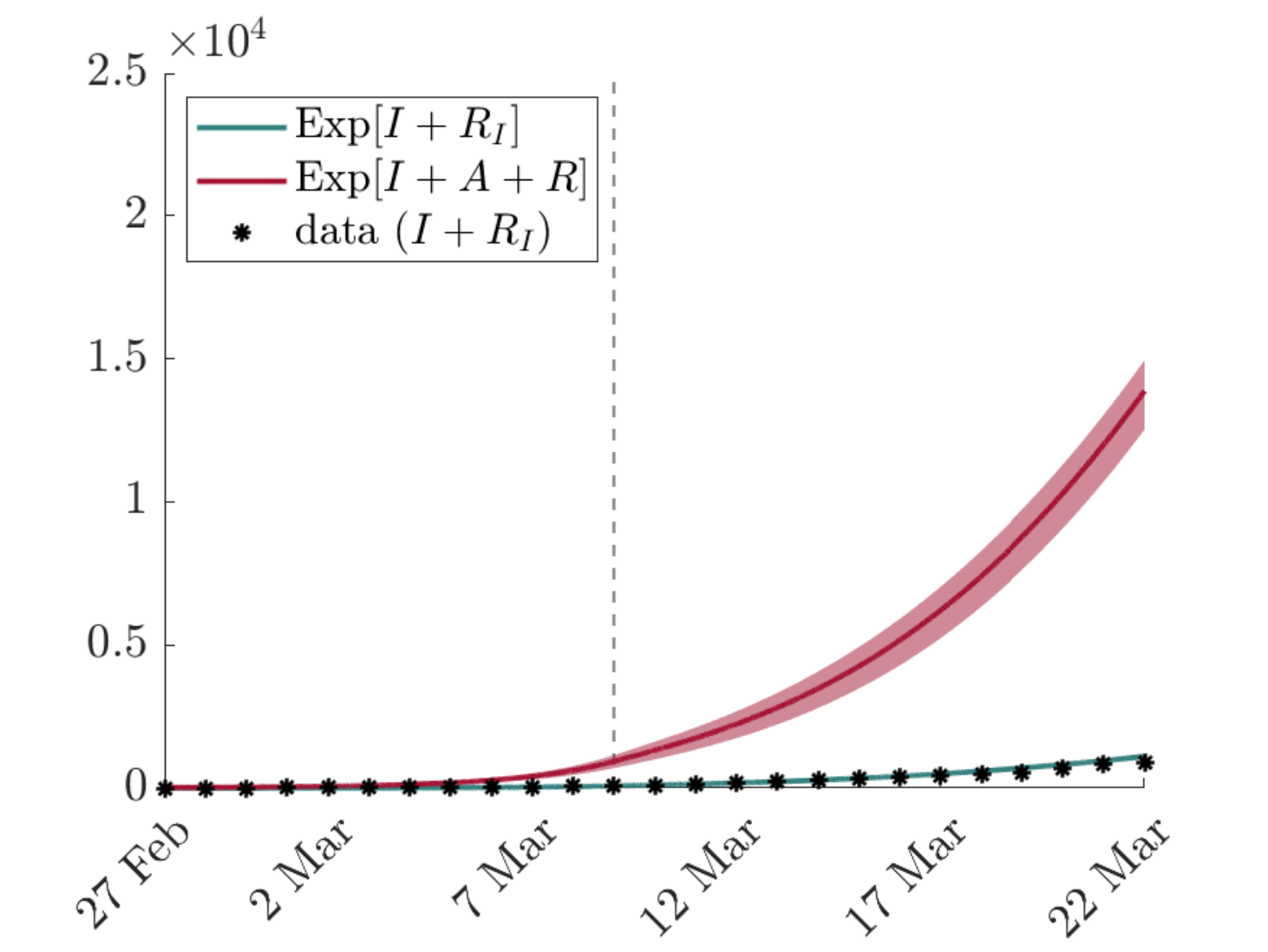}
\caption{Lecco}
\label{fig.LC_RIA}
\end{subfigure}
\begin{subfigure}{0.32\textwidth}
\includegraphics[width=1\linewidth]{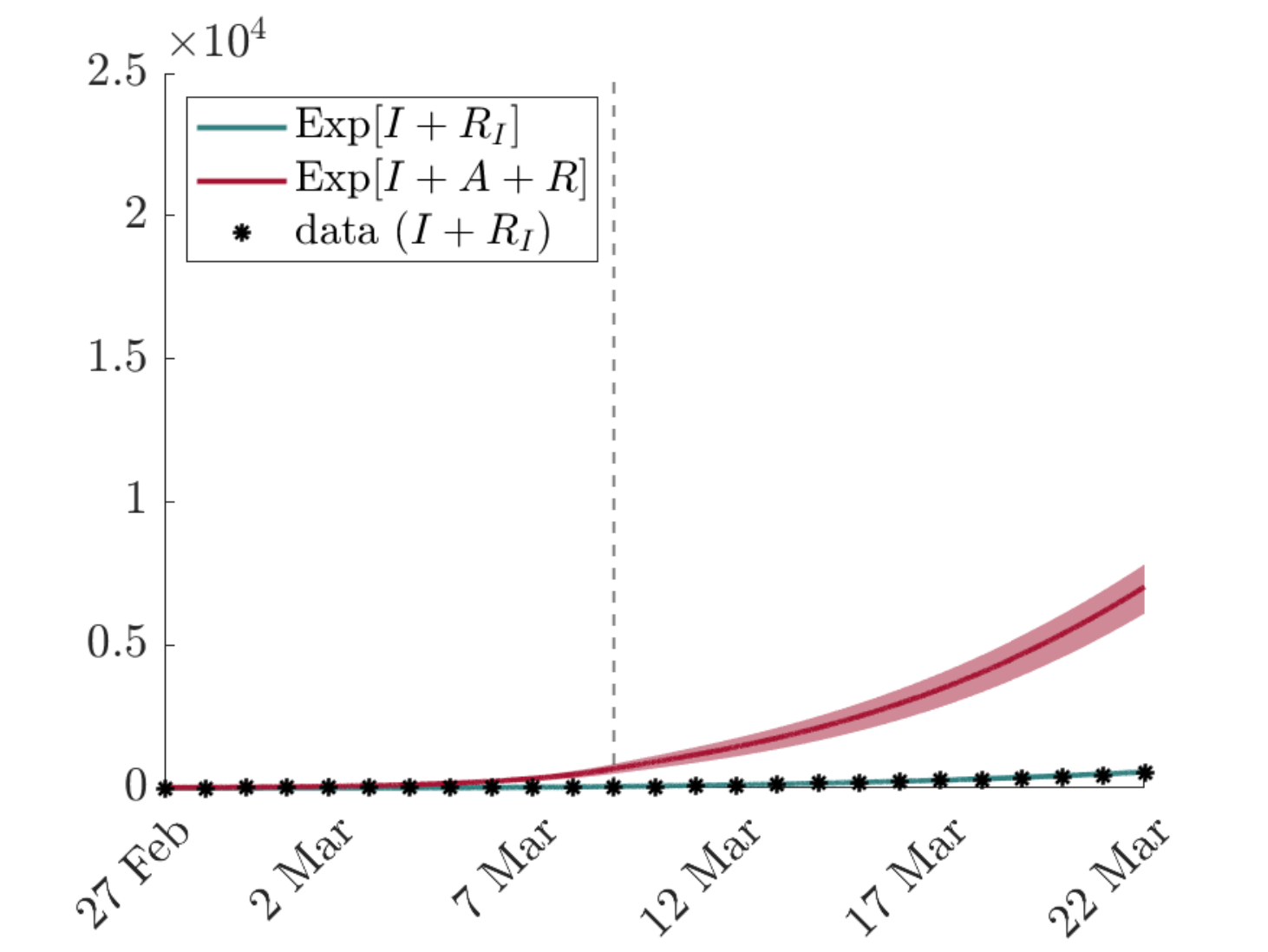}
\caption{Como}
\label{fig.CO_RIA}
\end{subfigure}
\begin{subfigure}{0.32\textwidth}
\includegraphics[width=1\linewidth]{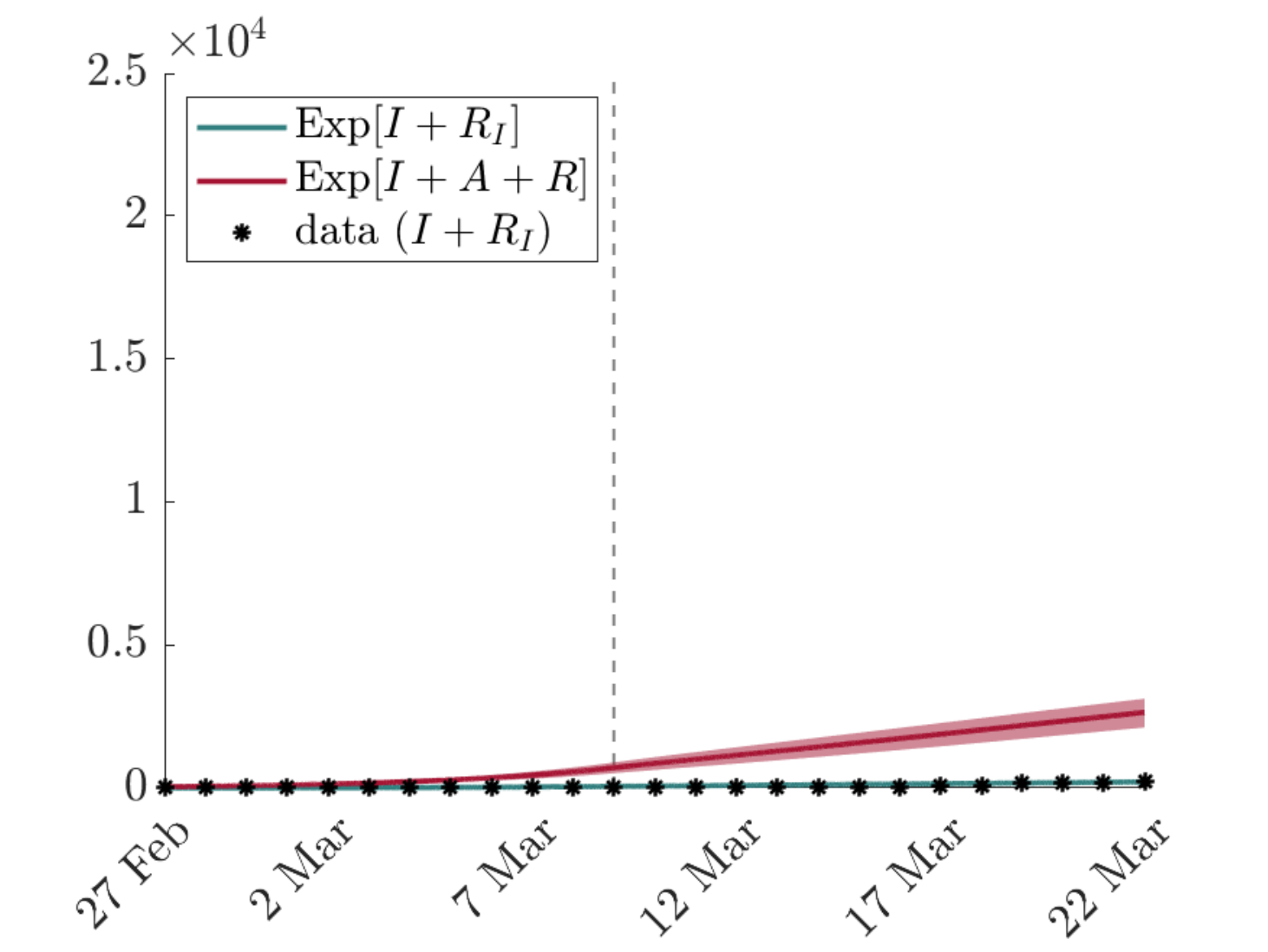}
\caption{Sondrio}
\label{fig.SO_RIA}
\end{subfigure}
\caption{Numerical results, with 95\% confidence intervals, of the simulation of the first outbreak of COVID-19 in Lombardy, Italy. Expected evolution in time of the cumulative amount of severe infectious ($I+R_I$) with respect to the effective cumulative amount of total infectious people, including asymptomatic and mildly symptomatic individuals ($I+A+R$). Data of cumulative infectious is taken from the COVID-19 repository of the Civil Protection Department of Italy \cite{prot_civile}. Vertical dashed lines identify the onset of governmental lockdown restrictions.}
\label{fig.results_Lombardy_RIA}
\end{figure}
\begin{figure}[t!]
\centering
\begin{subfigure}{0.48\textwidth}
\includegraphics[width=1\linewidth]{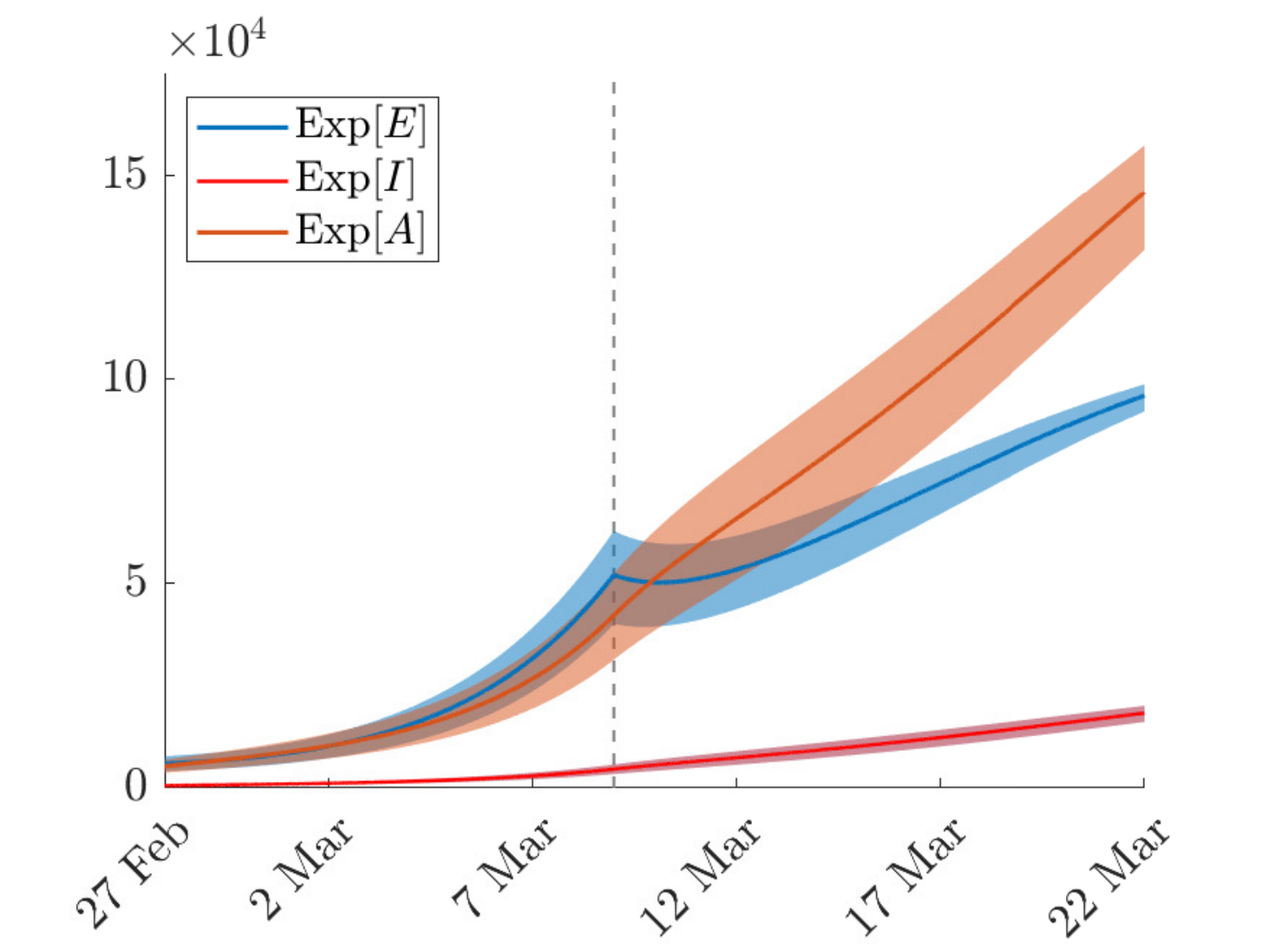}
\caption{}
\label{fig.Lombardia_EIA}
\end{subfigure}
\begin{subfigure}{0.48\textwidth}
\includegraphics[width=1\linewidth]{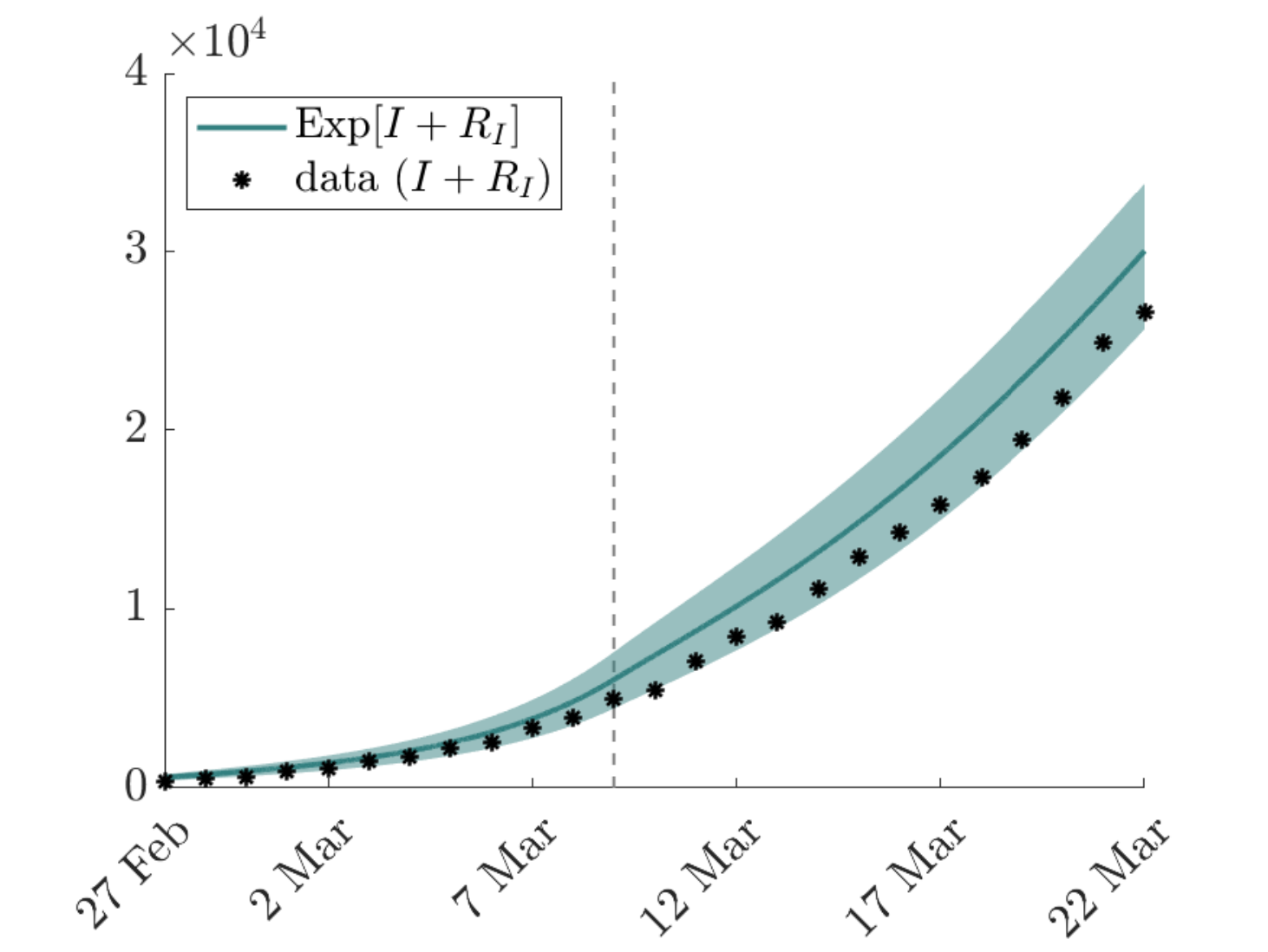}
\caption{}
\label{fig.Lombardia_RI}
\end{subfigure}
\begin{subfigure}{0.48\textwidth}
\includegraphics[width=1\linewidth]{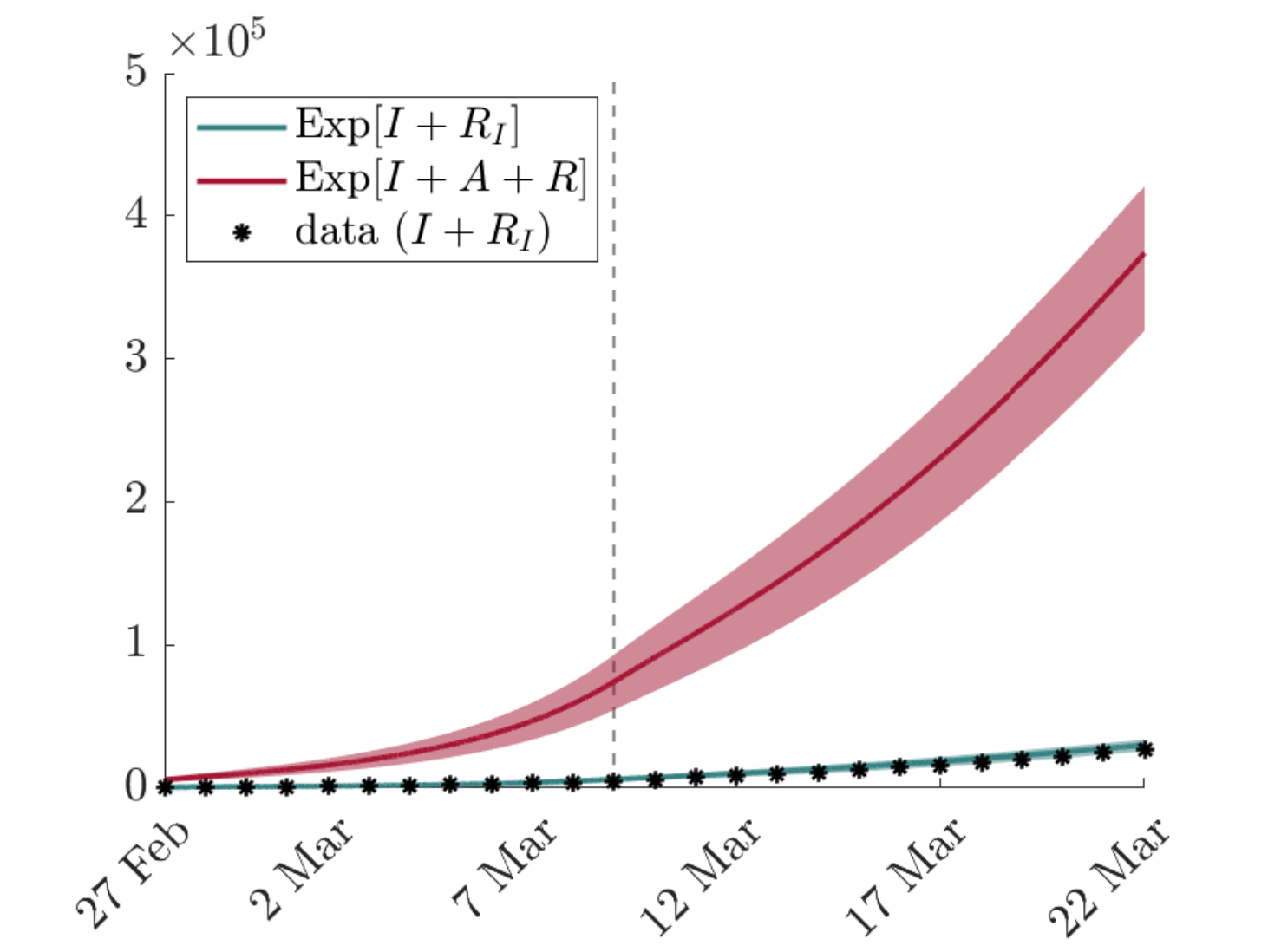}
\caption{}
\label{fig.Lombardia_RIA}
\end{subfigure}
\begin{subfigure}{0.48\textwidth}
\includegraphics[width=1\linewidth]{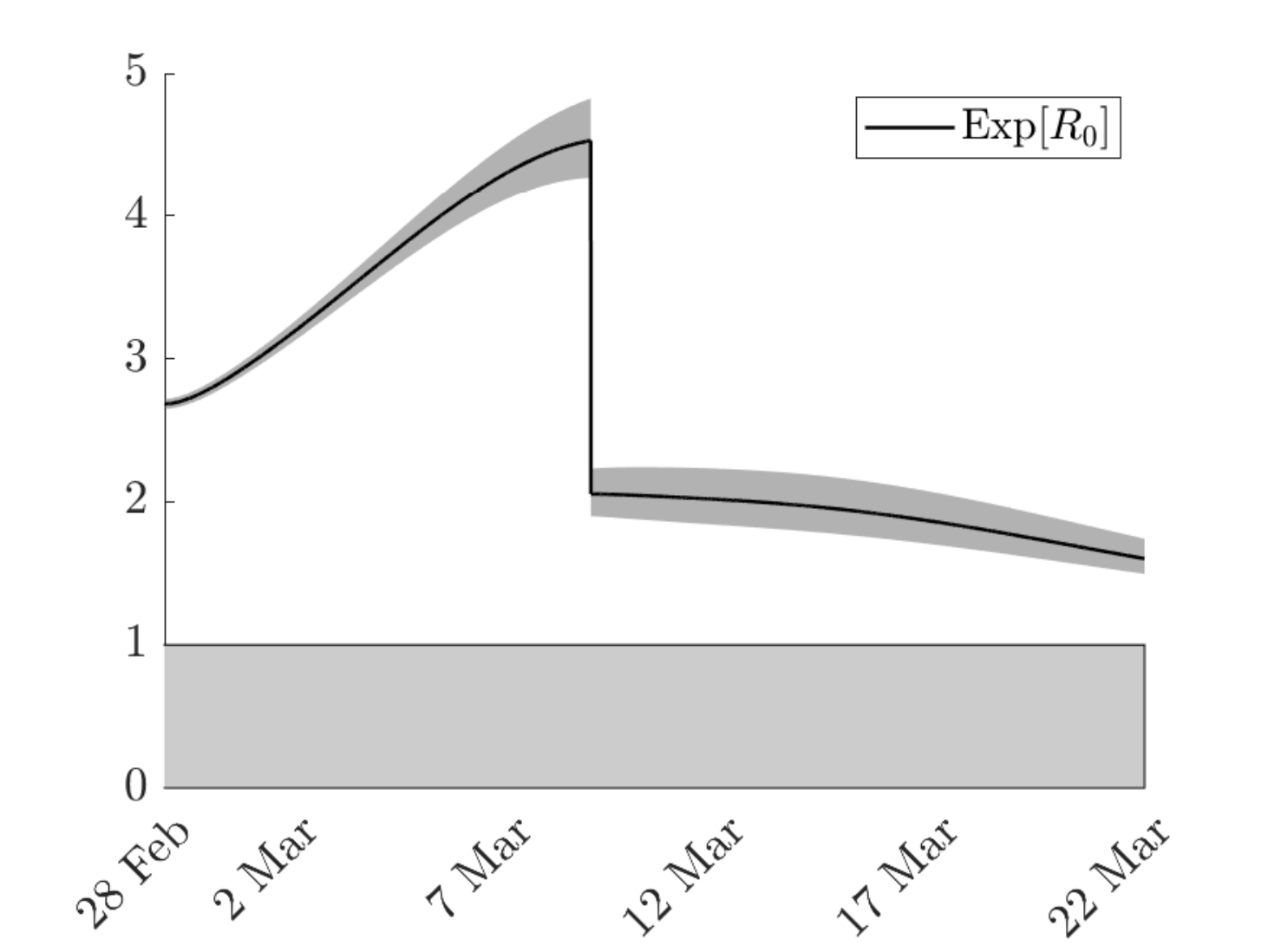}
\caption{}
\label{fig.Lombardia_R0}
\end{subfigure}
\caption{Numerical results, with 95\% confidence intervals, of the simulation of the first outbreak of COVID-19 in Lombardy, Italy. Expected evolution in time, for the whole Region, of: compartments $E$, $A$, $I$ (a); cumulative amount of severe infectious ($I+R_I$) compared with data of cumulative infectious (b); cumulative amount of severe infectious ($I+R_I$) with respect to the effective cumulative amount of total infectious people, including asymptomatic and mildly symptomatic individuals ($I+A+R$) (c); reproduction number $R_0(t)$ (d). Vertical dashed lines identify the onset of governmental lockdown restrictions.}
\label{fig.results_Lombardy_region}
\end{figure}
\begin{figure}[t!]
\centering
\begin{subfigure}{0.48\textwidth}
\includegraphics[width=1\linewidth]{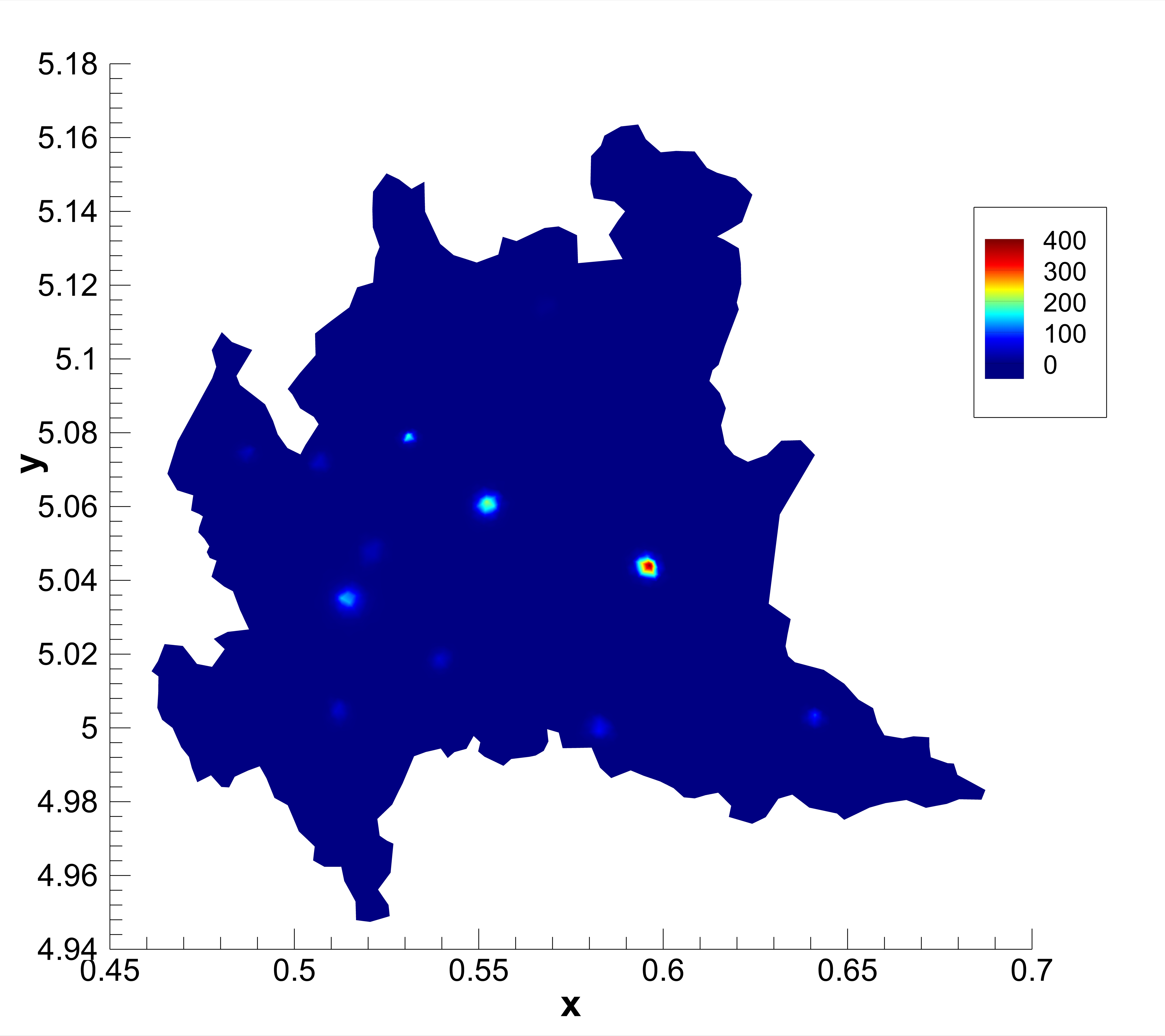}
\caption{Exp[$E_T(x,y)+I_T(x,y)+A_T(x,y)$]}
\label{fig.expEIA_final}
\end{subfigure}
\begin{subfigure}{0.48\textwidth}
\includegraphics[width=1\linewidth]{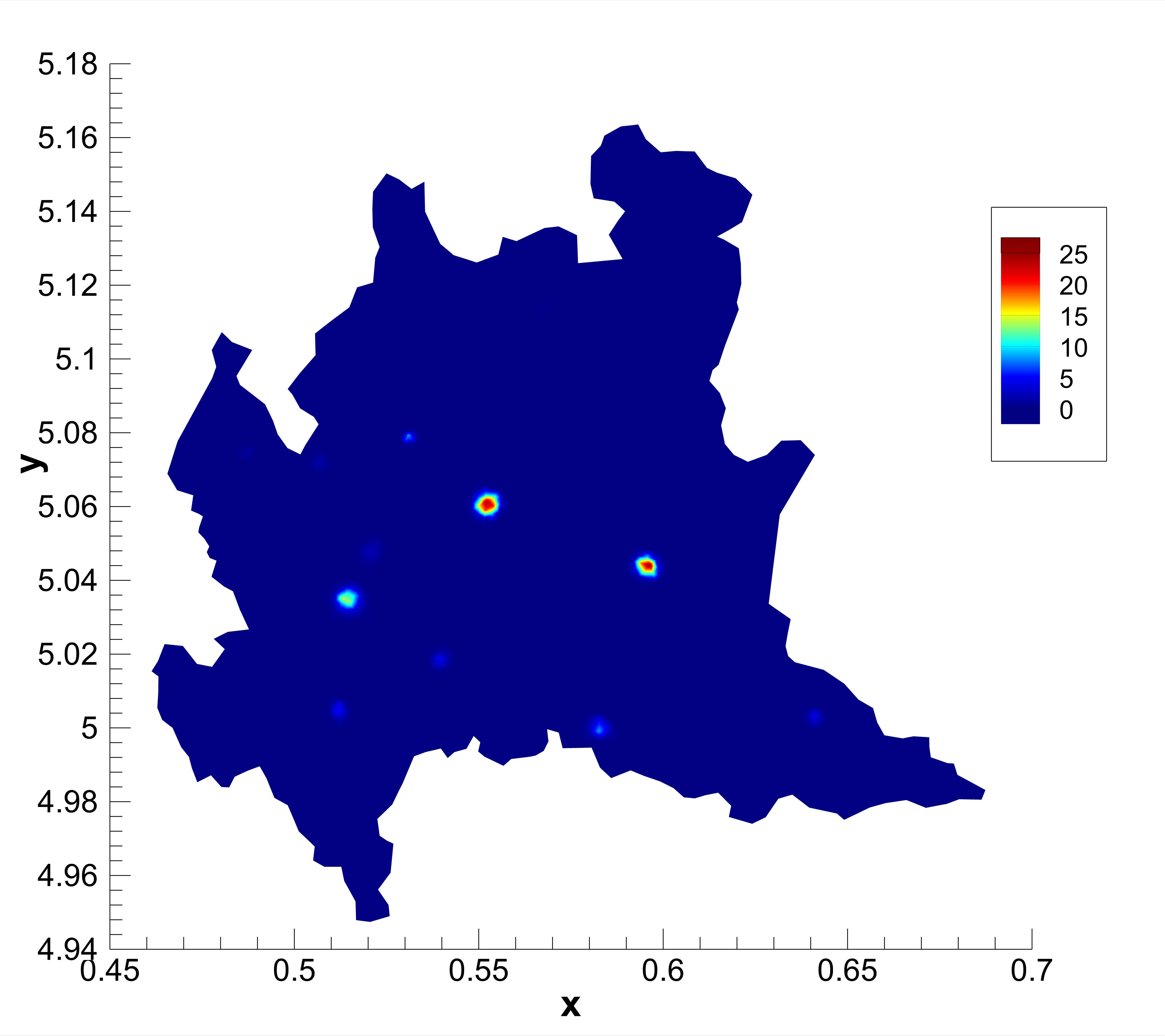}
\caption{Var[$E_T(x,y)+I_T(x,y)+A_T(x,y)$]}
\label{fig.varEIA_final}
\end{subfigure}
\begin{subfigure}{0.48\textwidth}
\includegraphics[width=1\linewidth]{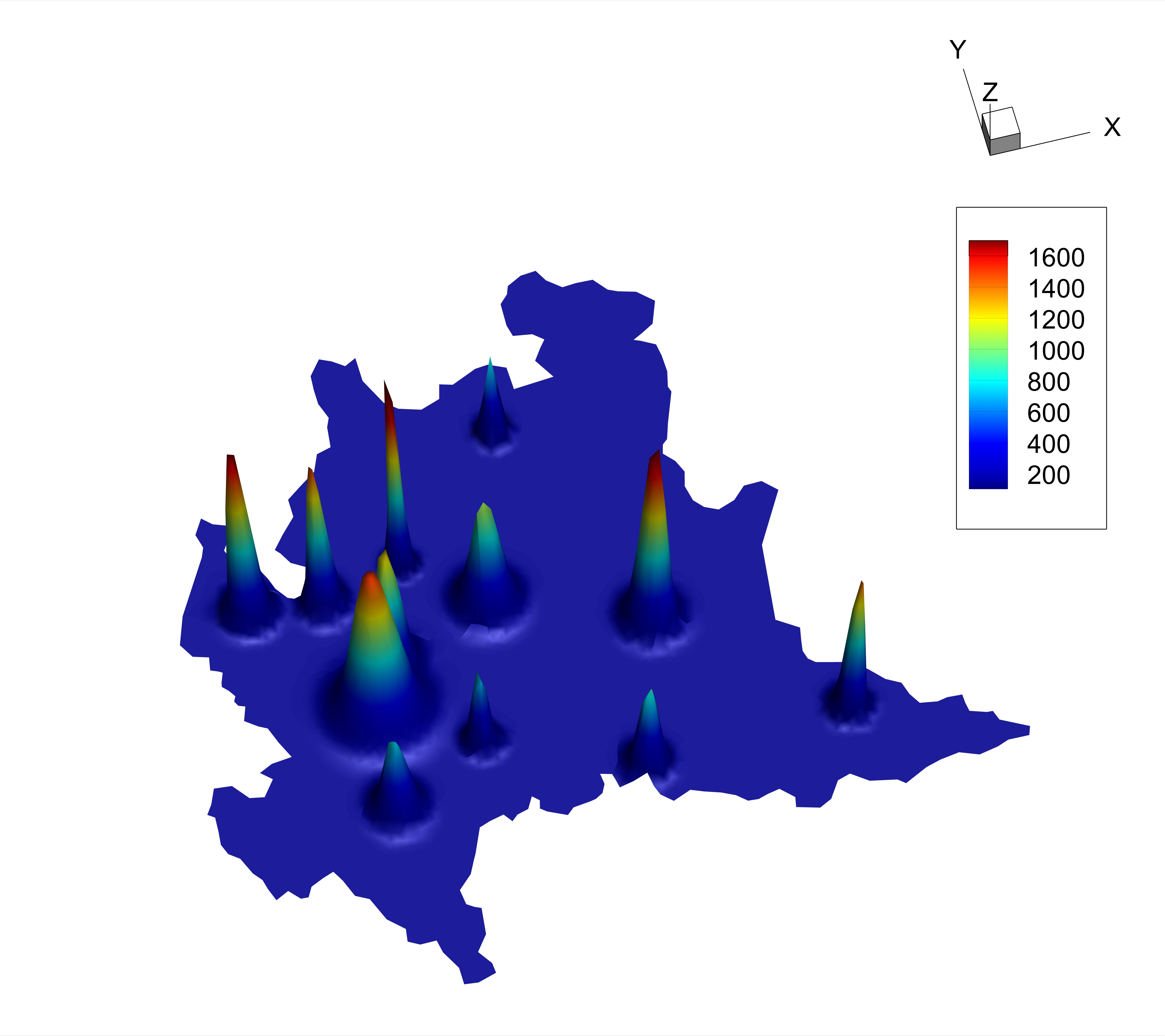}
\caption{Exp[$S_{T,0}(x,y)$]}
\label{fig.expS_IC}
\end{subfigure}
\begin{subfigure}{0.48\textwidth}
\includegraphics[width=1\linewidth]{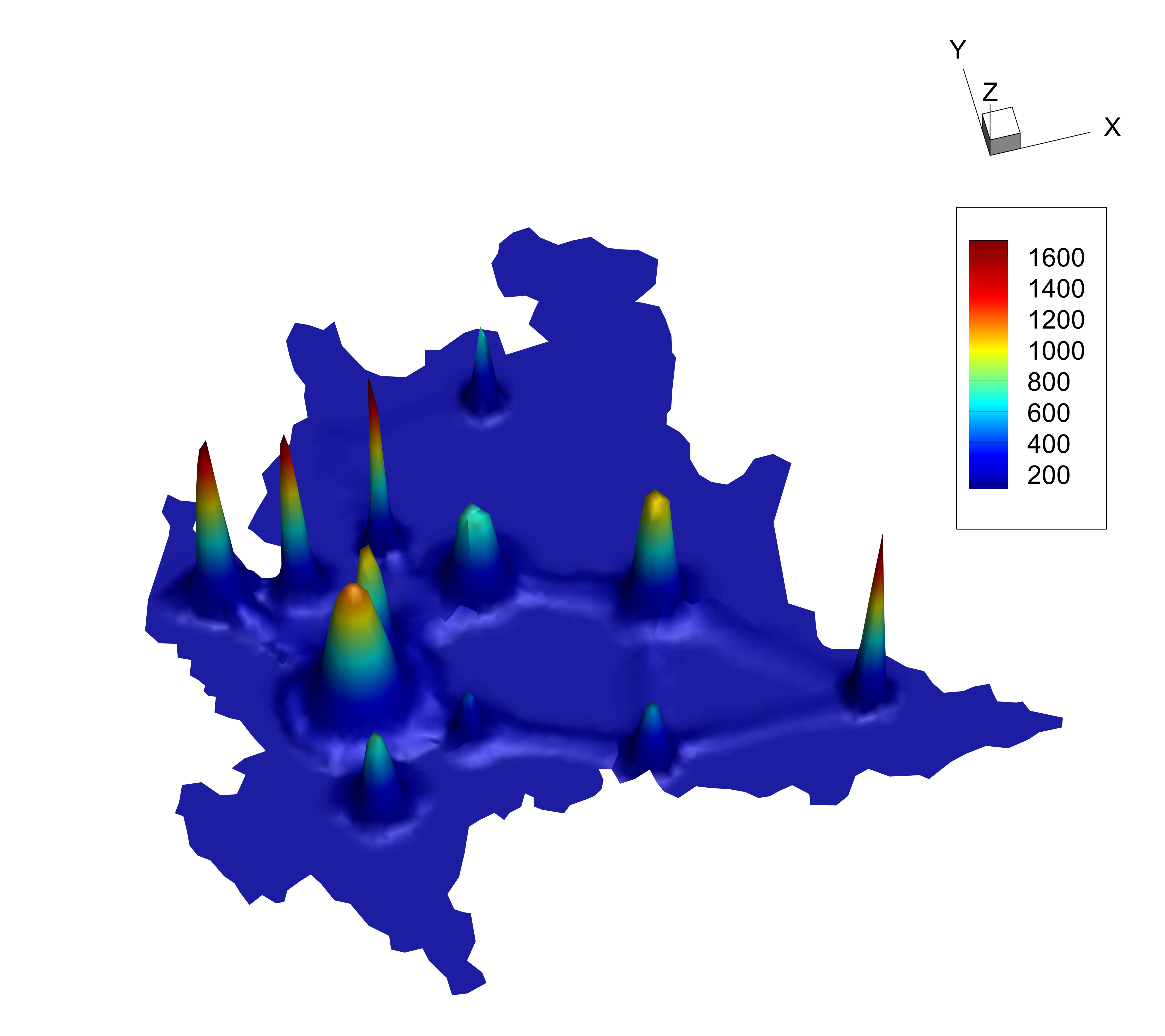}
\caption{Exp[$S_{T}(x,y)$]}
\label{fig.expS_final}
\end{subfigure}
\caption{Numerical results of the simulation of the first outbreak of COVID-19 in Lombardy, Italy. Top: expectation (a) and variance (b) of the cumulative amount of infected people $E_T+A_T+I_T$ at the end of the simulation (March 22, 2020). Bottom: expectation of the susceptible population $S_T$ on the initial day simulated (February 27, 2020) (c) and at the end of the simulation (March 22, 2020) (d).}
\label{fig.results_2D}
\end{figure}

\subsection{Results and discussion}
Numerical results of the test are reported in Figures~\ref{fig.results_Lombardy}--\ref{fig.results_2D}.
In Figure~\ref{fig.results_Lombardy}, the expected evolution in time of the infected individuals, together with 95\% confidence intervals, is shown for exposed $E$, highly symptomatic subjects $I$ and asymptomatic or weakly symptomatic people $A$, for each city of the Lombardy Region. Here it is already appreciable the heterogeneity of the diffusion of the virus. Indeed, from the different y-axis scales adopted for the plot of the provinces, it can be noticed that Milan, Bergamo and Brescia present a consistently higher contagion.

From Figure \ref{fig.results_Lombardy_RI} it can be observed that the lower bound of the confidence interval of the cumulative amount in time of highly symptomatic individuals is in line with data reported by the Civil Protection Department of Italy \cite{prot_civile}. As expected, due to the uncertainty taken into account, the mean value of the numerical result in each city is higher than the registered one.

The comparison between the expected evolution in time of the cumulative amount of severe infectious with respect to the effective cumulative amount of total infectious people, including asymptomatic and mildly symptomatic individuals, is shown in Figure \ref{fig.results_Lombardy_RIA}. From this figure it is clear that the number of infections recorded during the first outbreak of COVID-19 in Lombardy represents a clear underestimation of the actual trend of infection suffered by the Region and by Italy as a whole, and how the presence of asymptomatic subjects, not detected, has affected the pandemic evolution. 

Numerical results presented in terms of integrated variables for the whole Lombardy are shown in Figure \ref{fig.results_Lombardy_region}, together with the expected temporal evolution of the reproduction number $R_0(t)$, again with 95\% confidence bands. It is here highlighted that the drop of $R_0(t)$ on March 9 reflects the imposition of lockdown restrictions, as presented in the previous Section. Moreover, in this plot, results are reported starting from February 28, instead of February 27, to permit to the system to achieve a correct initialization of the commuters (who are totally placed in the origin location at the beginning of the simulation) during the first day simulated. 

In Figure \ref{fig.results_2D}, final expectation and variance of the cumulative amount of infected people, namely $E_T+A_T+I_T$, are reported in the 2D framework (first row). If comparing Figure \ref{fig.expEIA_final} with \ref{EIA}, it can be noticed that, at the end of March, the virus is no longer majorly affecting the province of Lodi and Cremona, but has been spread arriving to hit most of all Brescia, Milan and Bergamo. Finally, in the second row of Figure \ref{fig.results_2D}, the expectation of the susceptible population $S_T$ on the initial day simulated (February 27, 2020) is compared with the one obtained at the end of the simulation (March 22, 2020). Here it can be verified that the majority of the population does not leave their home city, as per real behavior, but there is only a small percentage of commuters who move within the domain, along the prescribed routes. Similar results are obtained for compartments $E$ and $A$, whose commuting part, even though small, strongly contributes to the spatial spread of the epidemic.

\section{Conclusions}\label{Conc}
In this paper we introduced a realistic model for the spatial spread of a virus with a focus on the case of COVID-19. Unlike models currently in the literature, which typically ignore spatial details or alternatively introduce them as simple diffusive dynamics, in our approach we have tried to capture the essential characteristics of the movements of individuals, which are very different if we consider commuting individuals who for work reasons move over long distances, from one city to another, to individuals who instead carry out their activities on an urban scale. The separation of individuals into these two classes, and the use of different spatial dynamics characterized by appropriate systems of transport and diffusive equations, allows in particular to avoid mass migration phenomena typical of models based on a single population and the instantaneous propagation of infectious disease over long distances. From the epidemiological point of view, modeling is developed in a compartmental context described by a SEIAR-type framework capable of describing the effect of exposed and asymptomatic individuals within the spatial spread of the disease. In addition, given the high uncertainty on the actual amount of individuals in the various compartments able to propagate the infection, the model was developed taking into account the presence of stochastic variables that therefore require an appropriate process of quantification. The resulting multiscale system of partial differential equations was then solved on realistic spatial geometries by a numerical method combining finite volume IMEX techniques for the deterministic part, with a non-intrusive collocation approach for the stochastic component. After a careful calibration of the model parameters based on the available data, an in-depth analysis of the results is reported in the case of the initial phase of the spread of COVID-19 in Italy occurred in the Lombardy region. The results show the ability of the model to correctly describe the epidemic dynamics and the importance of a heterogeneous spatial description and of the inclusion of stochastic parameters.

\section*{Acknowledgments} 
This work has been written within the
activities of the GNCS group of INdAM (National Institute of
High Mathematics). The support of MIUR-PRIN Project 2017, No. 2017KKJP4X “Innovative numerical methods for evolutionary partial differential equations and applications” is acknowledged. 

\appendix
\section{Supplementary material}
In this appendix, we report the details of the numerical scheme adopted to approximate the MK-SEIAR system \eqref{eq:kineticc}-\eqref{eq:diffuse} together with the data tables concerning the details of the population and of the commuting flows.

\subsection{Numerical method}
\label{appendix:NM}
We first give the details of the method in the case in which uncertainty is not present and successively we explain how the system  \eqref{eq:kineticc}-\eqref{eq:diffuse} is solved in the case of stochasticity. For the commuters, the numerical scheme for the deterministic case is based on a discrete ordinate method in velocity in which the even and odd parity formulation is employed \cite{DP,JPT,Klar}. The details of such approach are given in \ref{eo}. Then a finite volume method working on two-dimensional unstructured meshes \cite{BD-BGK,ArepoTN} for the discrete ordinate approximation of the commuters is introduced in \ref{fv}. The full discretization of the equations \eqref{eq:kineticc} is obtained through the use of suitable IMEX Runge-Kutta schemes \cite{Bos1, Bos2}. In particular, the above choices permit to obtain a scheme which consistently captures the diffusion limit from the kinetic system when the scaling parameters $\tau_{S,I,R}$ tends toward zero. This part of the method is discussed in \ref{RK}. Finally, the discretization of the stochastic part for the system \eqref{eq:kineticc}-\eqref{eq:diffuse} when uncertainty is present is explained in \ref{UQ}.

\subsubsection{Even and odd parities formulation}\label{eo}
We rewrite \eqref{eq:kineticc} by using the so-called even and odd parities formulation. In other words, we denote $v=(\eta,\xi)\in\mathbb{S}^1$ and then we obtain four equations with non-negative $\xi, \eta \geq 0$ for each compartment of the commuters. The change of variables reads, omitting the time and space dependence for simplicity, as \cite{JPT}
\[
\begin{split}
	r^{(1)}_i(\xi,\eta) &= \frac12(f_i(\xi,-\eta)+f_i(-\xi,\eta)),\quad 
	r^{(2)}_i(\xi,\eta) = \frac12(f_i(\xi,\eta)+f_i(-\xi,-\eta))\\
\end{split}
\]
while for the scalar fluxes one has
\[
\begin{split}
	j^{(1)}_i(\xi,\eta) &= \frac{\lambda_i}{2} (f_i(\xi,-\eta)+f_i(-\xi,\eta)),\quad 
	j^{(2)}_i(\xi,\eta) = \frac{\lambda_i}{2}  (f_i(\xi,\eta)+f_i(-\xi,-\eta))
\end{split}
\]
with $i=S,E,I,A,R$. An equivalent formulation with respect to \eqref{eq:kineticc} can then be obtained thanks to this change of variables and reads as 
\begin{equation}
	\begin{split}
		\frac{\partial r^{(1)}_S }{\partial t} +  \xi \frac{\partial j^{(1)}_S}{\partial x}-\eta \frac{\partial j^{(1)}_S}{\partial y} &= -F_I(r^{(1)}_S, I_T)-F_A(r^{(1)}_S, A_T) +\frac1{\tau_S}\left(S-r^{(1)}_S \right)\\
		\frac{\partial r^{(2)}_S }{\partial t} +  \xi \frac{\partial j^{(2)}_S}{\partial x}+\eta \frac{\partial j^{(2)}_S}{\partial y} &= -F_I(r^{(2)}_S, I_T)-F_A(r^{(2)}_S, A_T)+\frac1{\tau_S}\left(S-r^{(2)}_S \right)\\
		\frac{\partial r^{(1)}_E }{\partial t} +  \xi \frac{\partial j^{(1)}_E}{\partial x}-\eta \frac{\partial j^{(1)}_E}{\partial y} &= F_I(r^{(1)}_S, I_T)+F_A(r^{(1)}_S, A_T)-ar^{(1)}_E +\frac1{\tau_E}\left(E-r^{(1)}_E \right)\\
		\frac{\partial r^{(2)}_E }{\partial t} +  \xi \frac{\partial j^{(2)}_E}{\partial x}+\eta \frac{\partial j^{(2)}_E}{\partial y} &= F_I(r^{(2)}_S, I_T)+F_A(r^{(2)}_S, A_T)-ar^{(2)}_E +\frac1{\tau_E}\left(E-r^{(2)}_E \right)\\
		\frac{\partial r^{(1)}_I }{\partial t} +  \xi \frac{\partial j^{(1)}_I}{\partial x}-\eta \frac{\partial j^{(1)}_I}{\partial y} &= a\sigma r^{(1)}_E-\gamma_I r^{(1)}_I+\frac1{\tau_I}\left(I-r^{(1)}_I \right)\\
		\frac{\partial r^{(2)}_I }{\partial t} +  \xi \frac{\partial j^{(2)}_S}{\partial x}+\eta \frac{\partial j^{(2)}_I}{\partial y} &= a\sigma r^{(2)}_E-\gamma_I r^{(2)}_I+\frac1{\tau_I}\left(I-r^{(2)}_I \right)\\
		\frac{\partial r^{(1)}_A }{\partial t} +  \xi \frac{\partial j^{(1)}_A}{\partial x}-\eta \frac{\partial j^{(1)}_A}{\partial y} &= a(1-\sigma) r^{(1)}_E-\gamma_A r^{(1)}_A+\frac1{\tau_A}\left(A-r^{(1)}_A \right)\\
		\frac{\partial r^{(2)}_A }{\partial t} +  \xi \frac{\partial j^{(2)}_A}{\partial x}+\eta \frac{\partial j^{(2)}_A}{\partial y} &= a(1-\sigma) r^{(2)}_E-\gamma_A r^{(2)}_A+\frac1{\tau_A}\left(A-r^{(2)}_A \right)\\
		\frac{\partial r^{(1)}_R }{\partial t} +  \xi \frac{\partial j^{(1)}_R}{\partial x}-\eta \frac{\partial j^{(1)}_R}{\partial y} &= \gamma_I r^{(1)}_I+\gamma_A r^{(1)}_A+\frac1{\tau_R}\left(R-r^{(1)}_R \right)\\
		\frac{\partial r^{(2)}_R }{\partial t} +  \xi \frac{\partial j^{(2)}_R}{\partial x}+\eta \frac{\partial j^{(2)}_R}{\partial y} &= \gamma_I r^{(2)}_I+\gamma_A r^{(2)}_A+\frac1{\tau_R}\left(R-r^{(2)}_R \right)
	\end{split}
	\label{eq:eop1}
\end{equation}
and 
\begin{align}
\nonumber
		\frac{\partial j^{(1)}_S}{\partial t} + \vs^2 \xi \frac{\partial r^{(1)}_S}{\partial x} -\vs^2 \eta \frac{\partial r^{(1)}_S}{\partial y} &= - F_I(j^{(1)}_S, I_T)- F_A(j^{(1)}_S, A_T) -\frac1{\tau_S} j^{(1)}_S\\
		\nonumber
		\frac{\partial j^{(2)}_S}{\partial t} + \vs^2 \xi \frac{\partial r^{(2)}_S}{\partial x} +\vs^2 \eta \frac{\partial r^{(2)}_S}{\partial y} &= - F_I(j^{(2)}_S, I_T)- F_A(j^{(1)}_S, A_T) -\frac1{\tau_S} j^{(2)}_S\\
		\nonumber
		\frac{\partial j^{(1)}_E}{\partial t} + \ve^2 \xi \frac{\partial r^{(1)}_E}{\partial x} -\ve^2 \eta \frac{\partial r^{(1)}_E}{\partial y} &= \frac{\lambda_E}{\lambda_S}\left(F_I(j^{(1)}_S, I_T)+ F_A(j^{(1)}_S, A_T)\right) -a j^{(1)}_E -\frac1{\tau_E} j^{(1)}_E\\
		\nonumber
		\frac{\partial j^{(2)}_E}{\partial t} + \ve^2 \xi \frac{\partial r^{(2)}_E}{\partial x} +\ve^2 \eta \frac{\partial r^{(2)}_E}{\partial y} &=\frac{\lambda_E}{\lambda_S}\left( F_I(j^{(2)}_S, I_T) +F_A(j^{(2)}_S, A_T)\right)- a j^{(2)}_E -\frac1{\tau_E} j^{(2)}_E\\
		\nonumber
		\frac{\partial j^{(1)}_I}{\partial t} + \vi^2 \xi \frac{\partial r^{(1)}_I}{\partial x} -\vi^2 \eta \frac{\partial r^{(1)}_I}{\partial y} &= \frac{\vi}{\ve}a\sigma j^{(1)}_E-\gamma_I j^{(1)}_I  -\frac1{\tau_I} j^{(1)}_I\\
		\nonumber\\[-.5cm]
		\label{eq:eop2}\\[-.5cm]
		\nonumber
		\frac{\partial j^{(2)}_I}{\partial t} + \vi^2 \xi \frac{\partial r^{(2)}_I}{\partial x} +\vi^2 \eta \frac{\partial r^{(2)}_I}{\partial y} &= \frac{\vi}{\ve}a\sigma j^{(2)}_E-\gamma_I j^{(2)}_I  -\frac1{\tau_I} j^{(2)}_I.\\
	\nonumber
		\frac{\partial j^{(1)}_A}{\partial t} + \va^2 \xi \frac{\partial r^{(1)}_A}{\partial x} -\va^2 \eta \frac{\partial r^{(1)}_A}{\partial y} &= \frac{\va}{\ve}a(1-\sigma) j^{(1)}_E-\gamma_A j^{(1)}_A -\frac1{\tau_A} j^{(1)}_A\\
		\nonumber
		\frac{\partial j^{(2)}_A}{\partial t} + \va^2 \xi \frac{\partial r^{(2)}_A}{\partial x} +\va^2 \eta \frac{\partial r^{(2)}_A}{\partial y} &= \frac{\va}{\ve}a(1-\sigma) j^{(2)}_E-\gamma_A j^{(2)}_A  -\frac1{\tau_A} j^{(2)}_A\\
		\nonumber
		\frac{\partial j^{(1)}_R}{\partial t} + \vr^2 \xi \frac{\partial r^{(1)}_R}{\partial x} -\vr^2 \eta \frac{\partial r^{(1)}_R}{\partial y} &= \frac{\lambda_R}{\lambda_I}\gamma_I j^{(1)}_I+\frac{\lambda_R}{\lambda_A}\gamma_A j^{(1)}_A  -\frac1{\tau_R} j^{(1)}_R\\
		\nonumber
		\frac{\partial j^{(2)}_R}{\partial t} + \vr^2 \xi \frac{\partial r^{(2)}_R}{\partial x} +\vr^2 \eta \frac{\partial r^{(2)}_R}{\partial y} &= \frac{\lambda_R}{\lambda_I}\gamma_I j^{(2)}_I+\frac{\lambda_R}{\lambda_A}\gamma_A j^{(2)}_A -\frac1{\tau_R} j^{(2)}_R.
\end{align}
One can observe that due to symmetry, we need to solve these equations for $\xi$, $\eta$ in the positive quadrant only. Thus the number of unknowns in \eqref{eq:kineticc} and \eqref{eq:eop1}-\eqref{eq:eop2} is effectively the same. 

\subsubsection{Space discretization on unstructured grids}\label{fv}
We consider now a spatial two-dimensional computational domain $\Omega$ which is discretized by a set of non overlapping polygons $P_i, i=1, \dots N_p$. 
Each element $P_i$ exhibits an arbitrary number $N_{S_i}$ of edges $e_{j,i}$ where the subscripts indicates that this is the edge shared by elements $P_i$ and $P_j$. The boundary of the cell is consequently given by 
	$\partial P_i = \bigcup \limits_{j=1}^{N_{S_i}}{e_{ji}}$. 
The governing equations for the commuters rewritten in the odd and even formulation are then discretized on the unstructured mesh by means of a finite volume scheme which is conveniently rewritten in condensed form as 
\begin{equation}
	\label{PDE}
	\frac{\partial \Q}{\partial t} + \nabla_x \cdot \F(\Q) = \S(\Q), \qquad (x,y) \in \Omega \subset \mathds{R}^2, \quad t \in \mathds{R}_0^+, 
\end{equation} 
where $\Q$ is the vector of conserved variables
\begin{equation*}
	\Q = \left( r_i^{(1)} , \,  r_i^{(2)} ,  j_i^{(1)} , \,  j_i^{(2)}
	\right)^\top, \quad i=S,E,I,A,R
\end{equation*}
while $\F(\Q)$ is the linear flux tensor and $\S(\Q)$ 
represents the stiff source term defined in equations \eqref{eq:eop1}-\eqref{eq:eop2}.
As usual for finite volume schemes, data are represented by spatial cell averages, which are defined at time $t^n$ as  
\begin{equation}
	\Q_i^n = \frac{1}{|P_i|} \int_{P_i} \Q(\x,t^n) \, d\x,     
	\label{eqn.cellaverage}
\end{equation}  
where $|P_i|$ denotes the surface of element $P_i$. 
A first order in time finite volume method is then obtained by integration of the governing system \eqref{PDE} over the space control volume $|P_i|$
\begin{equation}
	\Q_i^{n+1} = \Q_i^n - \frac{\Delta t}{|P_i|}\sum \limits_{P_j \in \mathcal{N}_{S_i}} \,\, {\int \limits_{e_{ij}} 
		\F^n_{ij} \cdot \mathbf{n}_{ij}  \, d\mathbf{l}}
	+ \int \limits_{P_i} \S_i^{n} \, d\mathbf{x}. 
	\label{PDEfinal}
\end{equation} 
Higher order in space is then achieved by substituting the cell averages by piecewise high order polynomials. We refer to these polynomial reconstructions to as $\mathbf{w}_i(\x)$ which are obtained from the given cell averages \eqref{eqn.cellaverage}. In particular, our choice is to rely on a second order Central WENO (CWENO) reconstruction procedure along the lines of \cite{BD-BGK, ADER-CWENO}. We omit the details for brevity. The numerical flux function $\mathbf{F}_{ij} \cdot \mathbf{n}_{ij}$ is given by a simple and robust local Lax-Friedrichs flux yielding
\begin{equation}
	\mathbf{F}_{ij} \cdot \mathbf{n}_{ij} =  
	\frac{1}{2} \left( \F(\mathbf{w}_{i,j}^+) + \F(\mathbf{w}_{i,j}^-)  \right) \cdot \mathbf{n}_{ij}  - 
	\frac{1}{2} s_{\max} \left( \mathbf{w}_{i,j}^+ - \mathbf{w}_{i,j}^- \right),  
	\label{eqn.rusanov} 
\end{equation} 
where $\mathbf{w}_{i,j}^+,\mathbf{w}_{i,j}^-$ are the high order boundary extrapolated data evaluated through the CWENO reconstruction procedure. The numerical dissipation is given by $s_{\max}$ which is the maximum eigenvalue of the Jacobian matrix in spatial normal direction,  
\begin{equation}
	\label{eq:An}
	\mathbf{A}_{\mathbf{n}} = \frac{\partial \F}{\partial \Q}.
\end{equation}

Let us notice now that in the diffusion limit, i.e. as $(\tau_S,\tau_I,\tau_R)\to 0$, the source term $\S(\Q)$ becomes stiff, therefore in order to avoid prohibitive time steps we need to discretize the commuters system implicitly. To this aim, a second order IMEX method which preserves the asymptotic limit given by the diffusion equations \eqref{eq:diff} is proposed and briefly described hereafter.

\subsubsection{Time integration and numerical diffusion limit}\label{RK}
We consider again system \eqref{eq:kineticc} formulated using the parities \eqref{eq:eop1}-\eqref{eq:eop2}. We also assume $\tau_{S,I,R}=\tau$ and rewrite \eqref{eq:eop1}-\eqref{eq:eop2} in partitioned form as
\begin{equation}
	\begin{split}
		\frac{\partial \u}{\partial t} + \frac{\partial \f(\v)}{\partial x} + \frac{\partial \g(\v)}{\partial y} &= \E(\u) + \frac1{\tau}\left(\U-\u\right)\\
		\frac{\partial \v}{\partial t} + \boldsymbol{\Lambda}^2 \frac{\partial \f(\u)}{\partial x} +\boldsymbol{\Lambda}^2 \frac{\partial \g(\u)}{\partial y} &= \E(\v) - \frac1{\tau}\v,
	\end{split}
	\label{systcompactform}
\end{equation}
in which 
\begin{equation}
	\begin{split}
		&\u = \left(r_S^{(1)}, r_S^{(2)}, r_E^{(1)}, r_E^{(2)}, r_I^{(1)}, r_I^{(2)}, r_A^{(1)}, r_A^{(2)}, r_R^{(1)}, r_R^{(2)}\right)^T, \\
		&\v =\left(j_S^{(1)},  j_S^{(2)}, j_E^{(1)}, j_E^{(2)}, j_I^{(1)}, j_I^{(2)}, j_A^{(1)}, j_A^{(2)}, j_R^{(1)}, j_R^{(2)}\right)^T,\\ 
		&\f(\v) = \xi \v,\quad
		\g(\v) = \eta \J \v,\quad \J={\rm diag}\{-1,1,-1,1,-1,1,-1,1,-1,1\},\\
		&\E(\u)=\left(-F_I(r_S^{(i)},I_T)-F_A(r_S^{(i)},A_T), F_I(r_S^{(i)},I_T)+F_A(r_S^{(i)},A_T)-a r_E^{(i)}, a\sigma r_E^{(i)}-\gamma_Ir_I^{(i)},\right.\\
		&\left.a(1-\sigma)r_E^{(i)}-\gamma_Ar_A^{(i)},\gamma_Ir_I^{(i)}
		\gamma_Ar_A^{(i)}\right)^T,\ i=1,2\\
		&\U=\left(S,S,E,E,I,I,A,A,R,R\right)^T,\quad 
		\boldsymbol{\Lambda} ={\rm diag}\{\lambda_S,\lambda_S,\lambda_E,\lambda_E,\lambda_I,\lambda_I,\lambda_A,\lambda_A,\lambda_R,\lambda_R\},
	\end{split}
	\label{eq:variables}
\end{equation}
and $\f(\u)$, $\g(\u)$, $\E(\v)$ are defined similarly.
Now, following \cite{Bos2}, the Implicit-Explict Runge-Kutta (IMEX-RK) approach appid to system \eqref{systcompactform} reads as
\begin{equation}
	\begin{split}
		&\u^{(k)} = \u^n -  \Delta t \sum_{j=1}^{k} a_{kj} \left(\frac{\partial \f(\v^{(j)})}{\partial x} + \frac{\partial \g(\v^{(j)})}{\partial y}-\frac1{\tau}\left(\U^{(j)}-\u^{(j)}\right)\right) + \Delta t \sum_{j=1}^{k-1} \tilde{a}_{kj} \E\left(\u^{(j)}\right)
		\\
		&\v^{(k)} = \v^n -  \Delta t \sum_{j=1}^{k-1} \tilde{a}_{kj} \left(\boldsymbol{\Lambda}^2 \frac{\partial \f(\u^{(j)})}{\partial x} +\boldsymbol{\Lambda}^2 \frac{\partial \g(\u^{(j)})}{\partial y}-\E(\v^{(j)})\right)  + \Delta t \sum_{j=1}^{k} a_{kj} \frac1{\tau}\v^{(j)},
	\end{split}
	\label{eq.iterIMEX}
\end{equation}
where $\u^{(k)}, \v^{(k)}$ are the so-called internal stages. The  numerical solution reads
\begin{equation}
	\begin{split}
		&\u^{n+1} = \u^n -  \Delta t \sum_{k=1}^{s} b_{k} \left(\frac{\partial \f(\v^{(k)})}{\partial x} + \frac{\partial \g(\v^{(k)})}{\partial y}-\frac1{\tau}\left(\U^{(k)}-\u^{(k)}\right)\right) + \Delta t \sum_{k=1}^{s} \tilde{b}_{k} \E\left(\u^{(k)}\right)
		\\
		&\v^{n+1} = \v^n -  \Delta t \sum_{k=1}^{s} \tilde{b}_{k} \left(\boldsymbol{\Lambda}^2 \frac{\partial \f(\u^{(k)})}{\partial x} +\boldsymbol{\Lambda}^2 \frac{\partial \g(\u^{(k)})}{\partial y}-\E(\v^{(k)})\right)  + \Delta t \sum_{k=1}^{s} b_{k} \frac1{\tau}\v^{(k)}.
	\end{split}
	\label{eq.finalIMEX}
\end{equation}
In the above equations, the matrices $\tilde A = (\tilde a_{kj})$, with $\tilde a_{kj} = 0 $ for $ j\geq k$, and $A = (a_{kj})$, with $a_{kj} = 0 $ for $ j > k$ are $s \times s$ matrices, with $s$ number of Runge-Kutta stages, defining respectively the explicit and the implicit part of the scheme, and vectors $\tilde b = (\tilde b_1, ...,\tilde b_s)^T$ and $b = (b_1, ...,b_s)^T$ are the quadrature weights. 
Furthermore, we choose the Runge-Kutta scheme in such a way that the following relations hold true 
\begin{equation}\label{gsa}
	a_{kj} = b_j, \qquad j = 1,\ldots,s ,\qquad
	\tilde a_{kj} = \tilde b_j, \qquad j = 1,\ldots,s-1. 
\end{equation}

The scheme \eqref{eq.iterIMEX}-\eqref{eq.finalIMEX} treats implicitly the stiff terms and explicitly all the rest. Moreover, one can prove that the above scheme is a consistent discretization of the limit system in the diffusive regime. In fact, assuming for simplicity $D_{S,I,R}$ independent from space, the second equation in \eqref{eq.iterIMEX} can be rewritten as
\[
\tau\v^{(k)} = \tau\v^n -  \Delta t \sum_{j=1}^{k-1} \tilde{a}_{kj} \left(\tau\boldsymbol{\Lambda}^2 \frac{\partial \f(\u^{(j)})}{\partial x} +\tau\boldsymbol{\Lambda}^2 \frac{\partial \g(\u^{(j)})}{\partial y}-\tau \E(\v^{(j)})\right)  + \Delta t \sum_{j=1}^{k} a_{kj} \v^{(j)},
\]
therefore, assuming \eqref{eq:diffcf}, the limit $\tau\to 0$ yields
\begin{equation}
	\sum_{j=1}^{k} a_{kj} \v^{(j)} = \sum_{j=1}^{k-1} \tilde{a}_{kj} \left(2\D \frac{\partial \f(\U^{(j)})}{\partial x} +2\D \frac{\partial \g(\U^{(j)})}{\partial y}\right),
	\label{eq:vasym}
\end{equation}
where $\D={\rm diag}\left\{D_S,D_S,D_E,D_E,D_I,D_I,D_A,D_A,D_R,D_R\right\}$ and where we used the fact that from the first equation in \eqref{eq.iterIMEX} as $\tau\to 0$ we have $\u^{(j)}=\U^{(j)}$. Note also that \eqref{eq:vasym} implies that $j_{S,E,I,A,R}^{(1)}=j_{S,E,I,A,R}^{(2)}$ in $\v^{(j)}$, i.e. we restore perfect symmetry in direction of propagation of the information. Using now the identity $\u^{(j)}=\U^{(j)}$ into the first equation in \eqref{eq.iterIMEX} we get 
\begin{equation}
	\U^{(k)} = \U^n -  \Delta t \sum_{j=1}^{k} a_{kj} \left(\frac{\partial \f(\v^{(j)})}{\partial x} + \frac{\partial \g(\v^{(j)})}{\partial y}\right) + \Delta t \sum_{j=1}^{k-1} \tilde{a}_{kj} \E\left(\U^{(j)}\right),
	\label{eq:uasym}	
\end{equation}
and using \eqref{eq:vasym} into \eqref{eq:uasym} thanks to the definitions of $\f$ and $\g$ gives
\begin{equation}
	\begin{split}
		\U^{(k)} =& \U^n -  2\Delta t\D \sum_{j=1}^{k-1} \tilde{a}_{kj} \left(\xi^2 \frac{\partial^2 \U^{(j)}}{\partial x^2} + 2\xi\eta\J \frac{\partial^2 \U^{(j)}}{\partial x \partial y} + \eta^2 \frac{\partial^2 \U^{(j)}}{\partial y^2}\right)\\
		&+ \Delta t \sum_{j=1}^{k-1} \tilde{a}_{kj} \E\left(\U^{(j)}\right).
	\end{split}
	\label{eq:uasym2}	
\end{equation}
Finally, integrating over the velocity field one has
\begin{equation}
	\begin{split}
		S^{(k)} =& S^n -  \Delta t D_S \sum_{j=1}^{k-1} \tilde{a}_{kj} \left(\frac{\partial^2 S^{(j)}}{\partial x^2} + \frac{\partial^2 S^{(j)}}{\partial y^2}\right)- \Delta t \sum_{j=1}^{k-1} \tilde{a}_{kj} F(S^{(j)},I_T^{(j)}),\\
		I^{(k)} =& I^n -  \Delta t D_I \sum_{j=1}^{k-1} \tilde{a}_{kj} \left(\frac{\partial^2 I^{(j)}}{\partial x^2} + \frac{\partial^2 I^{(j)}}{\partial y^2}\right)+ \Delta t \sum_{j=1}^{k-1} \tilde{a}_{kj} \left(F(S^{(j)},I_T^{(j)})-\gamma I^{(j)}\right),\\
		R^{(k)} =& R^n -  \Delta t D_R \sum_{j=1}^{k-1} \tilde{a}_{kj} \left(\frac{\partial^2 R^{(j)}}{\partial x^2} + \frac{\partial^2 R^{(j)}}{\partial y^2}\right)+ \Delta t \sum_{j=1}^{k-1} \tilde{a}_{kj} \gamma I^{(j)}
	\end{split}
	\label{eq:RKdiffuse}	
\end{equation}
and thus, the internal stages correspond to the stages of the explicit scheme applied to the reaction-diffusion system \eqref{eq:diff}. To conclude the proof one has to notice that thanks to the choice \eqref{gsa}, the last stage is equivalent to the numerical solution. Thus, this is enough to guarantee that the scheme is a consistent discretization of the limit equation.

Note that the limit system is consistent with the discretization of the non commuters diffusive system \eqref{eq:diffuse}. In this latter case, we adopt the same finite volume setting for the unknowns
\begin{equation*}
	\Q^{u} = \left(\SO,\EO,\IO,\AO,\RO
	\right)^\top,
\end{equation*}
This simply reads
\begin{equation}
	\label{PDE1}
	\frac{\partial \Q^{u}}{\partial t} + \nabla_x \cdot \F^{u}(\Q^{u}) = \S^{u}(\Q^{u}), \qquad (x,y) \in \Omega \subset \mathds{R}^2, \quad t \in \mathds{R}_0^+, 
\end{equation} 
with \begin{equation*}
	\F^{u} = \left( \begin{array}{c}  -D^u_S \, (\SO)_x \\ -D^u_E \, (\EO)_x \\ -D^u_I \, (\IO)_x \\-D^u_A \, (\AO)_x \\-D^u_R \, (\RO)_x \end{array} \right.  
	\left. \begin{array}{c}  -D^u_S \, (\SO)_y \\-D^u_E \, (\EO)_y \\-D^u_I \, (\IO)_y \\ -D^u_A \, (\AO)_y \\ -D^u_R \, (\RO)_y \end{array} \right), \qquad 
	\S^{u} = \left( \begin{array}{c}  -F_I(\SO, I_T)-F_A(\SO, A_T) \\ F_I(\SO, I_T)+F_A(\SO, A_T) - a \EO \\a\sigma \EO -\gamma_I \IO\\a(1-\sigma) \EO -\gamma_A \AO\\\gamma_I \IO +\gamma_A \AO \end{array} \right).
\end{equation*}
Then, the same CWENO reconstruction and the same numerical local Lax-Friedrichs flux is employed where however, we account for a dissipation proportional to the diffusive terms. In other words, the numerical viscosity is given by the maximum eigenvalue of the viscous operator $s_{max}^V=\max \left( D^u_S, D^u_E, D^u_I,D^u_A,D^u_R \right)$. Concerning the time discretization, the explicit part of the Runge-Kutta scheme introduced in the previous paragraph is employed.

\subsubsection{Stochastic collocation method}\label{UQ}
In the case in which we deal with the stochastic system \eqref{eq:kineticc}-\eqref{eq:diffuse}, we employ a generalized Polynomial Chaos (gPC) expansion technique \cite{JHL,X}. We restrict to the case in which there is only one stochastic variable $z$ in the system. The extension to the case of a vector of random variables is straightforward. The probability density function of the single random input is supposed known: $\rho_z: \Gamma \rightarrow \mathbb{R}^+$. In this case, the approximate solution for the commuters $\Q_M(x,v,t,z)$ and the non commuters $\Q_M^{u}(x,t,z)$ are represented as truncations of the series of the orthonormal polynomials describing the random space, i.e.
\begin{equation}\label{eq:expansion}
	\Q_M(x,v,t,z)=\sum_{j=1}^{M} \hat{\bf{Q}}_{j}(x,v,t) \phi_j(z),\quad \Q^{u}_M(x,t,z)=\sum_{j=1}^{M} \hat{\bf{Q}}^{u}_{j}(x,t) \phi_j(z) 
\end{equation}
where $M$ is the number of terms of the truncated series and $\phi_j(z)$ are orthonormal polynomials, with respect to the measure $\rho_z(z)\, d z$. The expansion coefficients are obtained by
\begin{equation}
	\label{eq:exp_coeff_int}
	\hat{\bf{Q}}_{j}(x,v,t) = \int_{\Gamma} {\bf{Q}}  (x,v,t,z)\, \phi_j(z)\, \rho_z(z)\, d z,\ \hat{\bf{Q}}^{u}_{j}(x,t) = \int_{\Gamma} {\bf{Q}}^{u}  (x,t,z)\, \phi_j(z)\, \rho_z(z)\, d z, \ j=1,\ldots,M.
\end{equation}
Then, the exact integrals for the expansion coefficients in Eq.~\eqref{eq:exp_coeff_int} are replaced by a suitable quadrature 
formula characterized by the set $\{z_n, w_n \}_{n=1}^{N_p}$, where $z_n$ is the $n$-th collocation point, $w_n$ is the corresponding weight and $N_p$ represents the number of quadrature points. We then have
\begin{equation}
	\label{eq:exp_coeff_quad}
	\hat{\bf{Q}}_{j}(x,v,t) \approx \sum_{n=1}^{N_p} {\bf{Q}}(x,v,t,z_n)\, \phi_j(z_n)\, w_n,\ \hat{\bf{Q}}^{u}_{j}(x,t) \approx \sum_{n=1}^{N_p} {\bf{Q}}^{u}(x,t,z_n)\, \phi_j(z_n)\, w_n, \ j=1,\ldots,M
\end{equation}
where ${\bf{Q}}(x,v,t,z_n)$ and ${\bf{Q}}^{u}(x,t,z_n)$ with $n=1,\ldots,N_p$ are the solutions of the problem evaluated at the $n$-th collocation point for the commuters and non commuters. Thanks to the computation of the above coefficients than 
it is possible to compute all quantities of interest concerning the random variable. For example, the expectations 
are approximated as
\begin{equation}
	\label{eq:mean_apx}
	\mathbb{E}\left[{\bf{Q}}\right] \approx \mathbb{E}\left[{\bf{Q}}_M\right] =\int_{\Gamma} {\bf{Q}}_M(x,v,t,z)\, \rho_z(z)\, d z \approx \sum_{n=1}^{N_p}  {\bf{Q}}(x,v,t,z_n)\, w_n,
\end{equation}
and
\begin{equation}
	\label{eq:mean_apx2}
	\mathbb{E}\left[{\bf{Q}}^{u}\right] \approx \mathbb{E}\left[{\bf{Q}}^{u}_M\right] =\int_{\Gamma} {\bf{Q}}_M^{u}(x,t,z)\, \rho_z(z)\, d z \approx \sum_{n=1}^{N_p}  {\bf{Q}}^{u}(x,t,z_n)\, w_n .
\end{equation}
In the same way, all other quantities of interest such as the variance of $\Q$ and $\Q^{u}$ can be computed.

\subsection{Population and mobility data}
\label{appendix:tables}
In this appendix we report tables containing data on the initial distribution of populations in the various urban areas considered in the Lombardy region, see Table \ref{tab:IC_lombardia}, and data on the corresponding flows of commuters between cities, see Table \ref{tab:matrix_lombardia}.    

\begin{table}[h!] 
\caption{Setting of the Lombardy provinces: urban radius $r_c$, total inhabitants $M$ and initial amount of highly infectious individuals $I_{T,0}$, detected by February 27, 2020 (initial day of the simulation). The total population is given by ISTAT data of December 31, 2019 \cite{ISTATdemo}, while data of highly infectious correspond to those reported in the GitHub repository of the Civil Protection Department of Italy \cite{prot_civile}. Null values, listed with $^*$, in the simulation are substituted with 1 to permit an effective assignation of uncertain initial condition.} \label{tab:IC_lombardia} 
\centering
\vspace{0.2cm}
\begin{tabular}{ c | c c c }
Province &Urban radius $r_c$ [km] &Total population $M$ &Infectious $I_{T,0}$ \\
\hline
Pavia (PV) &3.24 &540376 &36\\
Lodi (LO) &2.04 &227412 &159 \\
Cremona (CR) &2.40 &355908 &91\\
Mantua (MN) &1.92  &406919 &0$^*$\\
Milan (MI) &5.76 &3265327 &15 \\
Bergamo (BG) &3.96 &1108126 &72 \\
Brescia (BS) &3.24 &1255437 &10\\
Varese (VA) &2.76 &884876 &0$^*$ \\
Monza-Brianza (MB) &3.24 &870193 &5\\
Como (CO) &2.40 &597642 &0$^*$\\
Lecco (LC) &3.24 &334961 &0$^*$\\
Sondrio (SO) &1.56 &180425 &3 \\
\hline
\end{tabular} 
\end{table} 
\begin{sidewaystable}
\caption{Matrix of commuters of the Lombardy Region (Italy). Departure provinces are listed on the first left column, while arrival provinces are reported in the following columns. Each entry is given as number of commuters of the origin province. The last column shows the amount of total commuters of the origin province and the corresponding percentage with respect to the total population of the city. This matrix is extracted from the origin-destination matrix provided by the Lombardy Region for the regional fluxes of year 2020 \cite{opendataLombardia}.} 
\label{tab:matrix_lombardia} 
\centering
\footnotesize
\vspace{0.2cm}
\begin{tabular}{ c | c c c c c c c c c c c c || c}
From/To &PV &LO &CR &MN &MI &BG &BS &VA &MB &CO &LC &SO &Total\\
\hline
PV &-- &9601 &--  &--  &83825 &-- &-- &-- &-- &-- &-- &-- &93426 (17.3\%)\\
LO &9169 &-- &13712 & &56717 &-- &-- &-- &-- &-- &-- &-- &79598 (35.0\%)\\
CR &-- &13264 &-- &11654 &23142  &12025  &17681 &-- &-- &-- &-- &-- &77766 (21.8\%) \\
MN &-- &-- &-- &1157 &--  &2267  &-- &22142 &-- &-- &-- &-- &35980 (8.8\%) \\
MI &82617 &55397 &21622 &1946 &-- &74168  &26709 &144681 &234682 &41575 &20801 &1000  &705198 (21.6\%) \\
BG &-- &-- &12016 &-- &76337  &-- &78348 &-- &14826 &-- &17611 &--   &199138 (18.0\%)\\
BS &-- &-- &16967 &21643 &26594  &70879  &-- &-- &-- &-- &-- &--   &136083 (10.8\%)\\
VA &-- &-- &-- &-- &143152  &--  &-- &-- &-- &33529 &-- &--   &176681 (20.0\%)\\
MB &-- &-- &-- &-- &247183  &14938  &-- &-- &-- &-- &37761 &--   &299882 (34.5\%)\\
CO &-- &-- &-- &-- &44412  &--  &-- &36249 &-- &-- &-- &--   &80661 (13.5\%)\\
LC &-- &-- &-- &-- &23621  &19392 &-- &-- &40317 &-- &-- &4851   &88181 (26.3\%)\\
SO &-- &-- &-- &-- &1227  &-- &-- &-- &-- &-- &4545 &--   &5772 (3.2\%)\\
\hline
\end{tabular} 
\end{sidewaystable} 

\end{document}